\documentclass[12pt]{article}
\pdfoutput=1
\usepackage{mheck}

\newcommand{\rem}[1]{} % Auskommentieren von Absaetzen
\def\C{\mathbb{C}}
\def\Z{\mathbb{Z}}

\def\R{\mathbb{R}}

\def\P{\mathbb{P}}

\def\Hirz[#1]{\mathbbm{F}_{#1}}
\def\o[#1]{\overline{#1}}

\newcommand{\hookuparrow}{\mathrel{\rotatebox[origin=c]{90}{$\hookrightarrow$}}}

\makeatletter
\newcommand\xleftrightarrow[2][]{%
  \ext@arrow 9999{\longleftrightarrowfill@}{#1}{#2}}
\newcommand\longleftrightarrowfill@{%
  \arrowfill@\leftarrow\relbar\rightarrow}
\makeatother

\newcommand{\Tt}{\mathcal{T}_3}
\newcommand{\Tf}{\mathcal{T}_4}

\institution{OXFORD}{\ Mathematical Institute, University of Oxford, Woodstock Rd, Oxford OX2 6GG, UK }
\institution{SCGP}{\ Simons Center for Geometry and Physics, SUNY, Stony Brook, NY, 11794-3636 USA}

\title{Towards Generalized Mirror Symmetry for Twisted Connected Sum $G_2$ Manifolds}
\authors{Andreas P. Braun \worksat{\OXFORD} \footnote{e-mail: {\tt andreas.braun@maths.ox.ac.uk}} and Michele Del Zotto \worksat{\SCGP} \footnote{e-mail: {\tt mdelzotto@scgp.stonybrook.edu}}}
\abstract{We revisit our construction of mirror symmetries for compactifications of Type II superstrings on twisted connected sum $G_2$ manifolds. For a given $G_2$ manifold, we discuss evidence for the existence of mirror symmetries of two kinds: one is an autoequivalence for a given Type II superstring on a mirror pair of $G_2$ manifolds, the other is a duality between Type II strings with different chiralities for another pair of mirror manifolds. We clarify the role of the B-field in the construction, and check that the corresponding massless spectra are respected by the generalized mirror maps. We discuss hints towards a homological version based on BPS spectroscopy. We provide several novel examples of smooth, as well as singular, mirror $G_2$ backgrounds via pairs of dual projecting tops. We test our conjectures against a Joyce orbifold example, where we reproduce, using our geometrical methods, the known mirror maps that arise from the SCFT worldsheet perspective. Along the way, we discuss non-Abelian gauge symmetries, and argue for the generation of the Affleck-Harvey-Witten superpotential in the pure SYM case. }

\begin{document}

\maketitle

\tableofcontents

\section{Introduction}

One of the most important features of superstring theory and M/F-theory are string dualities, which relate compactified string theories to one another, often making the very concept of space-time geometry ambiguous.\footnote{ This is similar to what happens in the context of field theory dualities, where often the notion of the strength of a coupling is made ambiguous (e.g. weakly coupled electric phases are dual to strongly coupled magnetic ones and vice versa).} The subject of string dualities has been vastly explored in the past two decades, and most results obtained in this context about toroidal compactifications and Calabi-Yau (CY) geometries are by now textbook material. While supersymmetric backgrounds with exceptional holonomy $G_2$ and $Spin(7)$ have been explored in the past,\footnote{ See e.g. \cite{Acharya:2004qe} for a nice review.} the corresponding stringy dualities are much less explored --- with the notable exceptions of few seminal works in the subject, that include e.g. \cite{Shatashvili:1994zw,Acharya:1996fx,Figueroa-OFarrill:1996tnk,Acharya:1997rh,Acharya:2000gb,Gukov:2002jv,Roiban:2002iv,Gaberdiel:2004vx,deBoer:2005pt}. One reason for this is the lack of supersymmetry for the corresponding models --- only very recently it has been shown that these backgrounds survive $\alpha^\prime$ corrections \cite{Becker:2014rea}. A further reason was the lack of examples of compact manifolds with holonomy $G_2$ and $Spin(7)$.

Recently, about 50 million novel examples of compact $G_2$ manifolds have been obtained as twisted connected sums (TCS) of pairs of asymptotically cylindrical CY three-folds\cite{MR2024648,MR3109862,Corti:2012kd}. This is a result with interesting implications in the context of compactifications of superstring theories and M-theory  \cite{Halverson:2014tya,Halverson:2015vta,Braun:2017ryx,Guio:2017zfn,Braun:2017uku}. In particular, this opens the question of the behaviour of these novel backgrounds with respect to stringy dualities. While some steps in these directions have been taken recently \cite{Braun:2017ryx,Braun:2017uku}, there is still lots of work to be done in this subject. The main purpose of this paper is to revisit and extend our exploration of the generalized mirror symmetry for Type II compactifications on TCS $G_2$ backgrounds \cite{Braun:2017ryx}.

\begin{figure}
\begin{center}
\includegraphics[scale=0.7]{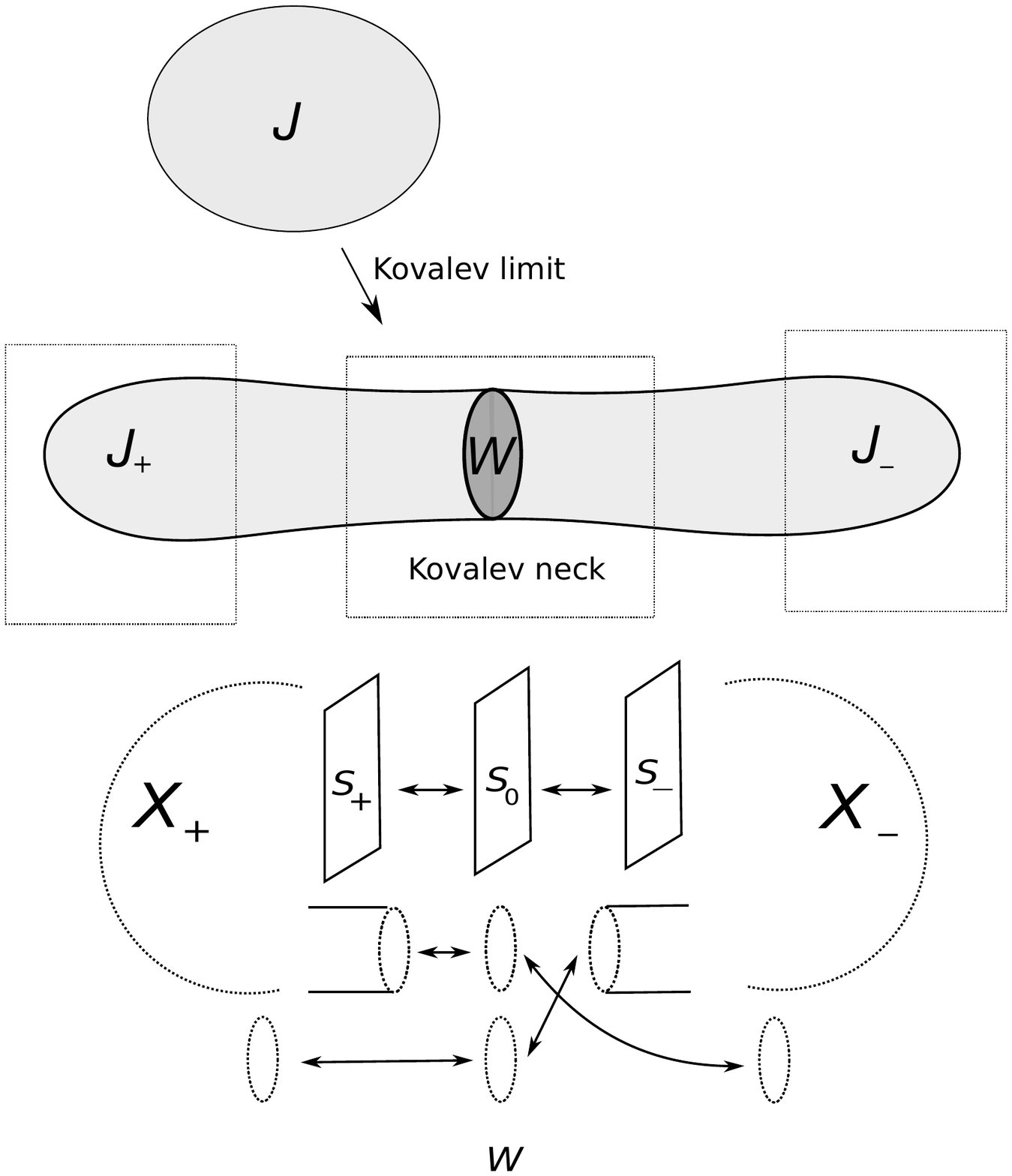}
\end{center}
\caption{A schematic description of the Kovalev limit for a $G_2$ manifold $J$. The K3 surfaces in the asymptotic cylinders of the two CYs $X_\pm$ are denoted $S_\pm$. The K3 surface in $W$ is denoted $S_0$.}\label{Fig:Kovalev}
\end{figure}

Aspects of stringy dualities have been understood for the Joyce orbifolds of $T^7$ and $T^8$\cite{joyce1996I,joyce1996II,joyce2000compact}. However, it is well-known that toroidal orbifolds enjoy many more special properties than generic supersymmetric backgrounds. Hence, the question of which among the observed stringy dualities were consequences of the Joyce examples being toroidal orbifolds, rather than just $G_2$ or $Spin(7)$ holonomy, is to answer.

Conjecturally, $G_2$ backgrounds can exhibit two types of generalized mirror symmetries which are close cousins of the more familiar mirror symmetries of CY compactifications. We denote the corresponding dualities $\Tt$ and $\Tf$, as these can be obtained in the same spirit of the SYZ mirror symmetry, by performing fiberwise $T$-dualities along calibrated associative and coassociative toroidal fibrations \cite{Acharya:1997rh}. The $\Tt$ duality gives an equivalence between Type IIA and Type IIB (and vice versa), while the $\Tf$ duality gives an equivalence between Type II strings of the same Type.

\begin{figure}
\begin{center}
\includegraphics[scale=0.8]{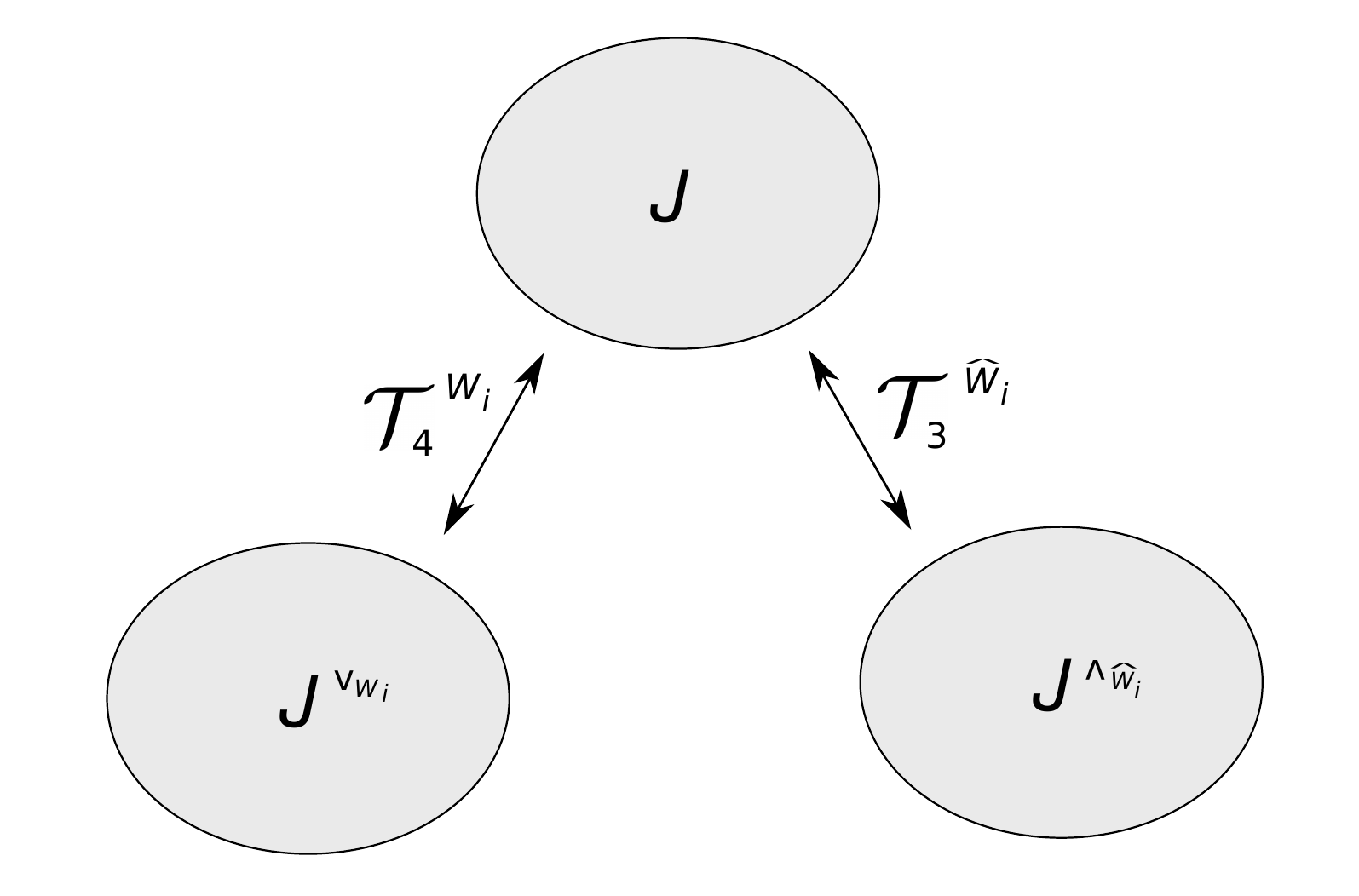}
\end{center}
\caption{Network of generalized $G_2$ mirror symmetry maps obtained from T-dualities.}\label{fig:wow}
\end{figure}

One of the main results of this paper is that TCS $G_2$ backgrounds can exhibit both types of behaviors.\footnote{ In our previous work on the subject only the map $\Tf$ case was addressed.} In order to better explain our results, let us briefly describe the TCS construction --- a more detailed review is found in Section \ref{sec:TCS_review}. A given $G_2$ manifold $J$ can be realized as a TCS if it admits a Kovalev limit. In a Kovalev limit, the manifold $J$ is stretched along one real direction developing a long neck that is locally isomorphic to a product of a K3, a torus and an interval $I$ (see Figure \ref{Fig:Kovalev}). Along the Kovalev neck, the $G_2$ manifold $J$ can be cut open along a codimension one hypersurface, denoted $W$ in the Figure, which is isomorphic to $K3 \times T^2$. We obtain a pair of manifolds with boundary, denoted by $J_+$ and $J_-$, such that $\partial J_\pm \simeq W$. In the limit of infinite Kovalev neck $J_+$ and $J_-$ are well approximated by asymptotically cylindrical CY three-folds, denoted by $X_+$ and $X_-$, times a circle. Recall that a given CY three-fold is said to be asymptotically cylindrical if outside of a compact submanifold it is isomorphic to a K3 surface times a cylinder. Crucially, the complex structures on the two asymptotic K3 surfaces $S_\pm$ induced from the complex structures on $X_\pm$ are different and related by a hyper-K\"ahler rotation $\Xi$, called a matching in the literature, that is induced by identifying $S_\pm$ with $S_0$, the K3 component of $W$. In order to preserve the $G_2$ structure of $J$, also the A and B cycles of the two-tori in the asymptotic regions of $J_\pm$ have to be swapped when identified with the $T^2$ component of $W$ (see Figure \ref{Fig:Kovalev}). Conversely, $G_2$ manifolds which allow such a limit can be constructed as a twisted connected sum from appropriate pieces. In particular, one has to provide a pair of asymptotically cylindrical CYs and a matching in between the corresponding asymptotic K3 fibers. Schematically, we denote the $G_2$ manifolds so obtained as
\begin{equation}
J \quad \longrightarrow \quad X_+ \cup_{\, \Xi} X_-
\end{equation}
to remind ourselves that $J$ has a codimension one locus in its moduli space where it can be described as a TCS of $X_+$ and $X_-$ with matching $\Xi$. Sometimes, we use also the notation 
\begin{equation}
J \quad \longrightarrow \quad X_+ \cup_{\, W} X_-\,.
\end{equation}
We find that the $\Tf$ mirror symmetry for TCS $G_2$ backgrounds $J \to X_+ \cup_\Xi X_-$ can be constructed whenever $X_+$ and $X_-$ both have CY mirrors, denoted respectively by $X^\vee_+$ and $X^\vee_-$. We obtain that the $\Tf$ mirror of $J$, denoted by $J^\vee$, is a TCS, 
\begin{equation}
J^\vee \quad \longrightarrow \quad  X^\vee_+ \, \cup_{\, \Xi^\vee} \, X^\vee_-\,,
\end{equation}
where $\Xi^\vee$ is the $\Tf$ mirror matching, obtained from $\Xi$ by K3 mirror symmetry. Similarly, the $\Tt$ mirror of $J$, denoted by $J^\wedge$, can be constructed if $X_-$ has a CY mirror and the K3 which fibers $X_+$ is elliptically fibered. We show that $J^\wedge$ is a TCS too,
\begin{equation}
J^\wedge \quad \longrightarrow \quad X_+ \, \cup_{\, \Xi^\wedge}\, X^\vee_-,
\end{equation}
where $\Xi^\wedge$ is the $\Tt$ mirror matching, obtained from $\Xi$ by K3 mirror symmetry as well.\footnote{ For the sake of simplicity, in this discussion we have dropped some useful technicalities that the interested readers can find in Section \ref{sec:GeneralMIR}.} It is very likely that a given $G_2$ background admits more than one Kovalev limit. If that is the case, to each such Kovalev limit can correspond a duality of $\Tf$ or $\Tt$ type, thus giving rise to a network of dualities, that we schematically represent in Figure \ref{fig:wow}.

The structure of the 3D $\cn=2$ theories corresponding to Type II compactifications on TCS $G_2$ backgrounds is as follows. The two CY sides contribute two subsectors with 3D $\cn=4$ supersymmetry, that are coupled together by a sector with 3D $\cn=2$ supersymmetry. The neck region locally looks like a 3D $\cn=8$ subsector, but there is no trace of $\cn=8$ multiplets left. This happens because the two 3D $\cn=4$ subsectors are coupled to the $\cn=8$ one preserving only a common $\cn=2$ supersymmetry subalgebra. In geometry this corresponds to the fact that the matching preserves only one of the three complex structures of the K3. For IIA, this is a circle reduction of the corresponding M-theory background, studied in \cite{Guio:2017zfn}.

The scalar manifold for a compactification of Type II supergravity on a $G_2$ manifold $J$ is a K\"ahler manifold with complex dimension
\begin{equation}
\text{dim } \mathcal{M}_J = 1 + b_2(J) + b_3(J).
\end{equation}
In the supergravity approximation, the generalized mirror symmetry predicts that 
\begin{equation}\label{eq:SVR}
\text{dim } \mathcal{M}_J = \text{dim } \mathcal{M}_{J^\ast}
\end{equation}
where we denote with $J^\ast$ a $G_2$ mirror of $J$, obtained with either the $\Tt$ or the $\Tf$ map. This is the supergravity analogue of the Shatashvili-Vafa relation \cite{Shatashvili:1994zw}. In our previous work we have showed that the $\Tf$ duality has the Shatashvili-Vafa property in all TCS examples that are built by means of dual pairs of tops \cite{Braun:2016igl}. The same strategy can be applied to the $\Tt$ duality in that context. In this paper, we drop the hypothesis that the building blocks are built from projecting tops, and show in full generality that both the $\Tt$ and $\Tf$ maps, if they exist, preserve the total integer cohomology group (forgetting the grading). Since
\begin{equation}
\dim H^\bullet(J,\mathbb{Z}) = 2 \, \text{dim } \mathcal{M}_J
\end{equation}
this shows that both dualities satisfy the Shatashvili-Vafa relation.

One of the main motivations to revisit our work \cite{Braun:2017ryx} was the confusing prescription that we proposed for the matching of the $B$-field in a TCS $G_2$ background. In this paper, we give the correct prescription to always obtain consistent  supergravity backgrounds, clarifying the role of the $B$-field in this context. This is extremely important because the $B$-field and its powers can be used to calibrate non-trivial even-dimensional cycles on $G_2$ backgrounds, thus giving rise to novel supersymmetric cycles that are neither associative, nor coassociative! Notice that this is a departure from the more familiar CY cases. In particular, in the CY case cycles calibrated by the $B$-field always coincide with holomorphic cycles in geometry, which are calibrated by the K\"ahler form and its powers.

We discuss some aspects of 3D $\cn=2$ non-Abelian gauge dynamics in the context of superstring compactifications on backgrounds with $G_2$ holonomy.  In particular, we discuss how the dynamically generated potential that lifts the Coulomb branch moduli arises in the context of IIA compactifications, reproducing some of the known results in the literature \cite{Affleck:1982as,Katz:1996th,deBoer:1997ka}. We stress that in order to obtain results which are consistent with the gauge theory predictions, the existence of two-cycles calibrated by the $B$-field is key. One of the features that we expect from $\Tt$ and $\Tf$ dualities is that they respect this property of the spectrum. A first step to show that this is indeed the case requires extending our results to the context of $G_2$ backgrounds with singularities.

Certain Joyce orbifolds admit a Kovalev limit,%\footnote{ If a given $G_2$ manifold admits a Kovalev limit its CGN invariant, ${\bar \nu}$, must vanish \cite{Crowley:1211wzk,Crowley:1505wzk}.} 
such as  for instance the $T^7/\Z_2^3$ orbifold discussed in $\S$ 12.2 of \cite{joyce2000compact}. This allows to directly compare our geometric results to the  well-known ones obtained from SCFTs \cite{Shatashvili:1994zw,Acharya:1996fx,Gaberdiel:2004vx}, thus providing a stringent consistency check for our methods. We conclude our paper by providing several examples for all of the above that are not Joyce orbifolds, obtained using the toric methods of \cite{Braun:2016igl}.

There is a caveat to all of our discussions, however: due to the lack of supersymmetry for compactifications of type II strings on $G_2$ manifolds, it is possible that these geometrical T-dualities can be spontaneously broken by quantum effects. It is important to stress that all of our arguments here are geometric in nature, therefore hold only in the large volume limit, where such stringy quantum corrections are negligible. For instance, to test the generalized mirror symmetry maps, it is important to verify that the corresponding K\"ahler potentials, superpotentials, BPS spectra and phase structures match. In this paper we make some first preliminary comments about the BPS spectrum, and the possible homological version of $G_2$ mirror symmetry. We find that the $B_n$-cycles, the $n$-cycles calibrated by the $B$-field, have a role to play here as well.

Vectors in 3D are dual to scalars, as is well-understood in the context of Abelian gauge groups. In the non-Abelian case the effect of this 3D duality is less transparent \cite{Nicolai:2003bp}. Our geometric engineering techniques can be used to shed some light about this interesting topic, and we take some first steps towards this interesting direction.

Another interesting consequence of these mirror symmetries is obtained by considering probe branes in the context of the proposed dualities. It is easy to argue that there are interconnections among these dualities and 3D $\cn=1$ dualities. Consider for instance the $\Tt$ generalized mirror symmetry map. It relates IIA on $J$ to IIB on $J^\wedge$. In particular, notice that by placing a D2-brane along the flat $\mathbb{R}^{2,1}$ component of the ten-dimensional background $\mathbb{R}^{2,1} \times J$, we obtain a worldvolume theory on the brane which has $\cn=1$ in three dimensions. Under the duality this configuration of branes is mapped to a D5-brane in Type IIB which is wrapped along the $T^3$ fiber. The coupling of the gauge theory on the D5 brane is inversely proportional to the volume of such a $T^3$, and as we are shrinking it to zero size, the corresponding 3D theory hence becomes strongly coupled. It is dual, however, to the weakly coupled theory on the stack of D2-branes probing the dual $G_2$ manifold. In contrast, the $\Tf$ generalized $G_2$ mirror symmetry maps relate superstrings of the same Type. Consider the IIA to IIA case. Here, a probe D2-brane is related to a D6-brane wrapping the coassociative $T^4$ fiber and vice versa, and a similar mechanism is in place. The precise nature of such 3D $\cn=1$ dualities along the lines of \cite{Gukov:2002er,Gukov:2002es} is currently under investigation.

It would be also interesting to study what happens in the context of heterotic compactification on TCS dual pairs in light of the recent progresses in this context \cite{delaOssa:2016ivz,delaOssa:2017pqy,delaOssa:2017gjq,Fiset:2017auc}.

\bigskip

The structure of this paper is as follows. In Section \ref{sec:superstring_back} we review the supergravity limit of $G_2$ compactifications, with special emphasis on the massless spectrum. In Section \ref{sec:TCS_review} we review the TCS construction, we discuss the structure of the massless spectra in these examples, and we clarify the role played by the $B$-field. Section \ref{sec:TCS_review} concludes with a brief review of the construction of \cite{Braun:2016igl}. In Section \ref{sec:GeneralMIR} we discuss the details of the $\Tt$ and $\Tf$ maps, we extend (and clarify) the heuristic arguments of \cite{Braun:2017ryx}, we show that the discrete torsion in cohomology is mapped consistently, we derive the Shatashvili-Vafa relation for these examples, and we generalize our construction to singular examples. Section \ref{sec:GeneralMIR} concludes with some speculations about possible generalizations of our construction outside the TCS class. Section \ref{sec:Joyce_ex} is devoted to a detailed study of a Joyce orbifold which can be realized as a TCS. There we compare the results in the SCFT literature with our geometric methods. Several explicit examples in the context of $G_2$ backgrounds obtained via projecting tops \cite{Braun:2016igl} are discussed in Section \ref{sec:EXAMPLS}. In this section we discuss explicit examples of the mirror manifolds which arises from this construction and revisit the examples discussed in \cite{Braun:2017ryx}, emphasizing the role played by $G_2$-manifolds with singularities.

\section{Type II superstrings on $G_2$ manifolds}\label{sec:superstring_back}
\subsection{Massless modes}\label{sec:massless_modz}
In this section we briefly review Type II superstring compactifications on a $G_2$ manifold $J$ in the supergravity approximation. Let us denote the Betti numbers of $J$ by $b_m \equiv b_m(J)$. Both Type IIA and  IIB supergravities compactified on $G_2$ manifolds preserve four supercharges giving rise to 3D $\cn=2$ supersymmetry. Let us split the 10d coordinates as follows 
\begin{equation}
X^M = (y^\mu,x^a) \qquad 0 \leq M \leq 9, \quad 0 \leq \mu \leq 2, \quad\text{and}\quad 1 \leq a \leq 7.
\end{equation}
Let us first consider the NS sector, which is common to both supergravities. The dilaton, the metric, and the $B$-field are the bosonic degrees of freedom. The KK reduction of the metric gives $b_3$ scalars, corresponding to the independent deformations of $g_{ab}$, and the dilaton gives a real scalar in 3D. The $B$-field can be expanded as
\begin{equation}\label{eq:BfieldKK}
B = \sum_{k=1}^{b_2} \omega^{(k)}_2 \beta_k(y)\, ,
\end{equation}
where $\omega^{(k)}_2$ is a basis of harmonic 2-forms for $J$ and the $\phi_k(y)$ are scalars. Notice that we discard the 2-form along ${\mathbb R}^{1,2}$, as well as the modes corresponding to one-forms on $J$ from the expansion of the $B$-field. The reason for the former is that in a $d+1$ spacetime $k$-forms with $k\geq d$ are non-dynamical and the reason for the latter is that the first and the sixth cohomology groups of a  $G_2$ holonomy manifold are trivial.

Let us now consider the RR sector of a IIB background. We find it convenient to consider a democratic formulation and impose the duality on the RR potentials after the reduction \cite{Bergshoeff:2001pv}. The 2-form RR potential $C^{(2)}$ has a mode expansion analogous to the $B$-field:
\begin{equation}\label{eq:C2formKK}
C^{(2)} = \sum_{k=1}^{b_2} \omega^{(k)}_2 \gamma_k(y)
\end{equation}
which gives additional $b_2$ scalars that combines with the $B$-field modes of Equation \eqref{eq:BfieldKK} into $b_2$ complex scalars. In a democratic formulation, one has to include the $C^{(6)}$ RR potential which gives vectors
\begin{equation}
C^{(6)} = \sum_{k=1}^{b_5} \omega^{(k)}_5 \wedge A_{1,(k)} + \dots
\end{equation}
where we have chosen our basis of harmonic forms on $J$ in such a way that $\star_7 \omega^{(k)}_5= \omega^{(k)}_2$, and the $\dots$ represent terms giving rise to higher forms in 3D that we discard. We have to impose the duality relation among the RR potentials
\begin{equation}
\star_{10}  d C^{(2)} = d C^{(6)}\,,
\end{equation}
which, using $\star_{10} = \star_3 \otimes \star_7$, gives
\begin{equation}
\star_3  d \gamma_k = d A_{1,(k)}
\end{equation}
for the massless 3D KK modes. Therefore the scalars $\gamma_k$ are dual to the vectors $ A_{1,(k)}$ obtained reducing the 6-form RR potential.  Consider now the self dual 4-form potential $C^{(4)+}$. It has the following decomposition
\begin{equation}
C^{(4)+} = \sum_{j=1}^{b_4} \left(\omega^{(j)}_4 \rho_j(y) + \omega^{(j)}_3 \wedge \tilde A_{1,(j)}\right) + \dots,
\end{equation}
where the dots indicate higher forms we are discarding, the $\rho_j(y)$ are scalars and the $\tilde A_{1,(j)}$ are 1-forms. The requirement that $F^{(5)} = d C^{(4)+}$ is self-dual implies that the vectors corresponding to the 1-forms $\tilde A_{1,(j)}$ are dual to the scalars $\rho_j(y)$. Indeed,
\begin{equation}
d C^{(4)+} = \sum_{j=1}^{b_4} \left(\omega^{(j)}_4  \wedge d \rho_j(y) + d \omega^{(j)}_3 \wedge \tilde A_{1,(j)}+\omega^{(j)}_3 \wedge d \tilde A_{1,(j)}\right) + \dots,
\end{equation}
and using the fact that $\star_{10} = \star_7 \otimes \star_3$ we get
\begin{equation}
\begin{aligned}
\star_{10}\, d C^{(4)+} &= \sum_{j=1}^{b_4} \left(\star_7\omega^{(j)}_4 \wedge \star_3 d \rho_j(y) + \star_7 \omega^{(j)}_3 \wedge \star_3 d  \tilde A_{1,(j)}\right) + \dots,\\
&= \sum_{j=1}^{b_4} \left(\omega^{(j)}_3 \wedge \star_3 d \rho_j(y) + \omega^{(j)}_4 \wedge \star_3 d  \tilde A_{1,(j)}\right) + \dots,
\end{aligned}
\end{equation}
where we have chosen our basis of harmonic forms on $J$ in such a way that $\star_7 \omega^{(j)}_3= \omega^{(j)}_4$, and the $\dots$ again represent terms giving rise to higher forms in 3D that we discard. Comparing the coefficients of the corresponding harmonic forms, we find that the KK modes associated with $C^{(4)+}$ are dual to each other
\begin{equation}
 \star_3 d  \tilde A_{1,(j)} = d \rho_j\,.
\end{equation}
We can combine the scalars $\rho_j$ with the scalars arising from the deformations of the metric into complex scalars. On top of these modes we have one complex scalar from the axio-dilaton field. Hence we obtain a scalar manifold parametrized by $1+b_2 + b_3$ complex modes.

We now come to the RR sector of a IIA compactification, again working in a democratic formulation. The 3-form potential can be expanded as follows:
\begin{equation}
C^{(3)} = \sum_{j=1}^{b_3} \omega^{(j)}_3 \rho_j(y) + \sum_{k=1}^{b_2} \omega^{(k)}_2 \wedge A_{1, \, (k)} \, ,
\end{equation}
while the 5-form potential gives 
\begin{equation}
C^{(5)} = \sum_{k=1}^{b_5} \omega^{(k)}_5 \, \gamma_k(y) + \sum_{j=1}^{b_4} \omega^{(j)}_4 \wedge \tilde{A}_{1\, (j)} + \dots 
\end{equation}
Now the requirement that these are dual
\begin{equation}
\star_{10} \, d C^{(3)} = d C^{(5)}
\end{equation}
reflects on the KK modes: indeed we obtain
\begin{equation}
\star_3 d \tilde{A}_{1, \, (j)} = d \rho_j\, \qquad  \star_3 d A_{1,\,(k)} = d \gamma_k\,,
\end{equation}
by suitably choosing our basis of harmonic forms. Of course, here we are using that $b_2 = b_5$ and $b_3 = b_4$ for any compact smooth 7-manifold. Hence these RR potentials give rise to $b_2 + b_3$ extra scalars which combine with the modes arising from the $B$-field and the metric giving a total of $b_2 + b_3$ complex scalars. The 1-form potential $C^{(1)}$ gives rise to a vector that can be dualized to a PQ scalar in 3D. The latter combines with the dilaton giving rise to one complex scalar field. Hence, the scalars manifold is $1+b_2 + b_3$ complex dimensional. 

The matter content of the IIA and IIB compactifications at this level looks identical, and one can be lead to conjecture that the compactifications of IIA and IIB on a $G_2$ manifold $J$ give rise to the same 3D $\cn=2$ physics \cite{Papadopoulos:1995da}. In order to establish quantum equivalence, however, the corresponding K\"ahler potentials, superpotentials, BPS spectra and phase structure have to match. This can happen, for instance, if a given $G_2$ manifold is self mirror, e.g. $J^\wedge = J$ --- as has been observed in the context of Joyce orbifolds by \cite{Gaberdiel:2004vx}.

\subsection{Non-Abelian gauge symmetries}\label{sec:nAB_IIA}

\begin{figure}
\begin{center}
\includegraphics[scale=0.5]{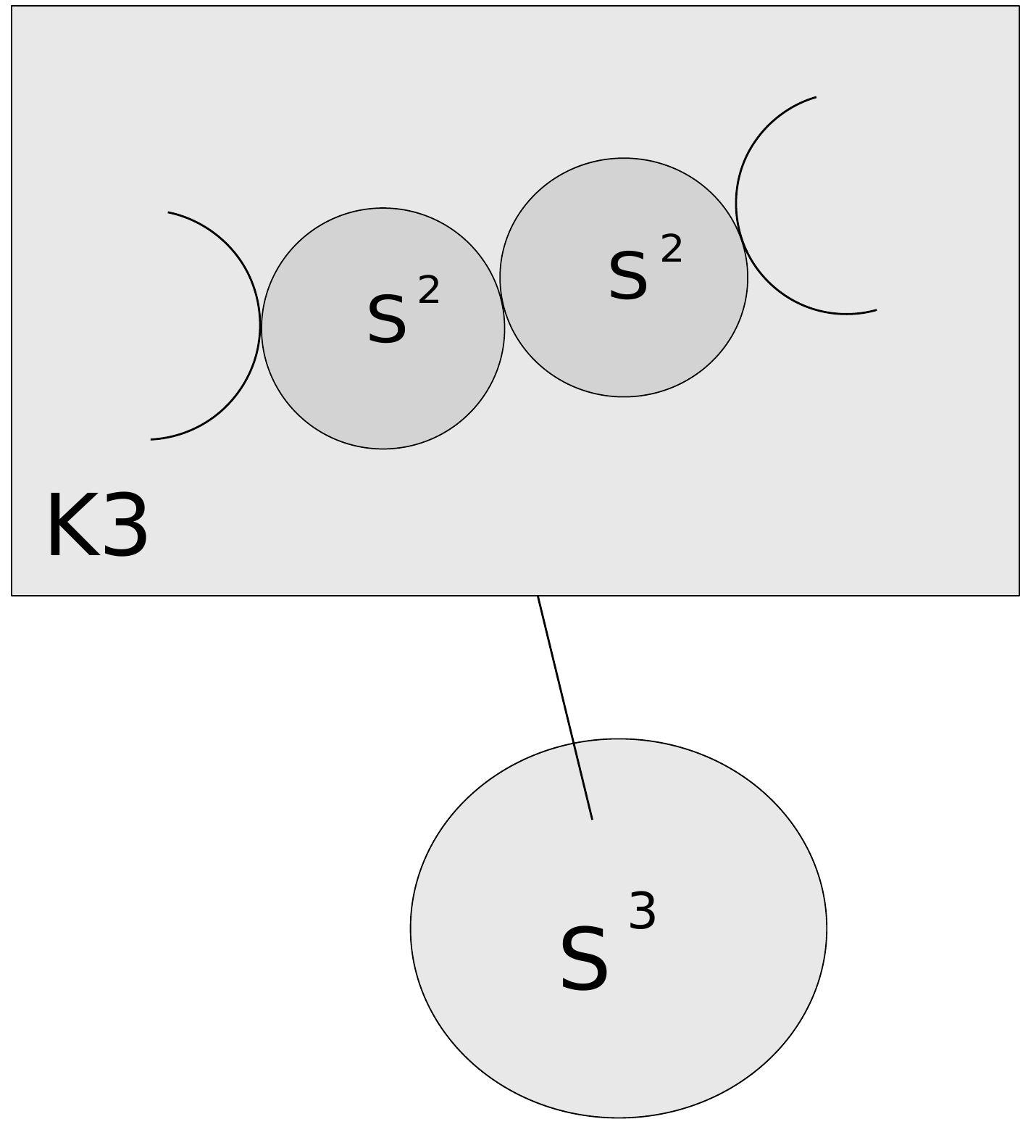}
\end{center}
\caption{Local model corresponding to an ADE gauge subsector in IIA compactifications. In this example, we depict an $A_2$ singularity that corresponds to an $SU(3)$ gauge subsector.}\label{fig:gauge_ordinary}
\end{figure}

It is well-known that in order to obtain a non-Abelian gauge symmetry specific types of singularities are needed in the context of geometric engineering. Let us consider the IIA case. A local model for the geometry that engineers a simple gauge group $G$ is given by a fibration of a $\mathbb{C}^2/\Gamma_G$ singularity over an associative $S^{\, 3}$ cycle.\footnote{ Here $\Gamma_G$ is a discrete subgroup of $SU(2)$, and we are considering the case of simply-laced gauge groups for simplicity. The case of non simply-laced gauge groups can be dealt with in the standard way by fibering the ADE singularity over $S^{\, 3}$ with monodromies that correspond to a folding of the ADE Dynkin diagram to its non-simply laced counterpart \cite{Bershadsky:1996nh}.} This can be argued by a circle reduction of the more familiar M-theory setup where gauge subsectors arise from singularities in real codimension four \cite{Acharya:2000gb,Acharya:2001gy,Witten:2001uq} --- see also \cite{Halverson:2015vta} for a recent discussion. In M-theory one obtains in this way a 4d $\cn=1$ SYM subsector. In particular, the singularity cannot be resolved in the M-theory moduli, which corresponds to the absence of scalars in $\cn=1$ vectormultiplets. In contrast to the M-theory case, however, the Type IIA backgrounds have a Coulomb phase. The additional modes that gives a resolution of the $\C^2/\Gamma$ singularity arise from the $B$-field in type IIA, and correspond to the classical Coulomb phase for the corresponding 3d $\cn=2$ theory. Resolving the ADE singularity $\mathbb{C}^2/\Gamma_G$ one obtains a collection of spheres $(S^{\, 2})_a$, $a=1,...,r\equiv \text{rank}(G)$, intersecting according to the corresponding ADE Dynkin graph (see Figure \ref{fig:gauge_ordinary}). On the resolved geometry, it is always possible to chose a basis of harmonic 2-forms $\omega_2^{(b)}$ as follows
\begin{equation}
\int_{(S^{\, 2})_a} \omega_2^{(b)} \equiv \delta_a^b. 
\end{equation}
The massless modes of the RR-potential $C^{(3)}$ along the $\omega_2^{(b)}$ directions give rise to the Cartan $U(1)_b$ subgroups of the gauge group, here $b=1,...,\text{rank}(G)$. The periods of the $B$-field give rise to the corresponding Coulomb branch scalars
\begin{equation}
\phi_a \sim \int_{(S^{\, 2})_a} B.
\end{equation}
D2-branes wrapped over connected collections of 2-cycles $(S^{\, 2})_a$ provide the massive W-bosons, leading to gauge group enhancement in the limit $\phi_a \to 0$, which is a well-known fact for geometric engineering in IIA, see e.g. the discussion in Section 2 of \cite{Katz:1997eq}. Notice that this geometric engineering setup is also consistent with the fact that pure SYM in 3d $\cn=2$ develops a non-trivial runaway superpotential \cite{Affleck:1982as}. For instance for a gauge group $SU(2)$ the superpotential is expected to be 
\begin{equation}
W \sim \text{exp}\left( - {\phi \over g_{YM}^2}\right)
\end{equation}
where $\phi$ is the Coulomb branch vev and $g_{YM}$ is the YM gauge coupling. A natural candidate to generate such a superpotential term is from wrapped euclidean D4 branes. Each such D4 brane is wrapping $(S^{\, 2})_a \times S^{\, 3}_{base}$ and it contributes to the effective action with the exponential of (minus) a factor
\begin{equation}\label{eq:thefactor}
A_{D4_a} \sim Vol(S^{\, 3}_{base}) \times \left(\int_{(S^{\, 2})_a} B\right) = {1 \over g_{YM}^2} \times \phi_a
\end{equation}
where the volume of the associative $S^{\, 3}$ is inversely proportional to $1/g^2_{YM}$. We claim that the superpotential $W$ receives contributions only from the Euclidean D4 branes wrapping a single $(S^{\, 2})_a$ at a time, because for others configurations of D4-branes wrapping multiple $(S^{\, 2})_a$ there are extra zero modes. We are thus lead to conjecture that
\begin{equation}
W \sim \sum_{a = 1}^r \text{exp} (-A_{D4_a}) = \sum_{a = 1}^r \text{exp} \left(- { \phi_a \over g_{YM}^2}\right),
\end{equation}
which agrees with the results in the literature \cite{Katz:1996th,deBoer:1997ka}. The IIB version of this setup can be worked out using the mirror map. We postpone this discussion to Section \ref{sect:sing_mirror}.

\section{Twisted connected sum $G_2$ backgrounds}\label{sec:TCS_review}

\subsection{Geometry}

\subsubsection{Review of the Kovalev construction}

A TCS $G_2$-holonomy manifold \cite{MR2024648,Corti:2012kd,MR3109862} is obtained from a pair $X_+$ and $X_-$ of CY threefolds which are asymptotically cylindrical. This means that outside of a compact submanifold, $X_\pm$ have the form $\mathbb{R}^+ \times S^{\, 1} \times S_{\pm}$, where $S_{\pm}$ are smooth K3 surfaces.\footnote{ See \cite{MR3109862}, Definition 2.4.} Let 
\begin{equation}
ds_{\pm}^2 = dt_{\pm}^2 + d\theta_{\pm}^2 + dS^{\, 2}_{S_{\pm}}, \qquad \omega_{\pm} =  dt_{\pm} \wedge d\theta_{\pm} + \omega_{S_{\pm}}, \qquad \Omega_{\pm} = (d\theta_{\pm} - i 
dt_{\pm})\wedge \Omega_{S_{\pm}}, 
\end{equation}
denote (respectively) the metric, K\"ahler form, and holomorphic top form on $X_\pm$ in the asymptotically cylindrical region. Here $t_{\pm}$ (resp. $\theta_{\pm}$) is a coordinate along the $\mathbb{R}^+$ (resp. $S^{\, 1}$) direction for $X_{\pm}$. The manifolds 
\begin{equation}
J_\pm \equiv (S^{\, 1})_\pm \times X_\pm
\end{equation}
inherit the standard $G_2$-structures
\begin{equation}\label{eq:PHI34}
\varphi_{\pm} \equiv d \xi_{\pm} \wedge \omega_{\pm} + \text{Re}(\Omega^{3,0}_{\pm}) \qquad \star \varphi_{\pm} \equiv \tfrac{1}{2} \omega_{\pm}^2 - d\xi_{\pm} \wedge \text{Im} (\Omega_{\pm}^{3,0}),
\end{equation} 
where we have denoted by $\xi_{\pm}$ the coordinate of the `outer' $S^{\, 1}$. Let us proceed by describing the construction of the corresponding TCS $G_2$ manifold. Consider the asymptotically cylindrical regions, fix an $\ell >0$ large enough, and let $t_\pm \in (\ell, \ell+1) \subset \mathbb{R}^+$. Denote by $J_{\pm,\ell}$ the submanifolds of $J_\pm$ over such intervals. We want to construct a diffeomorphism of $J_{\pm,\ell}$ that preserves the corresponding $G_2$-structures. The map
\begin{equation}
\Xi_\ell\colon J_{+,\ell}  \longrightarrow J_{-,\ell},
\end{equation}
which in local coordinates is given by
\begin{equation}\label{THADIFFEOH}
\Xi_\ell\colon (\xi_+, t_+ , \theta_+, z^+_1,z^+_2) \mapsto (\theta_-, \ell+1 - t_- , \xi_-, z^-_1,z^-_2), \qquad (z^-_1,z^-_2) \equiv g(z^+_1,z^+_2)
\end{equation}
does the job, provided $g: S_+ \to S_-$ is a hyperk\"ahler rotation, i.e. a diffeomorphism of $K3$ surfaces which induces
\begin{equation}\label{eq:HKrotation}
g^*I_{S_-}= - I_{S_+}, \qquad g^*R_{S_-}=\omega_{S_+}, \qquad g^*\omega_{S_-}=R_{S_+} \, ,
\end{equation}
where
\begin{equation}
I_S \equiv \text{Im}(\Omega_S) \qquad R_S \equiv \text{Re}(\Omega_S).
\end{equation}
The map $g$ in Equation \eqref{eq:HKrotation} is called a matching in \cite{Corti:2012kd}.  Notice that it follows from the definition that $\Xi_\ell$ is a diffeomorphism such that
\begin{equation}\label{eq:glueingdiffeo}
{\Xi_\ell}^*\varphi_- \equiv \varphi_+.
\end{equation}
Truncating both manifolds $J_\pm$ at $t_\pm = \ell+1$ one obtains compact manifolds with boundaries $(T^2)_\pm \times S_{\pm}$ which can be glued via the diffeomorphism $\Xi_\ell$. By Theorem 3.12 of \cite{Corti:2012kd}, for sufficiently large $\ell$, the manifold $J$ so obtained is a $G_2$-holonomy manifold, the twisted connected sum of $X_-$ and $X_+$. The Ricci-flat metric and three-form $\varphi$ of $J$ is a small perturbation of the Ricci-flat metrics on the asymptotically cylindrical Calabi-Yau manifolds $X_\pm$ and the three-forms \eqref{eq:PHI34}. As discussed in the Introduction, we use the following notation\begin{equation}
J \to X_+ \cup_{\, \Xi} X_-
\end{equation}
for a TCS $G_2$-holonomy obtained from $X_{\pm}$ with matching $\Xi$. Often in the discussion below, we are going to write
\begin{equation}
I_{S_-}= - I_{S_+} \qquad R_{S_-}=\omega_{S_+} \qquad \omega_{S_-}=R_{S_+}
\end{equation}
for the matching conditions, because we consider less abstract situations where the matching $\Xi$ is simply obtained from a restriction on the codimension one hypersurface $W$, as we discussed in the Introduction.

\subsubsection{Structure of a TCS Type II superstring background}\label{sec:back_structure}
From the perspective of the geometric engineering, the structure outlined above is reflected in the field content of the corresponding models. The compactification of IIA or IIB on a background of the form $S^{\, 1} \times X_\pm$ gives a 3D $\cn=4$ supersymmetric model, that oxidizes to an ordinary 4D $\cn=2$ CY compactification when the radius of the circle goes to infinity. Vector and hypermultiplet moduli are related in this context by the c-map \cite{Cecotti:1988qn}. These 3D $\cn=4$ subsectors are coupled to a 3D $\cn=8$ subsystem that corresponds to the Kovalev neck with topology $(\ell,\ell+1) \times T^2 \times K3$. The $\cn=8$ symmetry, however, is broken by the coupling with the two 3D $\cn=4$ subsectors in such a way to preserve only a common 3D $\cn=2$ supersymmetry.\footnote{ For IIA, this is a circle reduction of the corresponding M-theory background \cite{Guio:2017zfn}.} This gives rise to a model that has two 3D $\cn=4$ subsectors, that are coupled together by a common 3D $\cn=2$ subsector. The 3D $\cn=2$ modes decouple in the limit of infinite Kovalev neck. This structure is reflected in the massless KK spectrum when looking at the cohomology of TCS $G_2$ manifold in detail, which we do in Section \ref{sec:cohomology} below.

Another interesting angle on these compactifications is the following. The Kovalev neck for the compactified theory can be viewed as a 4D $\cn=4$ theory on an interval, where the matching plays the role of a codimension one defect. The CYs $X_\pm$ at the two ends of the Kovalev neck can be interpreted as end-of-the-world domain walls for such a 4D $\cn=4$ theory, chosen in such a way that the only propagating degrees of freedom are along the remaining three-dimensions and have $\cn=2$ supersymmetry.  For non-compact TCS $G_2$ holonomy backgrounds, the mirror maps we discuss in this paper should have a field theoretical interpretation along these lines. 

\subsection{Cohomology and the massless KK spectrum}\label{sec:cohomology} 
\subsubsection{Mayer-Vietoris argument and $H^\bullet(J,\mathbb{Z})$}
Consider a TCS $G_2$ manifold 
\begin{equation}
J \to  X_+ \cup_{\, \Xi} X_-,
\end{equation}
glued from two pieces $J_\pm = (S^{\, 1})_\pm \times X_\pm$, the intersection of which, $ J_+ \cap J_- = W \times I $, retracts to $W = T^2 \times S$. The integral cohomology of $J$ can be found by exploiting the exactness of the Mayer-Vietoris sequence:
\begin{equation}\label{eq:MVseq}
\begin{aligned}
\cdots\rightarrow &\, H^{m-1}(J) &\rightarrow &\hspace{.5cm} H^{m-1}(J_+) \oplus H^{m-1}(J_-) &\rightarrow &\,H^{m-1}(W) &\rightarrow& \\
\rightarrow &\, H^m(J) &\rightarrow &\hspace{.5cm} H^{m}(J_+) \oplus H^{m}(J_-) &\rightarrow &\,H^{m}(W) &\rightarrow &\,\,\cdots
\end{aligned}
\end{equation}
With the pull backs 
\begin{equation}
\gamma^m = \gamma_+^m \oplus \gamma_-^m :H^m(J_+)\oplus H^m(J_-)\rightarrow H^m(W) \, ,
\end{equation}
the cohomology groups of $J$ are hence given by
\begin{equation}
 H^m(J) =  \mbox{ker}(\gamma^m) \oplus \mbox{coker}(\gamma^{m-1}) \, .
\end{equation}
In order to describe the details of their structure a little bit of extra notation is necessary. Given an asymptotically cylindrical CY $X$ with asymptotic K3 fiber $S$, the embedding $S \hookrightarrow X$ induces a map in cohomology
\begin{equation}\label{eqn:rho}
\rho_X: H^2(X,\mathbb{Z}) \to H^2(S,\Z) \simeq \Gamma^{3,19}\,.
\end{equation}
The lattices
\begin{equation}
N \equiv \text{im } \rho_X, \qquad K \equiv \text{ker } \rho_X, \quad\text{and}\quad T \equiv N^\perp \subset H^2(S,\Z).
\end{equation}
are defined by this restriction map. The lattice $N$ is a sublattice of the Picard lattice of $S$ and in many examples the two are equal; If that is the case, the lattice $T$ is isomorphic to the transcendental lattice of $S$. The lattice $K$ contains those two-forms of $X$ which do not descend from two-forms of $S$. Let us denote the cohomology class of the 3-form $\alpha_S$, which is Poincar\'e dual to the fiber $S \subset W$, by $[\alpha_S]$. As detailed in \cite{Corti:2012kd}, the non-trivial cohomology groups of $J$ are determined by the following isomorphisms:
\begin{equation}\label{eq:bettinumbers}
\begin{aligned}
H^2(J,\mathbb{Z}) & \simeq  N_+ \cap N_- \oplus K_+ \oplus K_- \\
H^3(J,\mathbb{Z}) & \simeq \mathbb{Z}[\alpha_S] \oplus \Gamma^{3,19} /(N_+ + N_-) \oplus (T_+ \cap N_-) \oplus (N_+ \cap T_-)\\
& \hspace{1cm} \oplus H_+  \oplus H_+ \oplus H_- \oplus H_- \oplus K_+ \oplus K_- \\
H^4(J,\mathbb{Z}) & \simeq H^4(S) \oplus (T_+ \cap T_-) \oplus \ \Gamma^{3,19} /(T_+ + N_-) \oplus \Gamma^{3,19} /(N_+ + T_-) \\
& \hspace{1cm} \oplus  H_+ \oplus H_+\oplus H_-  \oplus H_- \oplus K^*_+ \oplus K^*_-
\end{aligned}
\end{equation}
Here, we have abbreviated $H_\pm\oplus H_\pm = H^3(X_+)/T_+$. 
The above is enough to compute all of the integral cohomology groups of $J$. Besides $H^1(J)=H^6(J) = 0$, we have that $b_5 = b_2$ and 
\begin{equation}\label{eq:torcoh}
\mbox{tor}\,  H^3(J,\mathbb{Z}) = \mbox{tor}\, H_2(J,\mathbb{Z})=\mbox{tor}\, H^5(J,\mathbb{Z}),
\end{equation}
so that all of the integral cohomology groups are determined. However, it will be convenient for us to find a different expression for $H^5$ by using \eqref{eq:MVseq}. Defining the restriction homomorphisms 
\begin{equation}
\beta^m_\pm : H^m (X_\pm) \rightarrow H^m(S \times S^{\, 1}) = H^m(S) \oplus H^{m-1}(S)\, ,
\end{equation}
the maps $\gamma^4$ and $\gamma^5$ needed for the computation of $H^5(J)$ can be written as
\begin{equation}
\begin{aligned}
\gamma^4 &=  \left(\begin{array}{cccc}
             \beta^4_+ &0& \beta^4_- & 0 \\
             0   & \beta^3_+ & 0 & \beta^3_-
            \end{array}\right): & H^{4}(X_+) \oplus H^{3}(X_+) \oplus H^{4}(X_-) \oplus H^{3}(X_-)\\
           & & \rightarrow H^4(S,\Z) \oplus H^{2}(S,\mathbb{Z}) \\
\gamma^5 &=      \left(\begin{array}{cc}
                       \beta^4_+ & 0 \\
                       0 & \beta^4_-
                  \end{array}\right): & H^4(X_+) \oplus H^4(X_-) \rightarrow H^4(S,\Z) \oplus H^4(S,\Z)
\end{aligned}
\end{equation}
As the image of $\beta^3_\pm$ is $T_\pm$ (see  \cite{Corti:2012kd} for details) it follows that 
\begin{equation}
\begin{aligned}
 \mbox{ker}(\gamma^5) &= K^*_+ \oplus K^*_- \\
 \mbox{coker}(\gamma^4) &= \Gamma^{3,19}/(T_+ + T_-)
\end{aligned}
\end{equation}
so that
\begin{equation}\label{eq:H5ofJ}
 H^5(J,\mathbb{Z})  = \Gamma^{3,19} /(T_+ + T_-) \oplus K^*_+ \oplus K^*_- \
\end{equation}
Note that $\text{tor}\,  H^3(J,\mathbb{Z}) = \text{tor}\,  H^5(J,\mathbb{Z})$ implies with the above results that 
\begin{equation}
\mbox{tor}\,\Gamma^{3,19} /(T_+ + T_-) = \mbox{tor}\, \Gamma^{3,19} /(N_+ + N_-).
\end{equation} 
Let us conclude by remarking that the total cohomology of a TCS $G_2$ manifold has the pleasingly symmetric form
\begin{equation}\label{eq:total_lattice}
\begin{aligned}
 H^\bullet(J)  = & \,\,H^0(J) \oplus H^7(J) \oplus \mathbb{Z}[\alpha_S] \oplus \mathbb{Z}[S]\\
 & \oplus K_+^{\oplus 4} \oplus K_-^{\oplus 4} \oplus H_+^{\oplus 4} \oplus H_-^{\oplus 4} \\
& \oplus N_+ \cap N_- \oplus \Gamma^{3,19} /(T_+ + T_-)\oplus   T_+ \cap T_- \oplus \Gamma^{3,19} /(N_+ + N_-) \\
& \oplus N_+ \cap T_- \oplus \Gamma^{3,19} /(T_+ + N_-) \oplus T_+ \cap N_- \oplus \Gamma^{3,19} /(N_+ + T_-)   \, .
\end{aligned}
\end{equation}
where we have exploited the isomorphisms $K_\pm\simeq K^*_\pm$. This is going to be very useful to establish the Shatashvili-Vafa relation for our conjectural $G_2$-mirror pairs.

A matching $\Xi$ is said to be orthogonal iff \cite{Corti:2012kd}
\begin{equation}\label{eq:orthogonorrea}
N_\pm = N_\pm \cap N_\mp \oplus N_\pm \cap T_\mp\,,
\end{equation}
where all lattices above are taken $-\otimes \mathbb{R}$. If that is the case,
\begin{equation}
\begin{aligned}
&b_2 = |K_+| + |K_-| + |N_+\cap N_-| \,,\\
&b_3 = 23 + 2 (|H_+| + |H_-|) + |K_+| + |K_-| - |N_+\cap N_-|\,,\\
\end{aligned}
\end{equation}
where we have used that $|\Gamma^{3,19}/(N_++N_-)| = 22 - |N_+| - |N_-| + |N_+ \cap N_-|$ together with \eqref{eq:orthogonorrea} to simplify the formula for the dimension of $b_3$.

\subsubsection{Massless spectrum of a TCS $G_2$ manifold}

We now discuss how the various subsectors of Type II strings on a TCS $G_2$ manifold arise from \eqref{eq:bettinumbers}. First consider the IIA setup. We work in a duality frame in which we choose the modes appearing in $C^{(3)}$ over those appearing in $C^{(5)}$. The modes corresponding to the sublattices $K_\pm$ of $H^2(J,\mathbb{Z})$ and $H^3(J,\mathbb{Z})$ give rise to $\cn=4$ vector multiplets in 3d: for each basis element in $K_+$ or $K_-$ we have an $\cn=2$ vector multiplet, from the decomposition of $C^{(3)}$, as well as an $\cn=2$ chiral multiplet, arising from the decomposition of the metric and the $B$-field. These combine to the 3 real scalars of an $\cn=4$ vector multiplet. These are remnants of the 4D $\cn=2$ vector moduli of the IIA compactification on the CYs $X_\pm$, which are inherited by the reduction on a TCS $G_2$ manifold. The modes corresponding to the sublattice $N_+ \cap N_-$ of $H^2(J,\mathbb{Z})$ give rise to 3D $\cn=2$ vectormultiplets instead. The sublattices $H^3(X_+)/T$ and $H^3(X_-)/T$ of $H^3(J,\mathbb{Z})$ are respectively isomorphic to $H_+ \oplus H_+$ and $H_-\oplus H_-$. The modes corresponding to the metric, which we identify with the periods of $\varphi_3$, as well as the modes arising from the decomposition of $C^{(3)}$ on $H_\pm\oplus H_\pm$ give 3D $\cn=4$ hypermultiplets. The remaining modes organize in 3D $\cn=2$ chiral multiplets.\footnote{ Even if there is no chirality in 3D, the notion of a chiral multiplet is defined --- see e.g. \cite{Hitchin:1986ea} for a review.} We summarize these findings in Table \ref{tab:typeII_TCS}

\begin{table}
\begin{center}
\begin{tabular}{c|c|c}
$H^\bullet(J)\otimes \mathbb{R}$ sublattice \phantom{\Big|} & IIA modes & IIB modes \\
\hline
\hline
\phantom{\Big|} $K_\pm$ & $\cn=4$ vector & $\cn=4$ vector \\
\hline
\phantom{\Big|} $H_\pm$ & $\cn=4$ hyper & $\cn=4$ hyper \\
\hline
\hline
\phantom{\Big|} $N_+ \cap N_- \simeq \Gamma^{3,19}/ (T_+\cap T_-)$ & $\cn=2$ vector & $\cn=2$ chiral \\
\hline
\phantom{\Big|} $T_+\cap T_- \simeq \Gamma^{3,19}/ (T_+\cap T_-)$&  $\cn=2$ chiral & $\cn=2$ vector \\
\hline
\phantom{\Big|} $N_+\cap T_- \simeq \Gamma^{3,19}/ (T_+\cap N_-)$&  $\cn=2$ chiral & $\cn=2$ vector \\
\hline
\phantom{\Big|} $T_+\cap N_- \simeq \Gamma^{3,19}/ (N_+\cap T_-)$&  $\cn=2$ chiral & $\cn=2$ vector \\
\hline
\hline
\phantom{\Big|} $\mathbb{Z}[\alpha_S] \simeq H^4(S)$ & $\cn=2$ chiral & $\cn=2$ vector\\
\hline
\end{tabular}
\end{center}
\caption{Structure of the massless spectra for type II compactifications on TCS. Here we show the spectrum in the $C^{(3)}$ duality frame for IIA. For IIB we chose to use the vectors arising from $C^{(4)+}$ and we work with $C^{(2)}$.}\label{tab:typeII_TCS}
\end{table}

We now discuss the IIB setup. In this case, the modes corresponding to the sublattices $K_\pm$ of $H^2(J,\mathbb{Z})$, $H^3(J,\mathbb{Z})$ and $H^4(J,\mathbb{Z})$ give rise to $\cn=4$ vectormultiplets in 3D: for each element in $K_+$ or $K_-$ we have an $\cn=2$ vector multiplet, where the vector arise from the decomposition of $C^{(4)+}$ and the corresponding scalar arise from the reduction of $C^{(2)}$, as well as an $\cn=2$ chiral multiplet, arising from the decomposition of the metric and the $B$-field. Together these form $\cn=4$ vector multiplets. Similarly, one obtains $\cn=4$ hypermultiplets from the reduction of $C^{(4)+}$ and the metric on the modes corresponding to the sublattices $H^3(X)/T$ of $H^4(J)$. Notice that we have picked a duality frame that matches our choice for the IIA setup.

From the structure of the cohomology groups, we conclude that there are no other $\cn=4$ multiplets that can be constructed out of the KK reduction of Type II on TCS $G_2$ manifolds. Notice that the structure of the spectrum that organizes in $\cn=4$ multiplets is identical. This is a low-energy manifestation of the fact that on a background $\mathbb{R}^{1,2} \times S^{\, 1} \times X$, where $X$ is a CY 3-fold, IIA and IIB are related by T-duality on the $S^{\, 1}$. In our setup, there is an unbroken subsector with enhanced supersymmetry for which this applies, corresponding to the components of the TCS $J$ that are not affected by the matching. The remaining parts of the massless KK spectrum are organized in terms of 3d $\cn=2$ multiplets, whose bosonic components are detailed in Section \ref{sec:superstring_back}. This completes our analysis of the structure of these backgrounds from Section \ref{sec:back_structure}.

It is amusing to remark that the 3D duality among scalars and vectors (both in IIA and IIB) relies on the following lattice identities 
\begin{equation}\label{eq:lattice_ids}
\begin{aligned}
&\dim N_+\cap N_- = \dim \Gamma^{3,19} /(T_+ + T_-)\,,\\
&\dim T_+\cap T_- = \dim \Gamma^{3,19} /(N_+ + N_-)\,,\\
&\dim T_+\cap N_- = \dim \Gamma^{3,19} /(N_+ + T_-)\,,\\
&\dim N_+\cap T_- = \dim \Gamma^{3,19} /(T_+ + N_-)\,,\\
\end{aligned}
\end{equation}
as well as on the isomorphisms $K_\pm \simeq K^*_\pm$. Changing duality frame in the democratic formulation of supergravity, correspond to dualizing vectors to scalars in the corresponding 3D model.

\subsection{Building blocks and tops: a lightning review}\label{sect:acylfrombb_and_tops}

An asymptotically cylindrical Calabi-Yau threefold $X$ can be obtained by excising a fibre $S_0$ from an appropriate compact $K3$ fibered algebraic threefold $Z$, commonly called a building block. Following \cite{MR3109862}, we call an algebraic threefold $Z$ a building block if $Z$ has a projection, 
\begin{equation*}
\begin{array}{ccc}
 S & \hookrightarrow & Z \\
 &&\downarrow_\pi \\
 && \P^1
\end{array}
\end{equation*}
the generic fibres of which are non-singular $K3$ surfaces and the anticanonical class of which is primitive in $H^{(2)}(Z)$ and equal to the class of the fibre, $S$:
\begin{equation*}
 [-K_Z] = [S] \,.
\end{equation*}
Picking a smooth and irreducible fibre $S_0$, there is a natural restriction map 
\begin{equation}\label{eq:defN}
\rho: H^2(Z,\mathbb{Z}) \rightarrow H^2(S_0,\mathbb{Z}) \cong \Lambda = (-E_8^{\oplus 2}) \oplus U^{\oplus 3}
\end{equation}
and the fibration is trivial in the vicinity of $S_0$. If we denote the image of $\rho$ by $N$, we demand that the quotient $\Lambda/N$ is torsion free, i.e. the embedding $N \hookrightarrow \Lambda$ is primitive. Finally, we demand that $H^3(Z,\mathbb{Z})$ has no torsion\footnote{ In fact, this is not strictly necessary as pointed out in \cite{Corti:2012kd}.}.  Under these assumptions, it follows that $Z$ is simply connected and the Hodge numbers $H^{1,0}(Z)$ and $H^{2,0}(Z)$ vanish. As $Z$ is a $K3$ fibration over $\P^1$, the normal bundle of the fibre, and in particular of $S_0$, is trivial. By excising the fibre $S_0$, we may form the open space 
\begin{equation}
 X \equiv Z \setminus S_0 \, .
\end{equation}
from a building block $Z$. If $Z$ satisfies the assumptions formulated above, $X$ is an asymptotically cylindrical Calabi-Yau threefold. For cohomology this gives, for instance, that
\begin{equation}\label{eq:3cohZ}
H^3(X,\mathbb{Z}) = H^3(Z,\mathbb{Z}) \oplus T\,,
\end{equation}
which simplifies the cohomology formulas in Equation \eqref{eq:bettinumbers}. In particular we have that
\begin{equation}
H_{\pm} \simeq H^{2,1}(Z_\pm,\mathbb{Z}).
\end{equation}
The TCS $G_2$ manifolds obtained by glueing asymptotically cylindrical CY arising from a pair of building blocks have the special feature that they turn out to be K3 fibrations over a base $S^3$ \cite{Braun:2017ryx}.

Building blocks may be realized from blowups of semi-Fano threefolds \cite{MR3109862}, but may also be realized as toric hypersurfaces. These have an elegant combinatorial description \cite{Braun:2016igl} which we will use extensively to make examples. We summarize this construction here for completeness, however for a detailed exposition we refer the reader to \cite{Braun:2016igl}. 

A dual pair of tops is a pair of four-dimensional lattice polytopes $\Diamond^\circ$ and $\Diamond$ embedded in a pair of dual lattices $N$ and $M$, respectively, such that 
\begin{equation}\label{eq:topsduality}
 \begin{aligned}
 & \langle \Diamond, \Diamond^\circ \rangle \geq -1 & \\
  \langle  \Diamond,\nu_e \rangle \geq 0 \hspace{.5cm} & &  \langle m_e, \Diamond^\circ\rangle \geq 0
\end{aligned}
\end{equation}
with the choice $m_e = (0,0,0,1)$ and $\nu_e = (0,0,0,-1)$. The lattice $\Diamond$ defines a compact toric variety through its normal fan $\Sigma_n(\Diamond)$, together with a family of hypersurfaces $Z_s$. Both the ambient variety and the corresponding hypersurfaces are generically singular. Each $k$-dimensional face $\Theta$ of $\Diamond$ is associated with a $(4-k)$ dimensional cone $\sigma(\Theta)$ of the normal fan $\Sigma_n(\Diamond)$. The resulting variety $Z_s$ has a crepant resolution into a smooth manifold $Z$ by refining the fan $\Sigma \rightarrow \Sigma_n$ using all of the lattice points on $\Diamond^\circ$ as rays. As shown in \cite{Braun:2016igl}, the resulting manifold $Z$ has all of the properties of a building block. Concretely, the defining equation of the resolved hypersurface is
\begin{equation}\label{HypSurf}
Z: \,\,\, \sum_{m \in \Diamond} c_m z_0^{\langle m, \nu_0 \rangle}\prod_{\nu_i \in \Diamond^\circ} z_i^{\langle m, \nu_i\rangle +1} = 0 \,  .
\end{equation}
Here, $m$ are lattice points on $\Diamond$, the $c_m$ are (generic) complex coefficients, the $z_i$ are homogeneous coordinates corresponding to the lattice points $\nu_i$ on $\Diamond^\circ$, while $z_0$ is the homogeneous coordinate associated with the ray through $\nu_0 = (0,0,0,-1)$.  
Note that $Z$ as defined in this way has first Chern class given by the fiber class, which equals $[z_0]$. 

The Hodge numbers of $Z$ as well the ranks of the lattices $N$ and $K$ can be described in purely combinatorial terms. The result is that $h^{i,0}(Z) =  0$ and \cite{Braun:2016igl,Braun:2017ryx}
\begin{equation}\label{eq:bbtopology}
\begin{aligned}
h^{2,1} (Z)& = \ell(\Diamond) - \ell(\Delta_F) + \sum_{\Theta^{[2]} < \Diamond} \ell^*(\Theta^{[2]})\cdot \ell^*(\sigma_n(\Theta^{[2]})) - \sum_{\Theta^{[3]} < \Diamond}\ell^*(\Theta^{[3]}) \cr 
h^{1,1} (Z) &= -4 + \sum_{\Theta^{[3]} \in \Diamond}  1 
+ \sum_{\Theta^{[2]}\in\Diamond} \ell^* (\sigma_n (\Theta^{[2]})) 
+ \sum_{\Theta^{[1]}\in\Diamond} \ell^* (\sigma_n (\Theta^{[1]}) +1) (\ell^* (\sigma_n(\Theta^{[1]}))) 
\cr  
|N| &=  \ell(\Delta_F^\circ) - \sum_{\Theta_F^{[2]} < \Delta_F^\circ} \ell^*(\Theta_F^{[2]}) - 4 + \sum_{\mbox{ve}\,\, \Theta_{F}^{\circ [1]} < \Delta_F^\circ}  \ell^*(\Theta_F^{[1]})\ell^*(\Theta_F^{\circ [1]}) \cr
|K| &= h^{1,1}(Z) - |N| -1 \cr 
&= \ell(\Diamond^\circ) - \ell(\Delta_F^\circ) + \sum_{\Theta^{\circ [2]} < \Diamond^\circ} \ell^*(\Theta^{\circ [2]})\cdot \ell^*(\sigma_n(\Theta^{\circ [2]})) - \sum_{\Theta^{\circ[3]} < \Diamond^\circ}\ell^*(\Theta^{\circ[3]}) \,.
\end{aligned}
\end{equation}
The quantities in these relations are defined as follows. $\Delta_F$ is the subpolytope of $\Diamond$, which is orthogonal to $\nu_e$. It is reflexive and determines the algebraic family in which the K3 fibres are contained via the construction of \cite{Batyrev:1994hm}. The $k$-dimensional faces of $\Diamond$ are denoted by $\Theta^{[k]}$ and $\sigma_n(\Theta)$ is the cone in the normal fan associated with each $\Theta$. The function $\ell$ counts the number of lattice points on a polytope, while the function $\ell^*$ counts the number of lattice points in the relative interior. In the case of $\ell^*(\sigma_n(\Theta^{[k]}))$ this refers to the points on $\Diamond^\circ$ in the relative interior of $\sigma_n(\Theta^{[k]})$. Finally,
in the expression for $|N|$, the sum only runs over vertically embedded (ve) faces $\Theta^{\circ[1]}$ of $\Delta_F^\circ$, which are those faces bounding a face $\Theta^{\circ[2]}$ of $\Diamond^\circ$ which is perpendicular to $F = \nu_e^\perp$, see \cite{Braun:2016igl,Braun:2017ryx} for a detailed exposition. The faces $\Theta^{\circ[1]}$ and $\Theta^{[1]}$ are the unique pair of faces on $\Delta_F,\Delta_F^\circ$ obeying $\langle \Theta^{[1]},\Theta^{\circ [1]} \rangle = -1$. 

\subsection{B-field}

In the context of the compactification of Type II superstrings, Equation \eqref{eq:HKrotation} must be supplemented with an instruction about the $B$-field. If we wish to have a string compactification which has a global geometric interpretation, these instructions must be such that $B$ descends from a two-form on $J$. In particular, for $J \to X_- \cup_{\,\Xi} X_+$ we must have
\begin{equation}\label{eq:glueBfield}
\Xi^*B_- = B_+.
\end{equation}
Such a two-form will pull back to a (unique) two-form on any K3 fibre $S$, so that the matching of Equation \eqref{eq:HKrotation}, must be extended as follows:\footnote{Notice the sign change with respect to  \cite{Braun:2017ryx}. We will show below that this action of $\Xi$ is indeed consistent with the mirror map.}
\begin{equation}\label{eq:Bfieldmatch}
g^*B_{S_-} = B_{S_+}.
\end{equation}
to include the $B$-field on the corresponding K3s. Possible choices for the $B$-field can be easily detected by examining the formula for $H^2(J)$, cfr. Equation \eqref{eq:bettinumbers}. The contributions $K_\pm$ are two-forms which stem from localized divisors on $X_\pm$ and we may hence think of components of the $B$-field along such cycles are localized as well. In particular, such components will not participate in the gluing \eqref{eq:glueBfield}. On the contrary, components of the $B$-field along $(N_+ \cap N_-)\otimes \mathbb{R}$ pull back to a two-form on every K3 fibre which are identified under the gluing $\Xi$. For TCS built from building blocks, elements of $N_\pm$ correspond to (not necessarily effective) curves in the fibres $S_\pm$ which are constant over the base $\P^1$s of the building blocks, i.e. they do not undergo monodromy. Hence we may think of two-forms of $J$ associated with $N_+\cap N_-$ as arising from two-forms of $S$ which are constant over the base $S^{\, 3}$ of $J$. 

On the K3 fibres $S_\pm$ in the asymptotically cylindrical region, the two-form $B$ hence becomes an element of
\begin{equation}
 B_{S_+} = B_{S_-} \subset (N_+\cap N_-) \otimes \R
\end{equation}
which nicely complements the constraints
\begin{equation}\label{eq:locationofHKstrinG319}
\begin{aligned}
 \omega_{S_+} = R_{S_-}&  \subset  (N_+ \cap T_-) \otimes \R \\
R_{S_+} =\omega_{S_-}  &  \subset  (T_+ \cap N_-) \otimes \R \\
I_{S_+} = - I_{S_-}&  \subset (T_+ \cap T_-) \otimes \R 
\end{aligned}
\end{equation}
which arise from the matching \eqref{eq:HKrotation}. We will later see how this structure is exploited under various mirror maps. 

In models with non-trivial torsion in $H^3(J,\mathbb{Z}) \cong  H_2(J,\mathbb{Z})$ there are furthermore discrete choices of the B-field, which correspond to discrete torsion in the corresponding CFT \cite{Vafa:1986wx}. By \eqref{eq:bettinumbers} and \eqref{eq:torcoh}, such components of the B-field restrict to elements in the torsion subgroup of $\Gamma^{3,19}/(N_+ + N_-)$ which is equal to the torsion subgroup of $\Gamma^{3,19}/(T_+ + T_-)$. It is precisely $\Gamma^{3,19}/(T_+ + T_-)$ which appears in the expression \eqref{eq:H5ofJ} for harmonic five-forms on $J$. These are Poincar\'e dual to two-cycles of $J$ which define the B-field background by integration. We are hence led to associate possible choices of the B-field to elements of $\Gamma^{3,19}/(T_+ + T_-) \oplus K_+ \oplus K_-$.

\section{Mirror maps for TCS $G_2$ manifolds}\label{sec:GeneralMIR}

The SYZ argument has a natural generalization in the context of $G_2$-holonomy manifolds \cite{Acharya:1997rh}. Recall that $G_2$-manifolds have two kinds of calibrated submanifolds: the three-dimensional associatives, which are calibrated by the 3-form $\varphi$, and the four-dimensional coassociatives, which are calibrated by $\star\varphi$ \cite{Harvey:1982xk}. The main difference among these two kinds of calibrated submanifolds is that the deformations of the former are obstructed, while the deformations of latter are not: a coassociative submanifold $M$ has a smooth moduli space of dimension $b_2^+(M)$, the number of self-dual harmonic 2-forms \cite{Mclean96}. Let $(J,J^\vee)$ (resp. $(J,J^\wedge)$) denote a putative $G_2$-mirror pair relating IIA to IIA (resp. IIA to IIB). In a compactification of IIA on $J$, a D$0$-brane has a moduli space which equals $J$, which must correspond to the moduli space of a wrapped D$p$-brane on $J^\vee$ (resp. $J^\wedge$). As we want a BPS configuration, in IIA the only option left is wrapping a coassociative $M\subset J^\vee$ with a D4-brane, in IIB instead we have to wrap an associative cycle $A\subset J^\wedge$ with a D3-brane. The $U(1)$ vector field on the brane gives rise to $b_1$ additional moduli whence the physical moduli space has dimension $b_1(M)+b_2^+(M)$ in the coassociative case, while $b_1(A) + x$ in the associative case. Here, $x$ is related to a counting of harmonic spinors with values in a particular vector bundle, whose definition may be found in \cite{Mclean96}. This is much harder and requires some additional (reasonable) assumptions.

For the moduli space of a wrapped D4-brane to coincide with the D$0$ brane moduli space we must require that $b_1(M)+b_2^+(M) = 7$. It is hence natural to conjecture that $M\simeq T^4$ \cite{Acharya:1997rh}. To determine $x$, following \cite{Acharya:1997rh}, we claim that $A$ is a $T^3$ which fibers the whole $J^\wedge$. If that is the case, the dimension of the moduli space of the wrapped D3 brane is again 7, and the argument of SYZ can be carried on. Four (resp. three) T-dualities along the cycles of such a $T^4$ (resp. $T^3$) map the wrapped D4-brane on $J^\vee$ (resp. the wrapped D3-brane on $J^\wedge$) to the D$0$-brane on $J$. Repeating the argument vice-versa entails that $J$ has an analogous $T^4$ (resp. $T^3$) fibration.

In what follows we are going to heuristically argue that this is indeed the case for TCS $G_2$ manifolds, provided they have some additional structure. Let us denote the asymptotically cylindrical CY which emerge in the Kovalev limit of $J$ by $X_+$ and $X_-$:
\begin{equation}
J \to X_+ \cup_{\, W} X_-,
\end{equation}
where $W = X_+ \times S^1 \cap X_- \times S^1  \simeq T^2 \times S_0$. If a given CY $X$ has a mirror $X^\vee$, let us denote the corresponding SYZ special lagrangian fibration $L_X$. Moreover, for the two asymptotic sides $J_\pm = (S^1)_\pm \times X_\pm$ we refer to $(S^1)_\pm$ as the `outer' circle. Then:
\begin{itemize}
\item $J$ admits a generalized $\Tf$ mirror map, if $X_{\pm}$ both have CY mirrors $(X_\pm)^\vee$ and $W$ is such that
\begin{equation}
(S^1)_\pm \times L_{X_\pm} \big|_W
\end{equation} 
are glued together to a coassociative toroidal fibration of $J$;
\item $J$ admits a generalized $\Tt$ mirror map, if $X_-$ admits a CY mirror $(X_-)^\vee$, while the asymptotic K3 surface $S_+$ of $X_+$ is elliptically fibered, with fiber $E_+$, and  $W$ is such that
\begin{equation}
L_{X_-}\big|_W \qquad\text{and}\qquad (S^1)_+ \times E_+\big|_W
\end{equation}
are glued together to form an associative toroidal fibration of $J$.\footnote{ It is natural to conjecture that such an associative $T^3$ fibration can be exploited to obtain a fiberwise duality relating Heterotic on TCS $G_2$-manifolds with M-theory on $Spin(7)$ manifolds that are K3 fibered. This construction hints towards extending the TCS construction to $Spin(7)$ manifolds.}

\end{itemize}
We remark that while the existence of a coassociative fibration is stable with respect to small deformations, this is not the case for an associative fibration. For this reason the SYZ associative argument holds only at a very specific point of moduli space. Of course, there are lots of subtleties we are not addressing here (which are in part related with the subtleties in the original SYZ argument \cite{Morrison:2010vf,Gross:2012rw} and also go beyond), but the arguments presented in Sections \ref{sec:T4hurx} and \ref{sec:T3hurx} are meant to be no more than a motivation to look for TCS $G_2$-mirror pairs that have relations like the ones we discuss in Section \ref{sect:cohom_action}.

In the discussion below, for the sake of simplicity, we are going to assume there is no discrete torsion. We deal with it in Section \ref{sec:discrete_tors}.

\subsection{$\Tf$ mirror maps from coassociative fibrations}\label{sec:T4hurx}
Let us begin by analyzing the generalized $\Tf$ mirror symmetry map in the context of TCS $G_2$ manifolds. As the Kovalev CYs $X_\pm$ admit mirrors by assumption, by the SYZ argument, these have two $T^3$ special lagrangian fibrations. Let us denote such fibrations by $L_\pm$. In the asymptotically cylindrical region of the manifold  $X_{\pm} \sim \mathbb{R}^+ \times S^{\, 1} \times S_{\pm}$, the SYZ special lagrangians must asymptote to $L_\pm \sim S^{\, 1} \times \Lambda_\pm$, where $\Lambda_{\pm}$ are the SYZ special lagrangian $T^2$ fibrations within the asymptotic K3s (with respect to the K3 complex structure induced by the ambient CY). In particular, they do not extend along the $\mathbb{R}^+$ direction. Let us fix the phase of the holomorphic top form on $X_\pm$ such that the SYZ special lagrangians are calibrated as follows
\begin{equation}\label{eq:coassociativecalibration}
\text{vol}_{L_\pm} = - \text{Im}\left( \Omega_{\pm}^{3,0}\right) |_{L_\pm}.
\end{equation}
Notice that a special lagrangian $L$  gives rise to a coassociative cycle $M_L \equiv S^{\, 1} \times L \subset S^{\, 1} \times X$ if and only if it satisfies \eqref{eq:coassociativecalibration}. In particular,
\begin{equation}
\text{Im} (\Omega_{\pm}^{3,0}) = d \theta_\pm \wedge I_{S_{\pm}} - dt_\pm \wedge R_{S_{\pm}}
\end{equation}
therefore, for our special lagrangians $L_\pm$ we have
\begin{equation}
\text{Im} (\Omega_{\pm}^{3,0})|_{L_\pm} =  - d \theta_\pm \wedge I_{S_{\pm}}.
\end{equation}
Now, Equation \eqref{eq:PHI34} entails that the four-cycles $ M_\pm \equiv \left(S^{\, 1}\right)_\pm \times L_\pm \subset J_\pm$ are such that
\begin{equation}
\star \varphi_\pm|_{M_\pm} = d \xi_\pm \wedge d\theta_\pm \wedge  \left( I_{S_{\pm}}\right)|_{\Lambda_{\pm}} = \text{vol}_{M_\pm}
\end{equation}
which entails these are coassociative submanifolds. Since
\begin{equation}
\begin{aligned}
\Xi^*(\star\,\varphi_-|_{M_-}) &=  \Xi^*(d \xi_- \wedge d\theta_- \wedge I_{S_-}) \\
&= - d\theta_+ \wedge  d\xi_+ \wedge - I_{S_+}\\
&= d\xi_+ \wedge d\theta_+\wedge  I_{S_+}\\
&= \star\,\varphi_+|_{M_+},
\end{aligned}
\end{equation}
the TCS glueing diffeomorphism $\Xi$ is such that $M_\pm$ are glued into a coassociative submanifold $M\subset J$ which has the topology of a $T^4$, which may become singular along loci in $J$. 

Performing three T-dualities along the $L_\pm$ SYZ fibres is mapping $X_{\pm}$ to their mirrors $X_{\pm}^\vee$ by construction. However, as the $S^{\, 1}$ within the asymptotic CY cylinders are swapped with the `outer' $S^{\, 1}$ by the matching along the Kovalev neck, these circles must have the same size and we have to necessarily perform four T-dualities along the corresponding $T^4$ coassociative. By construction $X^\vee_\pm$ are asymptotically cylindrical, and it is natural to ask whether $J^\vee$ is a twisted connected sum of the CY mirrors $X^\vee_\pm$ of $X_\pm$. In order to show this, it is sufficient to map the original matching $g$ across the mirror map. In the asymptotically cylindrical region, two of the four T-dualities occur along the $\Lambda_\pm$ special lagrangian SYZ fibrations of $S_\pm$. This induces mirror symmetries on the asymptotic K3 fibres in the glueing region. As the hyper-K\"ahler rotation induces a diffeomorphism, we can identify $\Lambda_+ = \Lambda_- \equiv \Lambda$ in the cylindrical region. Due to the matching \eqref{eq:HKrotation}, this means that $\Lambda$ is calibrated by $I_{S_+}$ on $S_+$, and by $-I_{S_-}$ on $S_-$. This leads to a several signs in the mirror maps. Furthermore, the $B$-field $B_\pm$ must be contained in $N_\pm$ on both sides both before and after the mirror map which constrains the hyper-K\"ahler structure. Explicitly, the K3 mirror maps act as:
\begin{equation}\label{eq:mirror_cal_I}
\begin{aligned}
 \omega_\pm^\vee &= \mp R_\pm \\
 R_\pm^\vee  &= \mp \omega_\pm \\
 I_\pm^\vee  &= \pm B_\pm + I_\pm - {I_\pm}_2 \\
 B_\pm^\vee &= \pm {I_\pm}_2
\end{aligned}
\end{equation}
where ${I_{S_\pm}}_2$ is the projection of $I_{S_\pm}$ to the sublattice $\Gamma^{2,18}$ perpendicular to $\Lambda$ and its dual $\Lambda_0$  (see Appendix \ref{sect:k3mirrorsyz} for our conventions about K3 mirror symmetry).
The matching needed for the TCS construction on the mirror side is obtained from the original matching by applying the mirror map of Equation \eqref{eq:mirror_cal_I} to Equations \eqref{eq:HKrotation} and \eqref{eq:Bfieldmatch} and then by applying the mirror map of Equation \eqref{eq:mirror_cal_I} to get back. One obtains:
\begin{equation}
\begin{aligned}
&\omega^\vee_\pm = \, \mp R_\pm = \, \mp \omega_\mp = R^\vee_\mp \, ,\\
&I_+^\vee =  -B_{+} + I_+ - I_{2+} = - B_{+} - I_+ + I_{2+} = -I_-^\vee \, ,\\
&B_+^\vee = I_{2+}  = -I_{2-} = B^\vee_- \,.
\end{aligned}
\end{equation}
This concludes our heuristic argument showing that $J^\vee$ is indeed a TCS $G_2$-holonomy manifold obtained as a twisted connected sum of the pair $(X^\vee_+,X^\vee_-)$, which are the CY mirrors of $(X_+,X_-)$.

\subsection{$\Tt$ mirror maps from associative fibrations}\label{sec:T3hurx}

Let us now turn to the generalized $G_2$ mirror map corresponding to fiberwise $T^3$ duality. In this case, only $X_-$ is required to have a mirror. Let us denote by $L_-$ the corresponding SYZ fibration. Rotating the phase of $\Omega_-$, we can choose to calibrate the SYZ special lagrangian fibers with $\text{Re}(\Omega_-)$. In this case,
\begin{equation}
\varphi_-|_{L_-} = \text{Re}(\Omega)|_{L_-} = d \theta \wedge R_{S_-}.
\end{equation}
In particular, this entails that $\Lambda_-$, the SYZ fibration of the asymptotic K3 fiber $S_-$, is calibrated by $R_{S_-}$. It is clear that $L_-$ gives rise to an associative fibration of $S^{\, 1}\times X_-$. Moreover, we require that the asymptotic K3 of $X_+$ is elliptically fibered. Let $E_+$ denote the corresponding elliptic fiber. By definition $E_+$ is holomorphic, meaning the it is a cycle calibrated by $\omega_{S_+}$. Therefore, $S^{\, 1} \times E_+ \subset S^{\, 1}\times X_+$ is an associative $T^3$ submanifold, and the elliptic fibration of $S_+$ gives rise to an associative $T^3$ fibration of $S^{\, 1} \times X_+$. Remarkably,
\begin{equation}
\Xi^*(d \theta_- \wedge R_{S_-}) = d \xi_+ \wedge \omega_{S_+}
\end{equation}
by definition. Hence, the $\Xi$ diffeomorphism glues the associative $T^3$ fibrations of $S^{\, 1} \times X_\pm$ constructed above into an associative $T^3$ fibration of $J$. Performing three fiberwise $T$-dualities transforms $J$ into its $\Tt$-mirror $J^\wedge$. The same argument above entails that $X_-$ is swapped with its mirror $(X_-)^\vee$ by performing the three SYZ fiberwise $T$-dualities, however the two fiberwise $T$-dualities along the holomorphic $T^2$ fibration of $S_+$ map $X_+$ back to itself. Without further ado, we claim that $J^\wedge$ is a TCS $G_2$ manifold that in the Kovalev limit degenerates to the TCS of $X_-^\vee$ and $X_+$. In this case, however, the SYZ fibration of $S_-$ is calibrated by $R_{S_-}$ while on $S_+$ we are T-dualizing on a cycle calibrated by $\omega_{S_+}$. We have worked out the associated mirror maps in Appendix \ref{sect:k3mirrorsyz}. Here, we again demand that the B-field is contained in $N_+ \cap N_-$, and is hence perpendicular to $\Omega$, both before and after application of the mirror map. The result is that 
\begin{equation}
\begin{aligned}
&R^\vee_+ = -I_+ = I_- = \omega_-^\vee \,, \\
&I_+^\vee = R_+ = \omega_-  = -I_-^\vee \,, \\
&B_+^\vee = \omega_{2+} = R_{2-} = B_-^\vee\,, \\
&\omega^\vee_+  = - B_{2+} + \omega_+ - \omega_{2+} = -B_{2-} + R_- - R_{2-}  = R_-^\vee \,.
\end{aligned}
\end{equation}
This concludes our argument that the $T^3$-mirror $J^\wedge$ is a TCS as well.\footnote{ As we have already stressed in \cite{Braun:2017ryx}, the same subtleties arising in the context of the original SYZ argument \cite{Morrison:2010vf,Gross:2012rw} arise here. This discussion is meant to be no more than a motivation to look for TCS $G_2$-mirror pairs where these structures are manifest.}

\subsection{Discrete torsion}\label{sec:discrete_tors}

In general, the B-field is not completely determined by giving a two-form in $H^2(J,\mathbb{R})$, but needs the extra data of discrete torsion  \cite{Vafa:1986wx}. This data is given by fixing a torsional element in $H_2(J,\mathbb{Z}) \cong H^3(J,\mathbb{Z})$, which in our case are isomorphic to the torsional elements of $\Gamma^{3,19}/(N_+ + N_-)$. Torsional cycles in $H^3(J,\mathbb{Z})$ hence restrict to cycles in $H^{2}(S,\mathbb{Z})$, which itself is torsion-free. As argued above, the B-field including discrete torsion is hence captured by periods of the B-field on elements of
\begin{equation}
\Gamma^{3,19}/(T_+ + T_-)
\end{equation}
on the asymptotic K3 fibres of a TCS $G_2$ manifold.

The appearance of torsion in cohomology in relation to discrete torsion phases is discussed for CY manifolds in the context of mirror symmetry in \cite{Aspinwall:1995rb,1998math......9072G}. For CY manifolds, the central observation are non-trivial torsional cycles in $\text{tor} H_2(X,\mathbb{Z})\cong \text{tor} H^3(X,\mathbb{Z})$ for a resolution $X$ of an orbifold theory including discrete torsion in the CFT \cite{Aspinwall:1995rb}. The discrete torsion in the CFT can be thought of as non-trivial discrete periods of the B-field over cycles in $\text{tor} H_2(X,\mathbb{Z})$. As the relation $\text{tor} H_2(X,\mathbb{Z})\cong \text{tor} H^3(X,\mathbb{Z})$ does not depend on the dimension of $X$, so we can make the same argument for $G_2$ manifolds. 

In the presence of discrete torsion, i.e. torsion in $\Gamma^{3,19}/(N_+ + N_-)$, we now longer need that $B_\pm \cdot \Omega_\pm = B_\pm^\vee \cdot \Omega_\pm^\vee = 0$ as $B$ restricted to $S_{ \pm}$ does not need to sit purely in  $(N_+ \cap N_-) \otimes \mathbb{R}$. Hence we can no longer conclude that the mirror map takes the rather restrictive forms \eqref{eq:RmirrorBorth}, \eqref{eq:ImirrorBorth} and \eqref{eq:omegamirrorBorth}. In the case that the torus used for two T-dualities is calibrated by the K\"ahler form $\omega$, this means that the associated elliptic fibration does no longer need to have a holomorphic section. 

Let us first discuss the mirror map related to associative fibrations. Assuming as before that the torus $E$ is calibrated by $\omega_+$ and $R_-$, we can use the general form \eqref{eq:grossmirror} adapted to these calibrations in order to find the action of the mirror maps on the matching. This yields
\begin{equation}
\begin{aligned}
\omega_+^\vee & = -B_+ + E_0 + \left( \tfrac12(R_{2+}^2-B_{2+}^2) + B_+ \cdot E_0 \right) E & \\
& =-B_- + E_0 + \left( \tfrac12(\omega_{2+}^2-B_{2-}^2) + B_- \cdot E_0 \right) E = R^\vee_- \\
R_+^\vee & = -I_+ - I_+\cdot(B_+ - E_0) E = I_- + I_-\cdot(B_- - E_0) E = \omega_-^\vee \\
I^\vee_+ & = R_+ + R_+\cdot(B_+ -E_0) E = \omega_- + \omega_-\cdot(B_- -E_0) E  = -I_-^\vee \\
B_+^\vee & = \omega_+ - E_0 + \omega_+\cdot(B_+ - E_0) E =  R_- - E_0 + R_-\cdot(B_- - E_0) E  = B_-^\vee 
\end{aligned}
\end{equation}
so that the gluing is preserved even in a completely general setting!

For mirror maps related to coassociative fibrations, the torus $E$ is calibrated by $\pm I$ on $S_{\pm}$. In the most general setting including components of the $B$-field which are not in  $(N_+ \cap N_-) \otimes \mathbb{R}$, the mirror map then acts as
\begin{equation}
\begin{aligned}
\omega_{\pm}^\vee & = \mp R_\pm \mp R_\pm \cdot(B_\pm -E_0) E  = \mp \omega_{\mp}  \mp \omega_{\mp} \cdot (B_- -E_0) E  = R^\vee_\mp \\
I_+^\vee &= -B_+ + E_0 + \left( \tfrac12(\omega_{2+}^2 - B_{2+}^2) + (B_+\cdot E_0) \right) E  & \\
	&= -B_- + E_0 + \left( \tfrac12(R_{2-}^2 - B_{2-}^2) + (B_-\cdot E_0) \right) E   = -I_-^\vee\\
B_+^\vee & = I_+ - E_0 + I_+\cdot(B_+-E_0) E = -I_- - E_0 - I_-\cdot(B_--E_0) E = B_-^\vee
\end{aligned}
\end{equation}
Note that we have the normalization of the various forms is such that $\omega^2 = R^2$, which implies $\omega_2^2 = R_2^2$, see Appendix \ref{sect:k3mirrorsyz}. 

We can hence conclude that even in the most general setting with discrete torsion, where $B$ is not contained purely in $(N_+ \cap N_-) \otimes \mathbb{R}$, the matching is preserved by both mirror maps considered in this work. In fact, we have seen that the matching conditions \eqref{eq:HKrotation} supplemented by the condition \eqref{eq:glueBfield} for the B-field, are respected by mirror maps for any choice of B-field on the K3 surface. 

\subsection{Mirror symmetry, cohomology, and Shatashvili-Vafa relation}\label{sect:cohom_action}
\subsubsection{Action of $\Tt$ and $\Tf$ on cohomology}
If one assumes to have only abelian gauge sectors in the 3d $\cn=2$ theory arising from a Type II background with $G_2$ holonomy, it is always possible to dualize the vectors to scalars. Hence, for establishing the duality among massless KK modes all that matters is thus the complex dimension of the corresponding scalar manifolds, which is given by $1+(b_2 + b_3)$ as we have discussed in Section \ref{sec:superstring_back}. One must check that the corresponding mirror $G_2$ manifolds satisfy the Shatashvili-Vafa relation
\begin{equation}
b_2(J) + b_3(J) = b_2(J^\vee) + b_3(J^\vee). 
\end{equation}
In this section, we will discuss the total cohomology $H^\bullet(J,\mathbb{Z})$ and show that it remains invariant under the mirror maps introduced above. This implies the invariance of $b_2+b_3$ under both mirror maps, because for a $G_2$ manifold $b_1(J) = b_6(J) =0$ implies 
\begin{equation}
\dim H^\bullet(J,\mathbb{Z}) = 2 + 2  b_2(J) + 2 b_3(J)\, .
\end{equation}
For models with non-Abelian gauge sectors, checking the duality is a bit more involved. We discuss this in Section \ref{sect:sing_mirror} below.

Now, for a $G_2$ manifold $J$ constructed from two asymptotically cylindrical Calabi-Yau threefolds $X_\pm = Z_\pm \setminus S_{\pm}$ with asymptotic $K3$ fibres $S_{\pm}$, the topology of $J$ is fixed in terms of the topologies of $X_\pm$ together with the matching $g: S_{+}\rightarrow S_{-}$ and the images $N_\pm$ of $\rho_\pm: H^{2}(Z_\pm) \rightarrow H^{2}(S_{\pm})$. 

To find the topological properties of the mirrors $X_\pm^\vee$, we may argue as follows \cite{Braun:2017ryx}. Assume that there is a compact Calabi-Yau threefold $X^c$ which is fibered over a base $\mathbb{P}^1$ by $K3$ surfaces from a lattice polarized family with polarizing lattice $N$, such that we can degenerate the base $\mathbb{P}^1$ of $X^c$ and obtain two asymptotically cylindrical Calabi-Yau threefolds $X_{i} = Z_{i}\setminus S_{i}$, $i=1,2$ with $N\equiv N_1 = N_2 $. Of course, gluing $X_1$ with $X_2$ along the asymptotic necks, so that $X_1 \cap X_2 = S_i \times S^1 \times I$, one obtains $X^c$ back. The Hodge numbers of $X^c$ are given by 
\begin{equation}\label{eq:Xfromx1x2hodge}
\begin{aligned}
 h^{1,1}(X^c) &= K(Z_1) + K(Z_2) + 1 + |N| \\
 h^{2,1}(X^c) &= h^{2,1}(Z_1) + h^{2,1}(Z_2) + 21 - |N|
\end{aligned}
\end{equation}
where $K_i = \mbox{ker}\, (\rho_i) / [S_{i}]$. These expressions can be found by using the Mayer-Vietoris sequence which is particularly easy for $h^{1,1}(X^c)=b^2(X^c)$. Here, there is a one-dimensional contribution from the cokernel of $\gamma^1$ and a contribution $ K(Z_1) + K(Z_2) + |N|$ from the kernel of $\gamma^2$. In the latter, the $|K_i|$ arise from the kernels of the maps $\gamma^2_i$, and $|N|$ is the dimension of the common image of both of the $\gamma^2_i$. The expression for $h^{2,1}(X^c)$ can now be found from $h^{1,1}(X^c)$ by expressing the Euler characteristic of $X^c$
\begin{equation}
\begin{aligned}
\chi(X^c) &= \chi(Z_1) + \chi(Z_2) - 48 =\chi(X_1) +\chi(X_2) \\
& = 2(h^{1,1}(X^c)-h^{2,1}(X^c)) \, ,
\end{aligned}
\end{equation}
in terms of the Euler characteristics of the $Z_i$: 
\begin{equation}
\chi(Z_i) = 2 + 2 h^{1,1}(Z_i) - 2h^{2,1}(Z_i) \, .
\end{equation}
Note that the $X_i$ are obtained by excising a K3 fibre with $\chi(K3) =24$ from $Z_i$ and $X_1 \cap X_2$ has Euler characteristic zero. Together with $h^{1,1}(Z_i) = K(Z_i) + 1 +|N|$, this implies \eqref{eq:Xfromx1x2hodge}. 

The mirror map on $X_i$ must also act as the mirror map on the asymptotic $K3$ fibres $S_{i}$. These are from a lattice polarized family with lattice polarization $N$, so that their mirrors are from a lattice polarized family with polarizing lattice $N_i^\vee$, such that $N_i \oplus N_i^\vee$ has a primitive embedding into $\Gamma^{2,18}$ \cite{Aspinwall:1994rg}. This implies that $|N| = 20 - |N^\vee|$ and both lattices share the same discriminant group and form \cite{MR525944}. This further implies that we can write the lattice $T$, which is the orthogonal complement of $N$ in $\Gamma^{3,19}$ as
\begin{equation}
T = U \oplus \tilde{T} 
\end{equation}
and $N^\vee = \tilde{T}$. This follows from $T$ sharing the same discriminant group and form with $N$ and hence with $N^\vee$. On the other hand, $U \oplus \tilde{T}$ is the unique lattice with the same rank, signature and discrimant group and form \cite{MR525944}, so that the existence of a mirror implies the above decomposition. We can hence think of the mirror map on lattice polarized families of K3 surfaces as exchanging the lattice $N$ with the lattice $\tilde{T}$ of rank $20-|N|$.

In order to study the action of the mirror map on the asymptotically cylindrical Calabi-Yau threefolds $X_1$ and $X_2$, it is now convenient to assume that 
 $Z \equiv Z_1 = Z_2$ which implies  $X \equiv X_1 = X_2$. The mirror map on the compact Calabi-Yau threefold $X^c$ must now exchange $h^{1,1}(X^c)$ with $h^{2,1}(X^c)$, so that $|N| = 20 - |N^\vee|$ now implies that $Z^\vee$ satisfies:
\begin{equation}
\begin{aligned}\label{eq:swapKh21}
 |K(Z^\vee)| &= h^{2,1}(Z) \\
 h^{2,1}(Z^\vee) &= |K(Z)| \, .
\end{aligned}
\end{equation}
We have hence found the action of the mirror map for any building block $Z$. Note that this is precisely the behaviour we have observed in our explicit construction of $Z^\vee$ in \cite{Braun:2017ryx}, but it is reassuring that such a relation can be derived regardless of how the threefolds $Z$ are realized. Furthermore, note that $|K_\pm|$ and $h^{2,1}_\pm$ precisely make up an $\cn=4$ subsector of the 3D effective theory, with $|K|$ and $h^{2,1}$ giving rise to the dimensional reduction of vector- and hyper-multiplets, respectively.\footnote{ Which contribution corresponds to which one of course depends on weather we consider  IIA or  IIB string theory, but as we have remarked above (c-map) in the circle reduction one can always find a duality frame in which the two ends up being identical.} Assuming as before that the mirror map preserves the structure of $J$ as being glued as a TCS hence forces \eqref{eq:swapKh21}.

In applications to mirror symmetry, it is convenient to rewrite the expression \eqref{eq:total_lattice} for the total cohomology in terms of $\Gamma^{4,20} = \Gamma^{3,19} \oplus U$. Instead of working with $N =  \text{im } H^2(Z,\mathbb{Z}) \rightarrow H^2(S,\mathbb{Z})$, we can lift the whole structure to
\begin{equation}\label{eq:nhat}
\begin{aligned}
\hat{N} &=  \text{im } H^\bullet(Z,\mathbb{Z}) \rightarrow H^\bullet(S,\mathbb{Z}) \\
\hat{T} &= \hat{N}^\perp \text{in } \Gamma^{4,20}
\end{aligned}
\end{equation}
so that $\hat{N} = N \oplus U$, with the extra summand given by $H^0(S,\mathbb{Z}) \oplus H^4(S,\mathbb{Z})$. We can then rewrite \eqref{eq:total_lattice}, as
\begin{equation}\label{eq:totallatticeG420}
\begin{aligned}
H^\bullet(J)  = & \oplus K_+^{\oplus 4} \oplus K_-^{\oplus 4} \oplus H_+^{\oplus 4} \oplus H_-^{\oplus 4} \\
& \oplus \hat{N}_+ \cap \hat{N}_- \oplus \hat{T}_+ \cap \hat{N}_- \oplus \hat{N}_+ \cap \hat{T}_- \oplus \hat{T}_+ \cap \hat{T}_- \\
& \oplus \Gamma^{4,20} /(\hat{N}_+ + \hat{N}_-) \oplus \Gamma^{4,20} /(\hat{T}_+ + \hat{N}_-) \oplus \Gamma^{4,20} /(\hat{N}_+ + \hat{T}_-) \oplus \Gamma^{4,20} /(\hat{T}_+ + \hat{T}_-) \, .
\end{aligned} 
\end{equation}
Note that the terms $H^0(J) \oplus H^7(J) \oplus \mathbb{Z}[\alpha_S] \oplus H^4(S,\mathbb{Z})$ are now contained in $\hat{N}_+ \cap \hat{N}_- \oplus \Gamma^{4,20} /(\hat{T}_+ + \hat{T}_-)$. 

We are now ready to study the behaviour of this expression under our mirror maps. In Appendix \ref{sect:k3mirrorsyz}, we have reviewed the mirror map for K3 surfaces. This is most naturally formulated in terms of a lift of the K\"ahler form $\omega$, the real and imaginary parts $R$ and $I$ of the holomorphic two-form $\Omega$ and the B-field $B$ to a positive norm four-plane $\Sigma_4$ in $\Gamma^{4,20}\otimes \R$. Our case of interest is in algebraic families, $\hat{\omega}$ and $\hat{B}$ are generic elements in $\hat{N} \otimes \R$, and $\hat{\Omega}$ is a generic element in $\hat{T}\otimes \R$. This means that we write $\hat{N} = \hat{\Omega}^\perp$ in $\Gamma^{4,20}$ and furthermore $\hat{T} = \langle \hat{\omega},\hat{B} \rangle ^\perp$ in $\Gamma^{4,20}$ and $\hat{N}^\perp = \hat{T}$ as usual. As the mirror map exchanges the roles of the two-plane spanned by $\hat{\Omega}$ with the two-plane spanned by $\hat{\omega}$ and $\hat{B}$, it follows that the mirror map acting on a $K3$ surface simply exchanges the two lattices $\hat{N}$ and $\hat{T}$.

Together with \eqref{eq:swapKh21} it then follows that the total cohomology, and hence $b_2 + b_3$, remains invariant both under the mirror map associated with four $T$-dualities, where both $X_\pm$ are traded for their mirror, and the mirror map associated with three $T$-dualities, for which only $X_-$ is replaced by $X_-^\vee$:
\begin{equation}
H^\bullet(J,\mathbb{Z}) =  H^\bullet(J^\vee,\mathbb{Z}) = H^\bullet(J^\wedge,\mathbb{Z}) \, .
\end{equation}
Note that the only non-trivial torsion classes are $\mbox{tor}\,H^3(J,\mathbb{Z})$ and $\mbox{tor}\,H^4(J,\mathbb{Z})$ and that these are independently conserved under the mirror map related to a coassociative $T^4$ fibration \cite{Braun:2017ryx}.

\subsubsection{Remark about D-branes and homological $G_2$ mirror maps}
We stress here that the quantum consistency checks we mentioned at the end of Section \ref{sec:massless_modz} are in order also for our mirror maps. While the fact that the resulting models have the same $b_2 + b_3$ entails that they have moduli spaces of the same dimensions neglecting non-perturbative corrections, for the duality to be true, it is important that all such corrections match. In particular, this entails that the corresponding K\"ahler potentials, superpotentials, BPS spectra and phase structure match across each proposed duality. 

In this Section we consider the behavior of branes and point out that it is slightly more interesting than one may naively expect. Indeed, it is tempting to claim that the possible BPS particles in these models arise from D3-branes wrapped on associative cycles in IIB, and D4-branes wrapped on coassociative cycles in IIA. As the BPS spectra on the two sides have to agree, which is the statement of homological mirror symmetry, this would lead one to conclude that the mirror symmetry map is swapping associatives with coassociatives and vice versa. This is not precise, however, because of the role played by the $B$-field, as seen for instance in the discussion in Section \ref{sec:nAB_IIA}. The point is that a $B$-field can give `quantum' volumes to additional cycles in the $G_2$ holonomy manifold, thus giving rise to additional BPS sectors that do not correspond to the ordinary calibrated submanifolds for $G_2$ holonomy that were analyzed, e.g. in \cite{Becker:1996ay}. In particular, it can happen that some vanishing two-cycles can be `puffed up' in the quantum moduli space, because the latter has a component corresponding precisely to the $B$-field periods.\footnote{ It would be interesting to carry over a detailed analysis of the quantum moduli space of the 2d worldsheet SCFTs in the spirit of \cite{Blumenhagen:2001jb,Roiban:2001cp}.}

We call $B_n$--cycles the non-trivial cycles of this sort. When $n$ is $2$ or $4$, the corresponding $B_n$--cycle is codimension $5$ and $3$, respectively quantum-calibrated by $B$ and $B\wedge B$. For instance, in a IIA setting there are BPS states corresponding to D2-branes wrapping $B_2$--cycles, as well as D4-branes wrapping $B_4$--cycles. And this is not the end of the story: in type IIB, the presence of non-trivial rigid $B_2$-cycles can give rise to instantons by wrapping them with euclidean D1-branes, that would end up correcting the prepotential.

Moreover, we can have non-trivial fibrations of $B_2$--cycles on associatives, which give rise to $B$-associatives cycles, or $B_5$--cycles, that have codimension 2, and are quantum-calibrated by $\varphi_3 \wedge B$. In type IIA the $B_5$--cycles wrapped by Euclidean D4-branes can give rise to non-trivial superpotential terms, as we have observed in Section \ref{sec:nAB_IIA}. One could also speculate about the existence of $B$-coassociatives, or $B_6$--cycles, but for compact non-singular $G_2$ manifolds it is a well-known theorem that there are no non-trivial 6-cycles, hence the latter are ruled out by topology, unless they trigger a supersymmetry breaking. Also, they could emerge in the context of flux compactifications, as well as in the case in which the hypothesis of compactness of the corresponding $G_2$-background is dropped.

These remarks make the homological version of $G_2$ mirror symmetry much more interesting than naively expected. In particular, the fact that the two spectra of BPS particles must agree in IIA and IIB gives a rather non-trivial prediction on the structure of associative, coassociative, and $B_n$--cycles across the $G_2$ mirror map. Notice that while coassociative are stable against perturbations of the $G_2$ structure, associatives are not and decay. In a IIB compactification, D3 branes wrapped on associatives give rise to BPS particles and vortexes in 3D. The fact that associatives decay corresponds to phase transitions in the BPS spectrum. D4 branes wrapped on coassociatives in IIA cannot reproduce this behavior, but we expect that it can be matched by D2 branes wrapped on $B_2$-cycles.

\subsection{Singular mirrors and non-Abelian gauge symmetry}\label{sect:sing_mirror}

In order to obtain gauge theory subsectors or any kind of non-trivial dynamics, we need that our backgrounds have singularities. For a Type IIA background, singularities in codimension 4,6,7 are understood and correspond respectively to non-Abelian gauge groups, matter organized in chiral plus antichiral $\cn=2$ multiplets, and matter organized in $\cn=2$ multiplets of given chirality for the resulting 3d theory. This follows by a circle reduction of the M-theory setup \cite{Acharya:2000gb,Gutowski:2001fm,Atiyah:2001qf,Witten:2001uq,Acharya:2001gy,Acharya:2004qe}. Using our mirror map we can translate this information to the Type IIB backgrounds. If dualities of the form
\begin{equation}
\begin{aligned}
&\Tf \quad \colon \quad IIA \text{ on } J \longleftrightarrow IIA \text{ on } J^\vee\\
&\Tt \quad \colon \quad IIA \text{ on } J \longleftrightarrow IIB \text{ on } J^\wedge
\end{aligned}
\end{equation}
hold, then the corresponding 3d $\cn=2$ theories must agree, which in particular entails that the corresponding gauge sectors must agree. Here we present some preliminary evidences for the existence of the relevant structures needed to establish that this is indeed the case, leaving a detalied analysis of this interesting question for the future.

\subsubsection{$\cn=4$ gauge subsectors and $G_2$ mirrors pairs}

\begin{table}
\begin{center}
\begin{tabular}{c|c|c}
$J$ \phantom{$\Bigg|$} & \phantom{$\Bigg|$}  $J^\wedge$ \phantom{$\Bigg|$}& \phantom{$\Bigg|$} $J^\vee$\\
\hline
\hline
$K_-$ \phantom{$\Bigg|$} & $H^\wedge_-$ & $H^\vee_-$ \\
\hline
$K_+$ \phantom{$\Bigg|$} & $K^\wedge_+$ & $H^\vee_+$ \\
\hline
$H_-$ \phantom{$\Bigg|$} & $K^\wedge_-$ & $K^\vee_-$ \\
\hline
$H_+$ \phantom{$\Bigg|$} & $H^\wedge_+$ & $K^\vee_-$ \\
\hline
\end{tabular}
\caption{Action of the mirror symmetry maps on $\cn=4$ subsectors.}\label{tab:schematic_mirror_N=4}
\end{center}
\end{table}

Let us begin by considering the $\Tt$ mirror map in between IIA on $J$ and IIB on $J^\wedge$. Our aim here is to show that the $\cn=4$ subsectors of the 3d $\cn=2$ theories so obtained agree. The Kovalev limit of $J$ is given by
\begin{equation}
J \to X_+ \, \cup_{\,W} \, X_-
\end{equation}
and the $\Tt$ mirror dual has Kovalev limit
\begin{equation}
J^\wedge \to X_-^\vee \,\cup_{\,W^{\wedge}} \, X_+.
\end{equation}
Recall from Section \ref{sec:massless_modz} that the $\cn=4$ subsectors arise from the KK modes corresponding to the sublattices $K_\pm$ and $H_\pm$ of cohomology. Under our mirror map we have that
\begin{equation}
K_-  = H^\wedge_- \,,\qquad K_+ = K^\wedge_+ \,, \qquad H_- =  K^\wedge_-  \,,\quad\text{and}\quad H_+ =  H^\wedge_+.
\end{equation}
On the $X_-$ side, the duality follows from the usual arguments in the context of compactifications on CY 3-folds \cite{Katz:1997eq}. Indeed $K_-$ plays the role of the K\"ahler moduli, while $H_-$ plays the role of the complex structure moduli in ordinary mirror symmetry. The 3D $\cn=4$ subsectors that arise from $K_-$ and $H_-$ have a 4D $\cn=2$ origin. The equivalence of IIA on $X_-$ and IIB on $X_-^\vee$ is sufficient to prove the equivalence of the latter 4D $\cn=2$ models. Then, to conclude, one has to remark that the outer $S^1$ on the $J_-$ side of the Kovalev limit is left untouched by the $\Tt$ mirror map. The 3D $\cn=4$ subsectors corresponding to $X_-$  and $X_-^\vee$ are obtained by the dimensional reduction on the outer $(S^1)_-$ at the same radius of the corresponding 4D $\cn=2$ models, and hence agree. 

Let us discuss the case of a single non-Abelian gauge group in detail. In type IIA geometric engineering, a given 4D $\cn=2$ gauge theory subsector with gauge group $G$ arises from a local CY three-fold $X$ by fibering a K3 with an ADE singularity along a base $\mathbb{P}^1$. The Cartan $U(1)$s correspond to the modes of $C^{(3)}$ along the K\"ahler classes that resolves the ADE singularity. If the gauge group is non-simply laced one obtains a fibration with monodromies implementing the corresponding outer automorphisms that from a simply-laced group map to a non-simply laced one \cite{Bershadsky:1996nh}. Resolving the singularities one obtains $-2$ curves in the K3 that intersect with intersection matrix that equals the Cartan matrix of the corresponding group. The W-bosons arise from wrapped D2 branes on such 2-cycles. In this case the 3D gauge coupling is governed by $S^1 \times \mathbb{P}^1_\text{base}$, that indeed is an associative in the $G_2$ structure corresponding to $J_-$.

In a Type IIB setup, the resulting local model for the mirror geometry is
\begin{equation}
e^Z + e^{-Z} + W_G(X,Y,U) = 0,
\end{equation}
where $W_G(X,Y,U)$ are canonical ADE singularities (orbifolded in the non-simply laced case). The complex structure deformation of such a CY hypersurface encodes the corresponding moduli. One obtains a bouquet of $2r$ rigid special Lagrangian $S^{\, 3}$'s that correspond to the Milnor fibration of such a complex singular hypersurface, where $r$ is the rank of the corresponding gauge group. It is always possible (up to Picard-Lefschetz monodromies) to chose a basis such that the anti-symmetric intersection matrix among such rigid $S^{\, 3}$'s is \cite{Cecotti:2010fi,Cecotti:2012gh}
\begin{equation}
\left(\begin{matrix}
C_G- C_G^t & C_G^t\\
-C_G & 0
\end{matrix}\right)
\end{equation}
where $C_G$ is the Cartan matrix of the corresponding gauge group. The W-bosons in this case are more tricky to obtain, because these arise as boundstates of monopoles and dyons, engineered by wrapping the D3 branes on the rigid $S^{\, 3}$ special lagrangians along the Milnor fiber, which is the IIB counterpart of the arguments in \cite{Fiol:2000pd}. Nevertheless, one can show that these are always present also in the Type IIB setup \cite{Cecotti:2012va, Cecotti:2012gh}. Adding $\cn=4$ hypermultiplets to this story one obtains systems that can be treated similarly, along the lines of \cite{Katz:1996xe,Cecotti:2012va,Cecotti:2013lda}. Since $J_- \simeq S^1 \times X_-$, for the modes that do not participate in the glueing, the equivalence of the corresponding 3D $\cn=4$ subsectors follows from the equivalence of their 4D $\cn=2$ avatars on $X_-$ and $X_-^\vee$.

Therefore, we conclude that on the $X_-$ side of the TCS the 3D $\cn=4$ theories agree by ordinary mirror symmetry. For the $\cn=4$ modes corresponding to $X_+$ one uses the fact that compactifications of IIA and IIB on $S^{\, 1} \times X_+$ are T-dual, hence give rise to the same 3d $\cn=4$ spectra (c-map \cite{Cecotti:1988qn}), as we remarked in Section \ref{sec:superstring_back}.

Let us now consider the case of the $\Tf$ mirror map. The $\Tf$ mirror of a $G_2$ background with Kovalev limit $J \to X_+ \, \cup_{\, W} \, X_-$ is given by
\begin{equation}
J^\vee \to X^\vee_+ \, \cup_{\, W^\vee} \, X^\vee_-.
\end{equation}
The 4D $\cn=2$ avatars of the $X_\pm$ and of the $X^\vee_\pm$ modes are indeed equivalent under mirror symmetry, but in this case the outer $S^1$ on $J_\pm$ undergoes a $T$-duality too, therefore we obtain that from the chain of equivalences
\begin{equation}
(\text{IIA on } S^1 \times X ) \leftrightarrow  (\text{IIB  on } S^1 \times X^\vee) \leftrightarrow   (\text{IIA on } (S^1)^\vee \times X^\vee)
\end{equation}
it follows that the $\cn=4$ subsectors of IIA on $J$ and IIA on $J^\vee$ have to agree. One can argue similarly for the IIB case.

\subsubsection{$\cn=2$ gauge subsectors and $G_2$ mirrors pairs}

\begin{table}
\begin{center}
\begin{tabular}{c|c|c}
$J$ \phantom{$\Bigg|$} & \phantom{$\Bigg|$}  $J^\wedge$ \phantom{$\Bigg|$}& \phantom{$\Bigg|$} $J^\vee$\\
\hline
\hline
$\hat N_-$ \phantom{$\Bigg|$} & $\hat T_-^\wedge$ & $\hat T_-^\vee$ \\
\hline
$\hat N_+$ \phantom{$\Bigg|$} & $\hat N_+^\wedge$ & $\hat T_+^\vee$ \\
\hline
$\hat T_-$ \phantom{$\Bigg|$} & $\hat N_-^\wedge$ & $\hat N_-^\vee$ \\
\hline
$\hat T_+$ \phantom{$\Bigg|$} & $\hat T_+^\wedge$ & $\hat N_+^\vee$ \\
\hline
\end{tabular}
\caption{Action of the mirror symmetry maps relevant to characterize $\cn=2$ subsectors.}\label{tab:schematic_mirror_N=2}
\end{center}
\end{table}

Consider now the case of the 3D $\cn=2$ gauge subsectors. As we have discussed in Section \ref{sec:superstring_back} the $\cn=2$ vectormultiplets arise from the matching of the K3 fibers for TCS $G_2$ manifolds. This entails that the corresponding $\cn=2$ gauge groups are encoded in the ADE sublattices of the asymptotic K3 fibers and in the way they intersect along $W$. Here, for the sake of simplicity, we neglect the effects of monodromies that can give rise to non-simply laced versions of the gauge symmetry groups. In a Type IIA engineering there are two possible sources for $\cn=2$ non-Abelian gauge groups. It is natural to expect that these arise whenever there are nontrivial ADE sublattices of the type
\begin{equation}
\Gamma_G^{(2)} \subseteq  N_+ \cap N_-
\end{equation} 
or
\begin{equation}
\Gamma_G^{(4)} \subseteq (T_+ \cap T_-) \oplus  \Gamma^{3,19} /(T_+ + N_-) \oplus \Gamma^{3,19} /(N_+ + T_-).
\end{equation} 
The notation $\Gamma^{(n)}_G$ is a shorthand to denote a $G$ type ADE sublattice for the second homology of $S_0$, the K3 fiber of W, while $n$ is a bookkeeping to remind ourselves of the degree of the cohomology group of the TCS in which the corresponding sublattices sit, from the Mayer-Vietoris argument. For instance $\Gamma_{A_2}^{(2)}$ would indicate that inside $N_+ \cap N_-$ there is an $A_2$ lattice, while $N_+ \cap N_-$ contributes to $H^{2}(J,\mathbb{Z})$. 

Back to our discussion now. The massless KK modes of $C^{(3)}$ (resp. $C^{(5)}$) give rise to vectors corresponding to the $N_+ \cap N_-$ sublattice of $H^2(J)$ (resp. to the $(T_+ \cap T_-) \oplus  \Gamma^{3,19} /(T_+ + N_-) \oplus \Gamma^{3,19} /(N_+ + T_-)$ sublattices of $H^4(J)$). The corresponding W-bosons arise from wrapped D2 branes and D4 branes respectively. The stringy instantons that generates the corresponding superpotentials arise by Poincar\'e duality. Indeed, for each elements in a $\Gamma^{(2)}_G$ lattice, there is a corresponding 2-form in the $G_2$ manifold, which in turn is Poincare' dual to a five cycle which, wrapped by euclidean D4 branes, give rise to a superpotential correction of the type we discussed in Section \ref{sec:nAB_IIA}. 
Similarly, the elements in $\Gamma^{(4)}_G$ correspond to three cycles which, wrapped by euclidean D2 branes, give rise to analogous superpotential corrections. In a Type IIB engineering non-Abelian gauge $\cn=2$ sectors can arise from ADE sublattices
\begin{equation}
\Gamma_G^{(3)} \subseteq  \Gamma^{3,19} /(N_+ + N_-) \oplus (T_+ \cap N_-) \oplus (T_- \cap N_+)
\end{equation} 
or
\begin{equation}
\Gamma_G^{(5)} \subseteq \Gamma^{3,19} /(T_+ + T_-)
\end{equation}
which correspond to vectors arising from $C^{(4)+}$ and $C^{(6)}$ respectively. The W-bosons in these cases arise from wrapped D3 and D5 branes respectively. The stringy instantons that generates the corresponding superpotentials arise by Poincar\'e duality here as well. \footnote{ A full-fledged analysis of these geometric engineerings is a rather interesting open problem that goes beyond the modest scope of this subsection.}

The effect of the generalized mirror symmetry maps on the elements of cohomology that are relevant for $\cn=2$ subsectors is summarized, schematically, in Table \ref{tab:schematic_mirror_N=2}. There we use the more transparent version that arises at the level of the $\Gamma^{4,20}$ lattices of the K3s. Of course, as we have emphasized in Section \ref{sect:cohom_action} the actual mirror map at the level of cohomology depends crucially on the splitting of the transcendental lattices $T_\pm$ in $U$ and its orthogonal complement $\tilde{T}_\pm$. For the sake of clarity, in the discussion in this Section we are sloppy about this fact, and assume that all the various $\Gamma_G^{(n)}$ sublattices are not affected by this splitting. Under this assumption, we can forget about the distinction in between $T$ and $\tilde{T}$ in the discussion that follows.

Let us begin by considering the $\Tt$ mirror map from IIA to IIB.  Considering the $C^{(3)}$ duality frame in IIA. In that frame the only source of non-Abelian gauge symmetry is an ADE sublattice $\Gamma_G^{(2)} \subseteq N_+ \cap N_-$. The $\Tt$ mirror of such a lattice is an ADE sublattice $\Gamma_G^{(3)} \subseteq N_+^\wedge \cap T_-^\wedge$. The latter in IIB gives rise to the an ADE gauge subsector choosing the duality frame corresponding to the vector moduli of the $C^{(4)+}$ RR potential. Hence wrapped D2 branes (resp. euclidean D4 branes) are mapped to wrapped D3 branes (resp. euclidean D3 branes). Consider now the $C^{(5)}$ duality frame in IIA. The modes arising from
\begin{equation}
\Gamma_G^{(4)} \subseteq (T_+ \cap T_-) \oplus \Gamma^{3,19} /(N_+ + T_-)
\end{equation} 
are mapped under the $\Tt$ mirror map to
\begin{equation}
\Gamma_G^{(3)} \subseteq (T_+^\wedge \cap N_-^\wedge) \oplus \Gamma^{3,19} /(N_+^\wedge + N_-^\wedge)
\end{equation}
which in IIB gives rise to the an ADE gauge subsector as discussed above. In particular wrapped D4 branes (resp. euclidean D2 branes) are mapped to wrapped D3 branes (resp. euclidean D3 branes). The modes arising from \begin{equation}
\Gamma_G^{(4)} \subseteq \Gamma^{3,19} /(T_+ + N_-)
\end{equation}
are mapped to 
\begin{equation}
\Gamma_G^{(5)} \subseteq \Gamma^{3,19} /(T_+^\wedge + T_-^\wedge)
\end{equation}
and hence give rise to vectors in the $C^{(6)}$ duality frame for IIB, wrapped D4 branes are mapped to wrapped D5 branes and wrapped euclidean D2 branes are mapped to wrapped euclidean D1s. This concludes our argument that the $\Tt$ mirror map respects non-Abelian gauge subsectors: for every possible realization of a non-Abelian gauge sector on one side of the duality, there is a corresponding duality frame on the other side that renders manifest the corresponding gauge symmetry.

Let us proceed by showing that this is indeed the case also for the $\Tf$ mirror map in between IIA on $J$ and IIA on $J^\vee$. Consider the $C^{(3)}$ duality frame on $J$. Again in that case the only source of non-Abelian gauge sectors is
\begin{equation}
\Gamma_G^{(2)} \subseteq N_+ \cap N_-.
\end{equation}
The latter is mapped to 
\begin{equation}
\Gamma_G^{(4)} \subseteq T_+^\vee \cap T_-^\vee.
\end{equation}
in the dual, that indeed corresponds to the same gauge group in the $C^{(5)}$ duality frame on $J^\vee$. Similar sublattices of $\Gamma^{3,19}/(N_\pm + T_\mp)$ are mapped to each other and give rise to a consistent self-dual subsector of the $C^{(5)}$ frame. Similarly for type IIB the ADE sublattices of $(T_\pm \cap N_\mp)$ are mapped to each other by the $\Tf$ map, while ADE sublattices
\begin{equation}
\Gamma_G^{(3)} \subseteq \Gamma^{3,19}/(N_++N_-)
\end{equation}
are mapped to 
\begin{equation}
\Gamma_G^{(5)} \subseteq \Gamma^{3,19}/(T_+^\vee+T_-^\vee).
\end{equation}
Hence vectors arising from $C^{(4)+}$ (resp. D3 branes, and dual euclidean D3 branes) for $\Gamma_G^{(3)} $ are mapped to vectors arising from $C^{(6)}$ (resp. D5 branes, and dual euclidean D2 branes) for $\Gamma_G^{(5)}$. This concludes our argument that the $\Tf$ mirror map respects non-Abelian gauge subsectors.

\subsubsection{Singular mirrors}
The physical reasoning above has the following apparently paradoxical mathematical counterpart: depending on the given TCS $G_2$ manifold $J$, its mirrors $J^\vee$ or $J^\wedge$ can end up being singular spaces. This happens as follows. First note that the matching in \eqref{eq:HKrotation} implies \eqref{eq:locationofHKstrinG319}, which in turn implies that
\begin{equation}
\begin{aligned}
(\Sigma_3)_\pm &\perp (N_+ \cap N_-)\\
\end{aligned}
\end{equation}
where $(\Sigma_3)_\pm$ is the space-like three-plane spanned by $\omega_\pm$, $R_\pm$ and $I_\pm$ in $\Gamma^{3,19} \otimes \mathbb{R}$. If we fix a point in the moduli space of Ricci-flat metrics of a K3 surface $S$, $S$ has ADE singularities if and only if there are vectors $\gamma$ of length $-2$, also called `roots', such that 
\begin{equation}
\Sigma_3 \perp \gamma \, . 
\end{equation}
The type of ADE singularities is then determined by the root lattice generated by such $\gamma$. If we try to construct a matching for which $N_+\cap N_-$ contains a root sub-lattice $G$ of ADE type, the asymptotic $K3$ fibres $S_{\pm}$ hence necessarily have the corresponding singularities.  As only the holomorphic two-form varies along the base $\P^1$s of the building blocks $Z_\pm$, this means that every $K3$ fibre of $Z_+$ and every $K3$ fibre of $Z_-$ in fact has singularities $G$ (as discussed in \cite{Guio:2017zfn,Braun:2017uku}). We expect such singularities to glue to an $S^{\, 3}$ worth of ADE singularities of the $G_2$ manifold $J$, thus giving rise to a non-Abelian $\cn=2$ gauge subsector.

In the following, let us focus on the mirror maps associated with four T-dualities for definiteness. There are mirror pairs for which $N_+ \cap N_-$ does not contain any roots but $N^\vee_+ \cap N^\vee_-$ does. However, our mirror map not only relabels $N_+\cap N_-$ as  $\tilde{T}^\vee_+ \cap \tilde{T}^\vee_-$ but also $B$ as $\pm \text{Im}(\Omega_\pm^\vee)$:
\begin{equation}
\begin{aligned}
  (N_+\cap N_-)\otimes\R \ \,\, \leftrightarrow \,\, (\tilde{T}^\vee_+ \cap \tilde{T}^\vee_-)\otimes\R  \\
  \hookuparrow \hspace{3.5cm}  \hookuparrow \hspace{.5cm}\\
  B   \hspace{1cm}  \leftrightarrow  \hspace{1cm}  \pm \text{Im}(\Omega_\pm^\vee)
\end{aligned}
\end{equation}
For a sufficiently generic choice of $X_\pm$, 
\begin{equation}
 \int_C \text{Im}(\Omega_\pm) \neq 0
\end{equation}
for any curve $C$ associated with a root in $\tilde{T}^\vee_+ \cap \tilde{T}^\vee_-$. Under the mirror map, this means that 
\begin{equation}
 \int_C B \neq 0
\end{equation}
for every rational curve in $N_+\cap N_-$. In other words, a compactification of IIA string theory on $J^\vee$ has no non-Abelian gauge symmetry and only the $U(1)$s associated with the Cartan subgroup are unbroken. Together with two-cycles of finite size, these are captured by the contribution $|N_+\cap N_-|$ to $b_2(J^\vee)$. If $J^\vee$ is indeed the mirror of $J$, we can hence argue that 
\begin{equation}
b_2(J^\vee) = |N_+^\vee \cap N_-^\vee| + |K_+^\vee| + |K_-^\vee| 
\end{equation}
and fix $b_3$ by
\begin{equation}
b_2(J) + b_3(J) = b_2(J^\vee) + b_3(J^\vee) 
\end{equation}
which is equivalent to applying the formula \eqref{eq:bettinumbers}. 

If, one the other hand, some of the roots in $N_+ \cap N_-$ do have vanishing periods of the $B$-field $B_\pm$, so that a non-Abelian gauge groups remains unbroken, the periods of $\text{Im}(\Omega^\vee)$ on the same roots vanish for the mirror. Hence the mirror has the same unbroken non-Abelian gauge groups. A similar discussion hold for the mirror map associated with three T-dualities. 

In general, for a fixed matching of geometry \eqref{eq:HKrotation} together with the B-field \eqref{eq:Bfieldmatch}, the non-Abelian gauge group is given by the lattice spanned by the roots $\gamma$ in $\Gamma^{4,20}$ orthogonal to $\hat{\Sigma}_4$:
\begin{equation}
 G = \langle\gamma \in \hat{\Sigma}_4^\perp \,\, |\,\, \gamma^2 = -2 \rangle \, .
\end{equation}
As this condition is manifestly invariant under the mirror map, it follows that the non-Abelian gauge groups are preserved.

\subsection{Multiple TCS mirrors and further generalization}\label{sec:general_mir}

The (co)associative fibrations used to construct our mirror maps restrict to calibrated fibrations of $S_\pm$ by tori, and one may wonder if it is possible to find different mirror maps (of type $\Tt$ or $\Tf$) by exploiting different calibrated $T^2$ fibrations on $S_\pm$ for a given TCS $G_2$ manifold $J$. In the light of \cite{Shatashvili:1994zw}, which conjectures that any two $G_2$ manifolds with same $b^2+b^3$ lead to mirror compactifications of type II strings, it becomes a particularly interesting question if this may lead to mirrors with different topologies. 

Let us first address this in the case $\Tt$. Let us hence assume that $S_+$ has several elliptic fibrations, calibrated by $\omega_+$, which are matched with special Lagrangian fibrations on $S_-$ calibrated by $\text{Re}(\Omega_-)$. Denoting the fibres of these fibrations on both sides by $E_\pm^i$, we furthermore assume that $E_-^i \times S^{\, 1}$ defines a special Lagrangian fibration on $X_-$ for some complex structure. We can then apply our mirror map $J \rightarrow J^{\wedge i}$ and deduce from the arguments given above that $b_2 + b_3$ stays invariant in each case. We can hence detect mirrors of different topology by comparing 
\begin{equation}
 b_2(J^{\wedge i}) = |N_+ \cap N_-^{\wedge i}| + |K_+| + |K_-^{\wedge i}| \, 
\end{equation}
for various choices of $E_i$. First note that the last term is equal to $h^{2,1}(X_-)$, so that it in particular does not depend on the fiber $E_\pm^i$ chosen. For the first term, note that the existence of an elliptic fibration on $S_+$ which is mapped to a special Lagrangian fibration calibrated by $\text{Re}(\Omega_-)$ on $S_-$ means that there is a primitive embedding
\begin{equation}\label{eq:embt2}
 U_{E^i} \hookrightarrow N_+ \cap T_- 
\end{equation}
which fixes the calibrated torus $E_i$. This implies that there is a decomposition
\begin{equation}
 N_+ \cap T_-  = U_{E^i} \oplus L(E_i) \, ,
\end{equation}
i.e. there exists a lattice $L(E_i)$ such that the above relation holds for every embedding \eqref{eq:embt2}. As the mirror map acts on $S_-$ such that $N_-^{\wedge i} = \tilde{T}_i$ where $T = U_{E^i} \oplus \tilde{T}_i$, we conclude that 
\begin{equation}
 N_+ \cap N_-^{\wedge i} =L(E_i) \, ,
\end{equation}
which always has the same rank $|N_+ \cap T_-|-2$ for a given TCS $G_2$ manifold $J$ irrespective of the choice of $E_i$. Hence we can conclude that
\begin{equation}
 b_2(J^{\wedge i}) = b_2(J^{\wedge j})
\end{equation}
for all choices of $E^i$, $E^j$. However, different choices of $E_i$ lead to different root sublattices in $L(E_i)$, which in turn leads to different singularities for the mirrors $J^{\wedge i}$. We leave a thorough exploration of this to future work.

The fact of having a Kovalev limit for $J$ is very interesting, but it is a strong requirement, and it is possible to drop it in favour of a slightly less restrictive setup. The key ingredients for our construction to work are the following. First, it should be possible to cut $J$ open along a hypersurface, $W$ in such a way that $J \setminus W$ has two connected components $J_+$ and $J_-$ each with boundary $W$. Identifying the common boundary, induces a diffeomorphism 
\begin{equation}
\Xi_W \colon J_+ \longrightarrow J_-
\end{equation}
such that it is possible to recover $J$ as a (twisted) connected sum
\begin{equation}
J = J_- \sqcup_{\, \Xi_W} J_+.
\end{equation}
Second, the two seven-manifolds $J_\pm$ should have some extra structure for generalized mirror symmetry to take place. In our examples 
\begin{equation}\label{eq:CY_asymptotes}
J_\pm \simeq (S^{\, 1})_\pm \times X_\pm,
\end{equation}
where $X_\pm$ are CYs with boundaries such that
\begin{equation}\label{eq:non_canonical}
(S^{\, 1})_\pm \times \partial X_\pm \simeq W.
\end{equation}
If $W \simeq T^2 \times S$ where $S$ is a K3, then we obtain a Kovalev limit for $J$, but this is not the case in general. For the case of Equation \eqref{eq:CY_asymptotes}, this isomorphism allows to reconstruct the $G_2$ structure of $J$ out of the CY structures of $X_\pm$, in particular we must have that
\begin{equation}
\varphi_J \big|_{J_\pm \setminus W} = d \xi_\pm \wedge \omega_\pm + \text{Re}(\Omega^{3,0}_\pm)\qquad\text{and}\qquad (\star \varphi)_J \big|_{J_\pm \setminus W} = \tfrac{1}{2} \omega_\pm \wedge \omega_\pm + d \xi_\pm \wedge \text{Im}(\Omega^{3,0}_\pm)
\end{equation}
where $\xi_\pm$ parametrize $(S^{\, 1})_\pm$. Assuming both $X_\pm$ admit CY mirrors $(X_\pm)^\vee$, we can repeat word for word our argument above and glue their SYZ $T^3$ fibrations into a coassociative $T^4$ fibration of $J$. Four fiberwise T-dualities thus lead to a generalized mirror map $J\to J^\vee$. Similarly, if $X_-$ admits a CY mirror, $X_+$ has an elliptic fibration with fiber $E_+$, and if the SYZ fibration of $X_-$ restricted to $W$ gets identified with the restriction of $(S^{\, 1})_+\times E_+$ to $W$, it is possible to glue them together to form an associative $T^3$ fibration of $J$. Then three fiberwise T-dualities lead to a generalized mirror map $J\to J^\wedge$.

Of course, there are more possibilities, corresponding to the fact that not all the generalized mirror symmetries for a given $G_2$ manifold are consequences of ordinary mirror symmetries for CYs. For instance, if $J_\pm$ are $G_2$-manifolds with $G_2$-structures $\varphi_\pm$ and $(\star \varphi)_\pm$ and boundaries
\begin{equation}
\partial J_\pm \simeq W
\end{equation}
such that
\begin{equation}
\varphi_J \big|_{J_\pm \setminus W} = \varphi_\pm\qquad\text{and}\qquad (\star \varphi)_J \big|_{J_\pm \setminus W} = (\star \varphi)_\pm,
\end{equation}
then it is still possible to glue coassociative and associative cycles of $J_\pm$ into coassociative and associative cycles of $J$. In particular, if $J_\pm$ admit coassociative $T^4$ (resp. associative $T^3$) fibrations it is possible to glue them together to form coassociative $T^4$ (resp. associative $T^3$) fibrations of $J$. As long as the cohomological properties of $J_\pm$ and $W$ are known, the Mayer-Vietoris argument still determines the cohomological properties of $J$. It is interesting to exploit this fact to check that the corresponding generalized $G_2$-mirror pair indeed satisfies the Shatashvili-Vafa relation \eqref{eq:SVR}, however this is technically slightly more involved than the TCS case and we leave it for future work.

\section{A Joyce Orbifold and its Smoothing as a TCS}\label{sec:Joyce_ex}

The first examples of compact manifolds with holonomy $G_2$ were constructed by Joyce \cite{joyce1996I,joyce1996II,joyce2000compact}, as resolutions of orbifolds $T^{\,7}/ \, \Gamma$ for appropriate groups $\Gamma$. Some of the Joyce examples have been analyzed in details using 2d SCFT techniques \cite{Shatashvili:1994zw,Acharya:1996fx,Gaberdiel:2004vx} and the relation among generalized $G_2$-mirror symmetry and T-duality has been addressed. In particular, the example considered in \cite{Gaberdiel:2004vx} can be realized as a twisted connected sum, and this gives rise to an interesting consistency check of the ideas discussed above to which we now turn.\footnote{ There are more examples of Joyce orbifolds that can be realized as TCS. A systematic analysis of the relation among the worldsheet results and the geometric construction presented in this paper goes beyond the scope of this work, and it will be discussed elsewhere.}

\subsection{Explicit realization of the TCS structure on the orbifold}

To present this example we follow the discussion in \S 12.3 of \cite{joyce2000compact}. We describe the seven-torus $T^{\, 7}$ using coordinates $x_i$ with identifications $x_i \sim x_i + 1$ and $i=1,\dots,7$. The flat $G_2$-structure on $\mathbb{R}^7$ descends to $T^{\, 7}$
\begin{equation}\label{eq:flat_G2_struct}
\begin{aligned}
\varphi_o &= dx_{123} + dx_{145} + dx_{167} + dx_{246} - dx_{257} - dx_{347} - dx_{356} \\
\star \varphi_o &= dx_{4567} + dx_{2367} + dx_{2345} + dx_{1357} - dx_{1346} - dx_{1256} - dx_{1247} 
\end{aligned}
\end{equation}
where $dx_{ijk} = dx_i \wedge dx_j \wedge dx_k$. Denote
\begin{equation}
r_i \equiv x_j \mapsto (1 - 2 \delta_{ij}) \, x_j \quad\text{and}\quad s_i \equiv x_j \mapsto  x_j + \delta_{ij} \, / \, 2.
\end{equation}
The Joyce orbifold group of $T^{\, 7}$ is generated by 
\begin{equation}\label{eq:orbiaction2}
\alpha \equiv r_4 \circ r_5 \circ r_6 \circ r_7,\qquad \beta \equiv   r_2 \circ r_3 \circ (s_6 \circ r_6) \circ r_7,\qquad \gamma \equiv  r_1 \circ r_3 \circ r_5 \circ (s_7 \circ r_7),
\end{equation}
acting on the coordinates as follows
\begin{equation}
\begin{aligned}
&\alpha \colon (x_1,x_2,x_3,x_4,x_5,x_6,x_6) \to (x_1,x_2,x_3,-x_4,-x_5,-x_6,-x_7)\\
&\beta \colon (x_1,x_2,x_3,x_4,x_5,x_6,x_6) \to (x_1,-x_2,-x_3,x_4,x_5,\tfrac12 -x_6,-x_7)\\
&\gamma \colon (x_1,x_2,x_3,x_4,x_5,x_6,x_6) \to (-x_1,x_2,-x_3,x_4,-x_5,x_6,\tfrac12-x_7)\\
\end{aligned}
\end{equation}
This action preserves the flat $G_2$ structure of Equation $\eqref{eq:flat_G2_struct}$, which descends to the quotient
\begin{equation}
J \equiv T^{\, 7} / \langle \alpha, \beta, \gamma\rangle.
\end{equation}
Let us proceed by showing that this singular $G_2$-manifold (as well as its resolutions) admit a Kovalev limit, adapting to this case an argument of \cite{Kovalev2010}. In order to do this, we need to have a better understanding of the action of the orbifold group $\langle \alpha, \beta , \gamma \rangle$ on $T^{ \,7}$. First of all, notice that all generators commute and square to the identity because of the identifications $x_i \sim x_i + 1$. This entails that 
\begin{equation}
\langle \alpha, \beta, \gamma \rangle \simeq (\Z_2)_\alpha \times (\Z_2)_\beta \times (\Z_2)_\gamma.
\end{equation}
Notice that the element $\beta \alpha = s_6 \circ r_2 \circ r_3 \circ r_4 \circ r_5$ acts freely on $T^{ \,7}$ because the $\tfrac12$-shift $s_6$ has no fixed points on $x_6$. The same argument holds for the elements $\gamma \alpha,  \gamma \beta$ and $\gamma \beta \alpha$. The elements $\alpha$, $\beta$ and $\gamma$, instead, have non-trivial fixed point sets, given by 16 copies of $T^{\, 3}$ each.\footnote{ Of course, $r_i$ fixes the points $x_i = 0$ and $x_i = \tfrac12$, while $s_i \circ r_i$ fixes the points $x_i = \tfrac14$ and $x_i = \tfrac34$. Moreover, each generator in Equation \eqref{eq:orbiaction2} acts only on four coordinates, leaving the others fixed.} The group $\langle \beta, \gamma \rangle$ (resp. $\langle \alpha, \gamma \rangle$) acts freely on the sixteen 3-tori with stabilizer $\langle \alpha \rangle$ (resp. $\langle \beta \rangle$), giving rise to orbits of four 3-tori each. The group $\langle \alpha, \beta \rangle $ however does not act freely on the sixteen 3-tori with stabilizer $\langle \gamma \rangle$: $\alpha \beta$ acts trivially, while $\alpha$ and $\beta$ gives rise to eight orbits of order two. Therefore the singular set of $J$ is given by eight copies of $T^3$ (four fixed by $\langle\alpha\rangle$ and four fixed by $\langle\beta\rangle$) and eight copies of $T^3/\langle \alpha\beta \rangle$ (fixed by $\langle\gamma\rangle$).

\begin{figure}
\begin{center}
   \scalebox{.5}{ 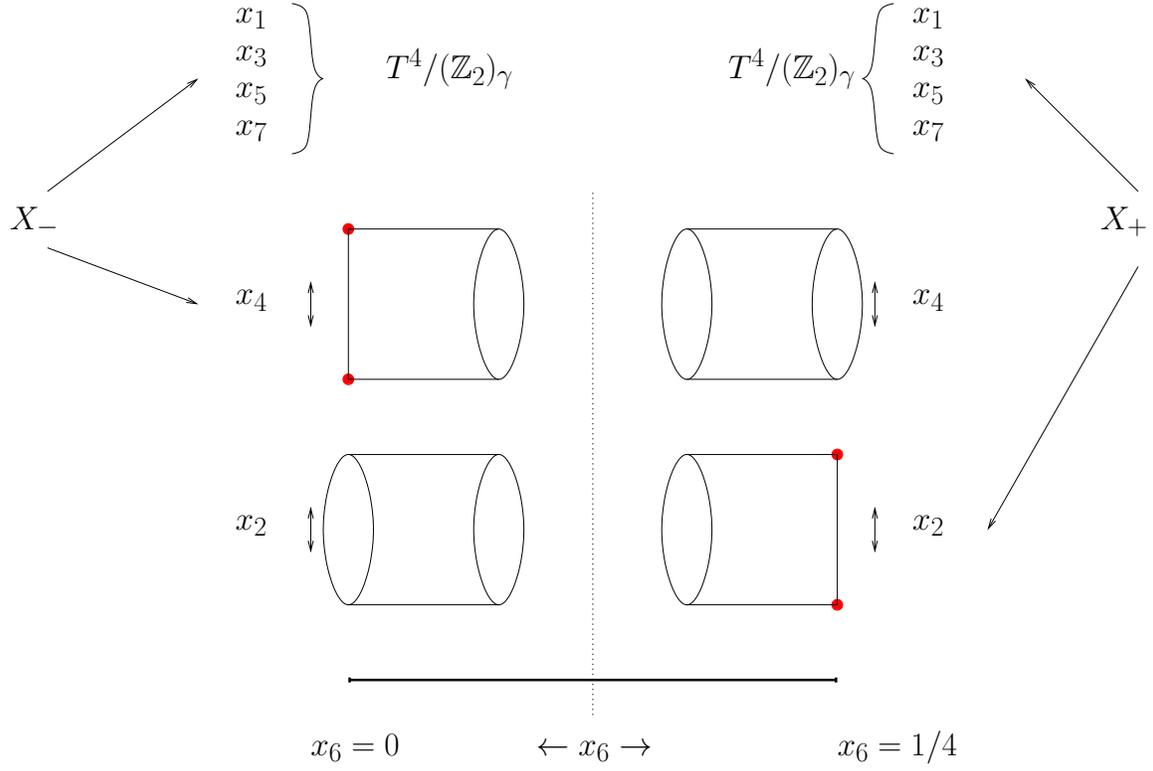 }
\caption{\label{fig:orbi1_2} Cutting along $x_6 = \tfrac18$, the orbifold $J$ is decomposed into two halves $S^{\, 1} \times X_-$ and
$S^{\, 1} \times X_+$ with isomorphic Calabi-Yau orbifolds $X_-$ and $X_+$. These are both realized as a (constant) fibration of a K3 orbifold $T^4/(\Z_2)_\gamma$ over a base which (topologically) forms an open $S^{\, 2}$ due to the degeneration of the $S^{\, 1}$ formed by $x_2$ (for $X_-$) and $x_4$ for ($X_+$). }
\end{center}  
\end{figure}

Consider the orbifold hypersurface $X_0$, defined as the image of the six-torus $\{x_6 = \tfrac18 \} \subset T^7$ under the orbifold group action. Since $x_6 = \tfrac18$ is fixed only by $\gamma$, we have the isomorphism
\begin{equation}
W \simeq \left(T^{\,4}\right)_{x_1,x_3,x_5,x_7} / (\Z_2)_\gamma \times \left(T^{\,2}\right)_{x_2,x_4} \times \{x_6 = \tfrac18\}
\end{equation}
Each of the eight copies of $T^3/\langle\alpha \beta\rangle$ in the singular locus of $J$ fixed by $\gamma$ intersects $W$ along a $T^2$, the other eight 3-tori in the singular locus do not intersect $W$. The complement of $W$ in $J$ is given by two connected components $M_-$ and $M_+$:
\begin{equation}
J\setminus W = M_- \, \sqcup \, M_+
\end{equation}
where $M_-$ is the connected component of $J \setminus W$ that contains $\{ x_6 = 0 \}$, while $M_+$ is the connected component that contains $\{ x_6 = \tfrac14 \}$. In particular, $M_-$ contains the four inequivalent 3-tori in the singular locus of $J$ fixed by $\alpha$ and $M_+$ contains the the four inequivalent 3-tori fixed by $\beta$. The two orbifolds $M_-$ and $M_+$ are diffeomorphic via the involution of $J$
\begin{equation} 
x_j \mapsto x_j + \delta_{6j} /4.
\end{equation}
Consider $M_-$. Its pre-image in $T^{\, 7}/\langle\alpha,\gamma\rangle$ has two connected components which are identified by $\beta$, following \cite{Kovalev2010} we identify
\begin{equation}
M_- \simeq \left(\left\{ -\tfrac18 < x_6 < \tfrac18\right\} \times \left(T^{\, 6}\right)_{x_1,x_2,x_3,x_4,x_5,x_7} \right) / \langle \alpha , \gamma\rangle
\end{equation}
This gives a diffeomorphism
\begin{equation}\label{eq:orcoddio}
M_- \simeq \left(\mathbb{R}_{x_6} \times \left(T^{\, 5} \right)_{x_1,x_3,x_4,x_5,x_7}\right) / \langle \alpha , \gamma\rangle \times \left(S^{\, 1}\right)_{x_2}
\end{equation}
because $\langle \alpha , \gamma\rangle$ acts on $x_2$ as the identity. Let us introduce three complex coordinates and one real coordinate:
\begin{equation}\label{eq:holcoordsZ_-}
\xi_- \equiv x_2, \qquad z^-_1 = x_1 - i x_3, \qquad z^-_2 = - x_5 + i x_7, \qquad z^-_3 = x_4 + i x_6.
\end{equation}
Using these coordinates we see that $M_- \simeq S^{\, 1} \times X_-$ where
\begin{equation}
X_- \simeq \left(\mathbb{R}_{x_6} \times \left(T^{\, 5} \right)_{x_1,x_3,x_4,x_5,x_7}\right) / \langle \alpha , \gamma\rangle
\end{equation}
and notice that the orbifold actions
\begin{equation}
\begin{aligned}
&\alpha \colon (z^-_1 ,z^-_2 ,z^-_3) \mapsto (z^-_1 ,-z^-_2 ,-z^-_3)\\
&\gamma \colon (z^-_1 ,z^-_2 ,z^-_3) \mapsto (- z^-_1 ,-z^-_2 + \tfrac{i}{2} ,z^-_3)\\
\end{aligned}
\end{equation}
are compatible with the K\"ahler form
\begin{equation}
\omega_- \equiv \tfrac{i}{2}(dz^-_1 \wedge d\bar{z}^-_1 +dz^-_2 \wedge d\bar{z}^-_2 +dz^-_3 \wedge d\bar{z}^-_3) = dx_{31} + dx_{75} + dx_{46},
\end{equation}
and with the top form
\begin{equation}\label{eq:orbif_om_CY_minus}
\begin{aligned}
\Omega^{3,0}_- &\equiv dz^-_1 \wedge dz^-_2 \wedge dz^-_3\\
& = - dx_{154} + dx_{374} - dx_{356} - dx_{176} + i \left( dx_{354} + dx_{174} - dx_{156} + dx_{376}\right),
\end{aligned}
\end{equation}
which gives to $X_-$ the structure of a CY orbifold. Moreover, for $x_6 > 0$, $X_-$ is isomorphic to a CY cylinder:
\begin{equation}
X_-\big|_{x_6 > 0} \simeq \mathbb{R}_{x_6 >0} \times \left(S^{\, 1}\right)_{x_4} \times \left(T^{\, 4}\right)_{x_1,x_3,x_5,x_7} / (\Z_2)_\gamma,
\end{equation}
where $t_- \equiv x_6$, $\theta_- \equiv x_4$, and $S_- \equiv \left(T^{\, 4}\right)_{x_1,x_3,x_5,x_7} / (\Z_2)_\gamma$ inherits the K3 structure:
\begin{equation}\label{eq:joyce_left}
\begin{aligned}
&\omega_{S_-} \equiv \tfrac{i}{2}(dz^-_1 \wedge d\bar{z}^-_1 +dz^-_2 \wedge d\bar{z}^-_2) = dx_{31} + dx_{75}, &\\
&\Omega_{S_-} \equiv dz^-_1 \wedge dz^-_2 \equiv  R_{S_-} + i I_{S_-},\\
&R_{S_-} \equiv - dx_{15} + dx_{37},\\
&I_{S_-} \equiv dx_{35} + dx_{17}.
\end{aligned}
\end{equation}
Notice that indeed, with these identifications
\begin{equation}
d\xi_- \wedge \omega_- + \text{Re}(\Omega^{3,0}_-) = \varphi_o.
\end{equation}
Now consider $M_+$. The discussion is identical to that for $M_-$ with the only difference that now the roles of $\alpha$ and $\beta$ are swapped. This suggest to choose the coordinates
\begin{equation}\label{eq:holcoordsZ_+}
\xi_+ \equiv x_4, \qquad z^+_1 = - x_1 + i x_5, \qquad z^+_2 =  x_3 + i x_7, \qquad z^+_3 = x_2 - i x_6
\end{equation}
to realize the isomorphism
\begin{equation}\label{eq:orcoddio2}
M_+ \simeq \left(\mathbb{R}_{x_6} \times \left(T^{\, 5} \right)_{x_1,x_3,x_2,x_5,x_7}\right) / \langle \beta , \gamma\rangle \times \left(S^{\, 1}\right)_{x_4}.
\end{equation}
Again, the orbifold action on the complex coordinates
\begin{equation}
\begin{aligned}
&\beta \colon (z^+_1 ,z^+_2 ,z^+_3) \mapsto (z^+_1 ,-z^+_2 ,-z^+_3- i \tfrac12)\\
&\gamma \colon (z^+_1 ,z^+_2 ,z^+_3) \mapsto (- z^+_1 ,-z^+_2 +  \tfrac{i}{2} ,z^+_3),
\end{aligned}
\end{equation}
is compatible with the K\"ahler form
\begin{equation}
\omega_+ \equiv \tfrac{i}{2}(dz^+_1 \wedge d\bar{z}^+_1 +dz^+_2 \wedge d\bar{z}^+_2 +dz^+_3 \wedge d\bar{z}^+_3) = - dx_{15} + dx_{37} - dx_{26},
\end{equation}
and the top form
\begin{equation}\label{eq:orbif_om_CY_plus}
\begin{aligned}
\Omega^{3,0}_+ &\equiv dz^+_1 \wedge dz^+_2 \wedge dz^+_3\\
& = - dx_{132} - dx_{572} + dx_{536} - dx_{176} + i \left( dx_{532} - dx_{172} + dx_{136} + dx_{576}\right).
\end{aligned}
\end{equation}
This endows the orbifold
\begin{equation}
X_+ \simeq \left(\mathbb{R}_{x_6} \times \left(T^{\, 5} \right)_{x_1,x_3,x_2,x_5,x_7}\right) / \langle \beta , \gamma\rangle
\end{equation}
of a CY structure. Moreover
\begin{equation}
X_+\big|_{x_6 < 0} \simeq \mathbb{R}_{x_6<0} \times \left(S^{\, 1}\right)_{x_2} \times \left(T^{\, 4} \right)_{x_1,x_3,x_5,x_7}/ (\Z_2)_\gamma
\end{equation}
has the structure of a CY cylinder, where $t_+ \equiv - x_6$, $\theta_+ \equiv x_2$. Along the fiber
\begin{equation}
S_+ \simeq \left(T^{\, 4} \right)_{x_1,x_3,x_5,x_7}/ (\Z_2)_\gamma
\end{equation}
the CY structure induces a K3 structure
\begin{equation}\label{eq:joyce_right}
\begin{aligned}
&\omega_{S_+}= - dx_{15} + dx_{37}\\
&R_{S_+} = dx_{31} + dx_{75}\\
&I_{S_+} = dx_{53} + dx_{71}.\\
\end{aligned}
\end{equation}
It is clear that

\begin{equation}
d\xi_+ \wedge \omega_+ + \text{Re}(\Omega^{3,0}_+) = \varphi_o.
\end{equation}
Moreover, to match $M_-$ and $M_+$ with $W$ it is necessary to map
\begin{equation}
\xi_+ = x_4 = \theta_- \qquad  -t_+ = x_6 = t_- \qquad \theta_+ = x_2 = \xi_-
\end{equation}
while the K3 structures on $S_-$ in Equation $\eqref{eq:joyce_left}$ and on $S_+$ in Equation $\eqref{eq:joyce_right}$ are identified precisely by a hyper-K\"ahler rotation --- see Figure \ref{fig:orbi1_2}.

\subsection{Generalized mirror symmetry from calibrated tori}

The mirror maps analyzed in \cite{Gaberdiel:2004vx} correspond to performing either three or four T-dualities. Equivalent T-dualities are obtained for $n$-uples in the following subsets:\footnote{ Note that we have adapted the conventions of \cite{Gaberdiel:2004vx} to ours, which are identical to the ones of \cite{joyce2000compact}. The translation is ($x_{8-i}|_{\text{\cite{Gaberdiel:2004vx}}} = x_{i}|_{\text{\cite{joyce2000compact}}}$).}
\begin{equation}\label{eq:orbif_T3}
\begin{aligned}
& {\mathcal I}_3^+ \equiv \{ (1,6,7) \, , \, (2,4,6) \, , \, (3,5,6) \}\\
& {\mathcal I}_3^- \equiv \{ (1,2,3) \, , \, (1,4,5) \, , \, (2,5,7) \, , \, (3,4,7)\}
\end{aligned}
\end{equation}
\begin{equation}\label{eq:orbif_T4}
\begin{aligned}
&{\mathcal I}_4^+ \equiv \{(1,2,4,7) \, , \, (1,3,5,7) \, , \, (2,3,4,5)\}\\
& {\mathcal I}_4^- \equiv \{(1,2,5,6) \, , \, (1,3,4,6) \, , \, (2,3,6,7) \, , \, (4,5,6,7)\}.
\end{aligned}
\end{equation}
According to our recipe, the maps above should be given by $T$-dualities along calibrated fibrations of $J$. A quick glance to the form of $\star \varphi_o$ and $\varphi_o$ in Equation \eqref{eq:flat_G2_struct} is sufficient to confirm this idea, at least at the level of the orbifold. Indeed, the forms $\star \varphi_o$ and $\varphi_o$ are left fixed by the orbifold group action on the coordinates. In particular each of the coordinates in the $T$-dualities of Equations \eqref{eq:orbif_T3} and \eqref{eq:orbif_T4} gives rise to an invariant form. In each equivalent subset of coordinates, at least two elements correspond to a calibrated associative or coassociative (see Table \ref{tab:orbif_caliber}).\footnote{ These cycles are singular in the orbifold limit. One should check that the $T$-dualities survive the desingularizations. We are dealing with this in Section \ref{sec:Joyce_Resolved} below.}

\begin{table}
\begin{center}
\begin{tabular}{|ccc|}
\hline
$\left(T^{\, 3}\right)_{x_1,x_6,x_7}$ , $\left(T^{\, 3}\right)_{x_2,x_4,x_6}$  $\phantom{\Bigg|}$ & ${\mathcal I}_3^+$  &  \\
\hline
$\left(T^{\, 3}\right)_{x_1,x_2,x_3}$ , $\left(T^{\, 3}\right)_{x_1,x_4,x_5}$ $\phantom{\Bigg|}$ & ${\mathcal I}_3^-$  & \\
\hline
$\left(T^{\, 4}\right)_{x_2,x_3,x_4,x_5}$ , $\left(T^{\, 4}\right)_{x_1,x_3,x_5,x_7}$ $\phantom{\Bigg|}$ & ${\mathcal I}_4^+$ & \\
\hline
$\left(T^{\, 4}\right)_{x_2,x_3,x_6,x_7}$ , $\left(T^{\, 4}\right)_{x_4,x_5,x_6,x_7}$ $\phantom{\Bigg|}$  & ${\mathcal I}_4^-$ & \\
\hline
\end{tabular}
\end{center}
\caption{The tori whose image in the orbifold quotient give rise to the calibrated cycles corresponding to the $T$-dualities of Equations \eqref{eq:orbif_T3}--\eqref{eq:orbif_T4}.}\label{tab:orbif_caliber}
\end{table}

Looking at the directions above, however, it is easy to see that all the maps in the sets ${\mathcal I}_4^-$, and ${\mathcal I}_3^+$ involve the $x_6$ direction.  This is precisely the one corresponding to the infinitely long neck of the Kovalev limit of $J$. This is incompatible with performing a $T$-duality along such a direction. There are other kinds of limits of $J$ in which such maps may be realized, but not the other ones along the lines of our discussion in Section \ref{sec:general_mir}. Before turning to that, we first consider the maps in the sets ${\mathcal I_4^+}$ and ${\mathcal I_3^-}$ involving four and three $T$-dualities.

Let us start with the case of four T-dualities and see that these indeed arise from different $T^3$ special lagrangian fibrations of $X_\pm$. Consider the top form of $X_-$ in Equation \eqref{eq:orbif_om_CY_minus}. We have
\begin{equation}
d\xi_- \wedge \text{Im}(\Omega^{3,0}_-) = dx_2 \wedge (dx_{345} - dx_{147} -dx_{156}-dx_{367}),
\end{equation}
while from Equation \eqref{eq:orbif_om_CY_plus} we obtain
\begin{equation}
d\xi_+ \wedge \text{Im}(\Omega^{3,0}_+) = dx_4 \wedge (dx_{127} +dx_{136}-dx_{235} -dx_{567}).
\end{equation}
From this follows that the maps ${\mathcal I}_4^+$ corresponding to four $T$-dualities along $(1,2,4,7)$ and $(2,3,4,5)$ are indeed realized according to our picture. 

Let us now turn to the case of three T-dualities. From Equation \eqref{eq:orbif_om_CY_minus}, we have that
\begin{equation}
\text{Re}(\Omega^{3,0}_-) =  dx_{145} + dx_{167} - dx_{347} - dx_{356}
\end{equation}
The $T^3$ on $(1,4,5)$ is indeed calibrated by $\text{Re}(\Omega^{3,0}_-)$ and corresponds to one of the realizations of the map ${\mathcal I_3^-}$. Within $M_+$ this same cycle corresponds to the product of $(S^{\, 1})_{\xi_+}$ with an holomorphic $T^2 \subset X_+$, which is calibrated by 
\begin{equation}
dx_{51} = \tfrac{i}{2}dz^+_1 \wedge d\bar{z}^+_1.
\end{equation}
This is precisely an elliptic fiber for the asymptotic K3 given by $S_+ \simeq \left(T^{\,4}\right)_{x_1,x_3,x_5,x_7} /(\Z_2)_\gamma$. Thus, ${\mathcal I}_3^-$ is instance of the generalized mirror map $J \to J^\wedge$ we discussed above.

In order to geometrically realize the $T$-dualities corresponding to ${\mathcal I}^+_3$ and ${\mathcal I}_4^-$, it is possible to cut $J$ open along different orbifold hypersurfaces. The various possible outcomes of this procedure have been considered in Section 7.3.5 of \cite{nordstroem_thesis}. In particular,
\begin{itemize}
\item A generic point along the $x_1$ direction is fixed by $\langle \alpha,\beta \rangle$. We can split $J$ apart along the hypersurface 
\begin{equation}
W = \{x_1 = \tfrac14\} \times \left(T^{\,6}\right)_{x_1,x_2,x_4,x_5,x_6,x_7} / \langle \alpha, \beta \rangle \simeq  \{x_1 = \tfrac14\} \times X_{19,19},
\end{equation}
where $X_{19,19}$ is the famous split bicubic Calabi-Yau. In this case $M_\pm$ are $G_2$-manifolds with a CY boundary. Similarly, a generic point along the $x_2$ (resp. $x_4$) coordinate is fixed by the subgroup $\langle\alpha,\gamma\rangle$ (resp. $\langle \beta, \gamma\rangle$) and can be treated similarly, also giving rise to $M_\pm$ that are $G_2$-manifolds with the same CY boundary (but with a different topology).
\item A generic point along the $x_3$ coordinate is fixed by the subgroup $\langle \alpha, \beta \gamma\rangle$ of the orbifold group. Therefore one can cut $J$ along the hypersurface
\begin{equation}
W = \{x_3 = \tfrac14\} \times \left(T^{\,6}\right)_{x_1,x_2,x_4,x_5,x_6,x_7} / \langle \alpha, \beta \gamma \rangle,
\end{equation}
which is a double cover of $T^2 \times K3$. The two sides $M_\pm$ in this case are asymptotically cylindrical $G_2$ manifolds. The case of $x_5$ is identical, as it is fixed by $\langle \beta , \alpha \gamma\rangle$.
\item A generic point along the $x_7$ coordinate is fixed by $(\Z_2)_{\alpha\beta}\equiv\langle \alpha \beta \rangle$. In this case $(\Z_2)_{\alpha\beta}$ acts without fixed points, hence
\begin{equation}
W = \{x_7 = \tfrac18\} \times \left(T^6\right)_{x_1,x_2,x_3,x_4,x_5,x_6}
\end{equation}
is a regular hypersurface. Moreover we have that $M_- \simeq S^{\, 1} \times X_-$ where $X_-$ is an asymptotically cylindrical CY, and $M_+$ is a $G_2$ manifold which is isomorphic to $(T^3 \times dP_9)/\Z_2$.  
\end{itemize}
It is precisely stretching $J$ along $x_7$ that one can realize the $T$-dualities corresponding to ${\mathcal I}^+_3$ and ${\mathcal I}_4^-$ geometrically using the ideas outlined above. For example, exploiting the fact that $dP_9$ is elliptically fibered one can easily realize ${\mathcal I}^+_3$ explicitly by glueing a special lagrangian $T^3$ fibration of $X_-$ with the $T^3$ fibration of $M_+$ obtained out of the elliptic fiber of $dP_9$ times one of the `outer' circles.

\subsection{Smoothing the Joyce orbifold}\label{sec:Joyce_Resolved}

Recall that the singular locus of $J$ is given by eight copies of $T^3$ (four fixed by $\langle \alpha\rangle$ and four fixed by $\langle \beta\rangle$) and eight copies of $T^3/\langle \alpha \beta\rangle$ (fixed by $\langle\gamma\rangle$). The orbifold group acts also in the normal directions of such loci giving rise to conifold singularities. We have eight singularities of the form
\begin{equation}
T^3 \times \mathbb{C}^2 / \mathbb{Z}_2
\end{equation}
and eight singularities of the form
\begin{equation}
(T^3 \times \mathbb{C}^2 / \mathbb{Z}_2)/\mathbb{Z}_2.
\end{equation}
The former can be resolved by smoothing $ \mathbb{C}^2 / \mathbb{Z}_2$ using the Eguchi-Hansen metric for each of the eight 3 tori in the singular set. The idea is that we cut out a tubular neighbourhood which deformation retracts to $T^3$ and glue in a resolution $T^3 \times Y$ where $Y \simeq T^* \mathbb{P}^1$ is the blowup of $ \mathbb{C}^2 / \mathbb{Z}_2$ at the origin, and deformation retracts to $T^3 \times S^{\, 2}$. Each of these operations increases the Betti numbers of $J$ by the difference between the Betti numbers of $T^3 \times S^{\, 2}$ and those of $T^3$. Each such singularity therefore contributes to the resolution
\begin{equation}
\Delta(b_2,b_3,b_4,b_5) = (1,3,3,1).
\end{equation}
For the latter type of singularities there are two inequivalent choices of resolutions $Y_+$ and $Y_-$ that are distinguished by the action of $\alpha \beta$ on the locus fixed by $\gamma$. Choosing $(T^3 \times Y_+)/\Z_2$ one obtains
\begin{equation}
\Delta(b_2,b_3,b_4,b_5) = (1,1,1,1).
\end{equation}
Choosing $(T^3 \times Y_-)/\Z_2$ instead one obtains
\begin{equation}
\Delta(b_2,b_3,b_4,b_5) = (0,2,2,0).
\end{equation}
We therefore find
\begin{equation}
\begin{aligned}
&b_2(J_\ell) = 8 \times 1 + \ell \times 1 = 8 + \ell \\
&b_3(J_\ell) = 7 + 8 \times 3 + \ell \times 1 + (8-\ell) \times 2 = 47 - \ell,
\end{aligned}
\end{equation}
where $\ell$ is the number of $Y_+$ and $8-\ell$ is the number of $Y_-$, $\ell = 0,...,8$, we have chosen to obtain a manifold $J_\ell$. It is shown in \cite{Gaberdiel:2004vx} that the signature $l$ of $\alpha \beta$ is related to discrete torsion phases and that 
\begin{equation}\label{eq:GKmaps} 
{\mathcal I}_4^+: J_\ell \leftrightarrow J_\ell \qquad {\mathcal I}_3^-: J_\ell \leftrightarrow J_{8-\ell}. 
\end{equation}
Here we would like to reproduce this result using our language. Moreover, in \cite{Gaberdiel:2004vx} it is shown that 
\begin{equation}\label{eq:GKmaps2} 
{\mathcal I}_3^+: J_\ell \leftrightarrow J_\ell \qquad {\mathcal I}_4^-: J_\ell \leftrightarrow J_{8-\ell},
\end{equation}
whence they identify ${\mathcal I}_3^+$ with the duality proposed in \cite{Papadopoulos:1995da}. This is an instance of self-mirror $G_2$ manifold, of the kind mentioned at the end of Section \ref{sec:massless_modz}.

\subsection{Smoothing in the language of TCS}\label{sect:ex1topZ}

Let us first discuss the geometry of the smoothing obtained by blowing up $X_-$. It is convenient to view $X_-$ as half of the well-known compact Calabi-Yau orbifold $T^6/(\Z_2\times\Z_2)$, for which the two $\Z_2$s act as
\be
\begin{array}{cccc}
&z_1 & z_2 & z_3 \\
\hline
\Z_2^a:&-z_1 & -z_2 & z_3 \\
\Z_2^b:&z_1 &\tfrac12 -z_2 & -z_3
\end{array}\, .
\ee
The resolution $X_{19,19}$ of this orbifold has Hodge numbers $(19,19)$ and it can be described as a constant $K3$ fibration with four (isomorphic) singular fibres \cite{Curio:1997rn}. The generic fibre is a Kummer surface of product type, the resolution of $(T^2 \times T^2)/\Z_2$. It has a Picard lattice of rank $18$. Over four points in the $\P^1$ base, the fibre degenerates into $k$ components. The corresponding monodromy action is inherited from the second $\Z_2$, it is the non-symplectic involution with invariants $(r,a,\delta) = (10,8,0)$ in Nikulin's classification \cite{Nikulin86discretereflection}. In particular, this means that the rank of the even sublattice of the Picard lattice of the generic fibre is ten-dimensional. We can hence compute the Hodge number $h^{1,1}(X_{19,19})$ as $19 = 10 + 1 + 4(k-1)$, so that $k=3$ follows. We may also compute the Euler characteristic of $X_{19,19}$ by excising the bad fibres to find the contribution $\chi_f$ of each degenerate fibre
\be 
\chi(X_{19,19}) = 0 = 24\cdot (2-4) + 4 \cdot \chi_f \, ,
\ee
giving $\chi_f = 12$. 

The (resolution of the) threefolds $X_\pm$ has a compactification to a threefold $Z_\pm$ (a `building block' in the language of \cite{Corti:2012kd}) by adding a single $K3$ fibre. Note that the monodromy action on each of the singular $K3$ fibres of $X_\pm$ is the same $\Z_2$, so that this can be done consistently. Note also that $Z_\pm$ is very similar to $X_{19,19}$: it just has two degenerate fibres instead of four. Hence we can immediately compute
\be
\chi(Z_\pm) = 24\cdot (2-2) + 2 \cdot \chi_f = 24 \, . 
\ee
On the other hand, $\chi(Z_\pm) = 2 + 2\cdot h^{1,1}(Z_\pm) - 2 \cdot h^{2,1}(Z_\pm)$ and $h^{1,1}(Z_\pm) =  1 + 10 + 2\cdot (k-1) = 15$. As $k=3$, it hence follows that $h^{2,1}(Z_\pm) = 4$. The building blocks $Z_\pm$ are hence characterized by the topological numbers 
\begin{equation}
h^{2,1}(Z_\pm) = 4\,, \qquad |K(Z_\pm)| = 4 \,, \qquad |N(Z_\pm)| = 10 \, .
\end{equation}

Before smoothing, we can think of each of the two building blocks $Z_\pm$ as a constant fibration of a Kummer surface $(T^2\times T^2)/\Z_2$ over $\P^1$ which only encounters $\Z_2$ monodromy when going on a loop around two points in the base $\P^1$. This remains true if we completely resolve $Z_\pm$ as discussed above. We may however, consider slightly more general smoothings in which we still blow up the singularities localized over $x_6=0,\tfrac14$, but allow more general smoothings of the (constant) $K3$ fibre. This is the strategy we will use in the following to describe the various smoothings of the orbifold \eqref{eq:orbiaction2} as a TCS. 

\begin{figure}[tbp]
\begin{center}
   \scalebox{.5}{ 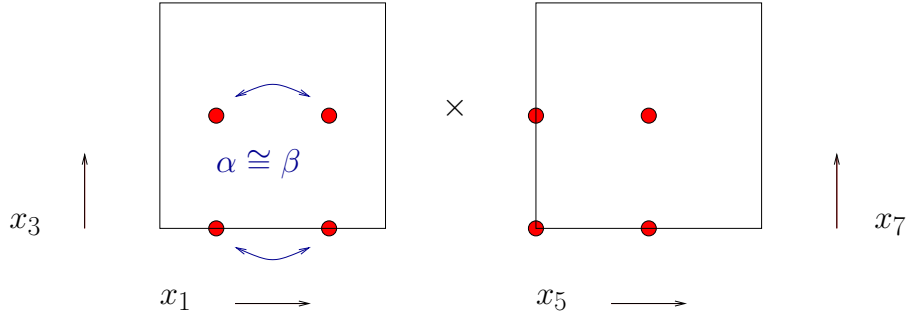 }
\caption{\label{fig:t4modz2ex2} The generic fibre of the fibration of $T^7/\Z_2^3$ is $T^4/\Z_2$. We have shown the $T^4$ as a product of two $T^2$'s. The fixed points of the action of $\gamma$ are shown in red and we have also indicated the action of $\alpha$ on the fixed points. The action of $\beta$ is identical. }
\end{center}  
\end{figure}

First, let us discuss the action of $\alpha$ and $\beta$ on the $16$ fixed points of $\gamma$ on the $K3$ fibre. These $16$ fixed points are at any of the $2^4$ combinations of $x_1=(\tfrac14,\tfrac34)$, $x_3=(0,\tfrac12)$, $x_5=(0,\tfrac12)$, and $x_7=(0,\tfrac12)$. We may hence label the $16$ nodes as $n_{1357},n_{\bar{1}357},\cdots,n_{\bar{1}\bar{3}\bar{5}\bar{7}}$, where e.g. $n_{1357}$ is the node at $x_1=\tfrac14$, $x_3=x_5=x_7=0$ and 
$n_{\bar{1}3\bar{5}7}$ is the node at $x_1=\tfrac34$, $x_3=0$, $x_5=\tfrac12$ and $x_7=0$, and so on. The action of $\alpha$ and $\beta$ on the $16$ nodes 
is the same on $Z_+$ and $Z_-$, in that both $\alpha$ and $\beta$ swap the position of the fixed points in the $x_1$ direction: $\alpha,\beta: n_{1357}\mapsto n_{\bar{1}357}$, see Figure \ref{fig:t4modz2ex2}.
This means the sixteen fixed points in the fibre K3 form eight orbits under $\alpha$ and $\beta$. The transverse space is the $T^3/Z_2$ with coordinates $x_2, x_4,x_6$; note that the action of $\alpha\beta$ leaves the fixed points invariant and acts as $(\frac12 + x_2,-x_4,-x_6)$ on the $T^3$. Hence we see again that the fixed point set is four $T^3$ from each building block and 8 times $T^3/\Z_2$ along the $K3$ fibre. 

For the fixed $T^3$s which are localized on $Z_-$ and $Z_+$, the same smoothing as before can be applied, i.e. we can simply blow them up \cite{joyce1996II}. The remaining singularities, which are the eight fixed points of $\gamma$, each have two different smoothings, i.e. we can blow them up or deform them. Such choices influence the action of $\alpha$ and $\beta$ on the exceptional cycles \cite{Gaberdiel:2004vx}.  Let us digress to discuss this in some detail. First of all, we like to describe the resolution process in terms of deformations of the K\"ahler forms $\omega$ and the holomorphic two-forms $\Omega$ of the $K3$ fibres. Of course, such deformations have to respect the TCS gluing conditions \eqref{eq:HKrotation} and need to have the right behaviour under $\alpha$ and $\beta$, i.e. for $X_\pm$ to be Calabi-Yau threefolds, $\omega_-$ ($\omega_+$) must be even and $\Omega_-$ ($\Omega_+$) must be odd under $\alpha$ ($\beta$). As the discussion is essentially the same for both $Z_+$ and $Z_-$, we only discuss the action of $\alpha$ on the smoothing of the fibre of $Z_-$. 

When we smooth any of these nodes $n_{jklm}$ either by deformation of blow-up on the $K3$ fibre, an exceptional cycle with the topology of an $S^{\, 2}$ appears. We denote the homology class of such cycles by the symbol $\eta_{jklm}$. Together with the six even forms $dx_{ij}$ for $i,j=\{1,3,5,7\}$ of $T^4$ under the action of $\gamma$, there are then $22$ independent two-forms spanning $H^2(K3,\mathbb{Q})$. The inner form among those cycles is such that
\begin{equation}
\begin{aligned}
\eta_{jklm}^2 &= -2 \\
dx_{ij} \cdot dx_{kl} &= 2 \epsilon_{ijkl}
\end{aligned}
\end{equation}
with all other products vanishing. Here we have defined $\epsilon_{1357} = 1$ and the remaining values of $\epsilon_{ijkl}$ are fixed by it being totally antisymmetric. The full lattice $H^2(K3,\mathbb{Z})$ furthermore contains appropriate half-integral linear combinations of the $\eta_{jklm}$ and the $dx_{ij}$ which are of the form
\begin{equation}\label{eq:halfintcombosKummer}
\tfrac12\left( dx_{ij} + \sum_{ij} \eta_{ijkl} \right) \, ,
\end{equation}
see e.g. \cite{Nahm:1999ps,Braun:2009wh} for a detailed discussion.

The sixteen nodes are all modelled on $\C^2/\Z_2$, which is equivalent to a simple surface singularity of type $A_1$. Using the coordinates $z_1^-$ and $z_2^-$, see \eqref{eq:holcoordsZ_-}, the action of $\alpha$ is $z_2^- \rightarrow -z_2^-$. The blow up of $\C^2$ at the origin is described by 
\begin{equation}\label{eq:blowup}
z_1^- \xi_2 =  z_2^- \xi_1 \, .
\end{equation}
The action of the involution $\alpha$ then extends the homogeneous coordinates $[\xi_1:\xi_2]$ on the exceptional $\P^1$ as $[\xi_1:\xi_2]\rightarrow [\xi_1:-\xi_2]$. This map is nothing by rotation by $180^\circ$ on the two-sphere and, in particular, leaves the volume form invariant. If we choose to blow up $n_{1klm}$ and $n_{\bar{1}klm}$, the action of $\alpha$ is hence
\begin{equation}
 \alpha: \eta_{1klm} \mapsto \eta_{\bar{1}klm} \, .
\end{equation}

To study the deformation, we start by describing $\C^2/\Z_2$ as a hypersurface singularity by introducing the (invariant) coordinates
\begin{equation}
w_1 + w_2 = (z_1^-)^2\, ,\hspace{.5cm} w_1 - w_2 = (z_2^-)^2\, , \hspace{.5cm} w_3 = z_1^- z_2^- \, ,
\end{equation}
which satisfy the identity 
\begin{equation}
 w_1^2 + w_2^2 = w_3^2 \, .
\end{equation}
The action of $\alpha$ is now $ w_3 \mapsto - w_3$. A deformation is given by
\begin{equation}\label{eq:def}
 w_1^2 + w_2^2 = w_3^2 + \epsilon \, .
\end{equation}
and we can think of the class of the exceptional $S^{\, 2}$ as being the real two-sphere formed by $\Re(w_1), \Re(w_2)$ and $\Im(w_3)$, taking $\epsilon$ to be real. The involution $w_3 \mapsto - w_3$ now corresponds to an antipodal map on $S^{\, 2}$. This map does not respect the volume form on the $S^{\, 2}$ and correspondingly
\begin{equation}
 \alpha: \eta_{1klm} \mapsto - \eta_{\bar{1}klm} \, .
\end{equation}
follows for a (complex structure) deformation. The same analysis can be repeated for $S_+$. Here, the nodes are described by $\C^2/\mathbb{Z}_2$ with coordinates $z_1^+,z_2^+$ and the action of $\beta$ is $z_2^+ \rightarrow z_2^+$. 

Summarizing, the action of $\alpha$ and $\beta$ is:
\begin{equation}
\begin{aligned}
\mbox{resolution}:\, \eta_{1klm}  \mapsto &\,\,\eta_{\bar{1}klm} \\
\mbox{deformation}:\,\eta_{1klm}  \mapsto & \,\,- \eta_{\bar{1}klm} 
\end{aligned}
\end{equation}
The action of $\alpha\beta$ keeps all of the sixteen nodes $n_{jklm}$ fixed, but precisely acts with a sign whenever a node is deformed from the perspective of one of the building blocks and resolved from the perspective of the other. Out of the eight linear combinations
\begin{equation}
\eta_{klm} \equiv \eta_{1klm} +  \eta_{\bar{1}klm} 
\end{equation}
there may be $l$ cycles for which $\alpha\beta$ acts as $+1$ and $8-l$ for which it acts as $-1$, so that $l$ out of the cycles $\eta_{klm}$ give a contribution to $b_2$ and $8-l$ give a contribution to $b_3$. This distinguishes between different smooth $G_2$ manifolds $J_l$ for which
\begin{equation}\label{eq:bettig2l}
b_2(J_l) = 8+ l \, ,\hspace{.5cm} b_3(J_l) = 47-l \, , 
\end{equation}
together with the remaining cycles independent of this choice \cite{joyce1996II}. Let us use the notation
\begin{equation}
\begin{aligned}
\eta_i^+ &= \eta_{1klm} +  \eta_{\bar{1}klm} \\
\eta_i^- &= \eta_{1klm} -  \eta_{\bar{1}klm} \, .
\end{aligned}
\end{equation}
i.e. the index $i=1..8$ runs over the eight choices $\{(3,5,7),(\bar{3},5,7),\cdots (\bar{3},\bar{5},\bar{7})\}$ for $(klm)$.

As the threefolds $X_\pm$ must become (open) Calabi-Yau threefolds, the holomorphic two-forms of the K3 fibres $S_\pm$ must be odd under the action of $\alpha$ and $\beta$, whereas the K\"ahler forms must be even. Due to the matching, nodes which are resolved on one side are necessarily deformed on the other side. We can hence realize a situation where $l$ nodes have eigenvalue $+1$ under $\alpha\beta$ by deforming all of the $16$ nodes on $Z_+$, but blowing up $8-l$ pairs of nodes on $Z_-$. This means that 
%putting
% \begin{equation}\label{ex2_o_O}
% \begin{aligned}
%  \omega_0 &= \Re(\Omega_{1/4}) =  dx_1 \wedge dx_3 + dx_5\wedge dx_7 + \sum_{i=  \, l}^{8} \eta_i^+ \\
% \Im(\Omega_{0}) &= -\Im(\Omega_{1/4}) = dx_1 \wedge dx_7 + dx_3\wedge dx_5 + \sum_{i=0}^{l-1} \eta_i^+ \\
% \Re(\Omega_0) &= \omega_{1/4} = dx_1 \wedge dx_5 - dx_3\wedge dx_7 
% \end{aligned}
% \end{equation}
% Using this, the contribution of the cycles $\eta_i^\pm$ to $N_0$ and  $N_{1/4}$ is given by the linear combintations
\begin{equation}\label{eq:ex2n0n14}
\begin{aligned}
N_- \supset \langle \eta_i^+ \rangle_{i = l+1..8} +  \langle \eta_i^- \rangle_{i = 1.. l} \, ,\\
N_{+}\supset\langle \eta_i^- \rangle_{i = l+1..8} +  \langle \eta_i^- \rangle_{i= 1 .. l} \, .
\end{aligned}
\end{equation}
% The $\eta^\pm_i$ do not generate all of $N_\pm$ as these lattices contain further cycles of the form $dx_i\wedge dx_j$ as well as 
% certain half-integral combinations between the $\eta^\pm_i$ and appropriate $dx_i\wedge dx_j$.
It now follows that the rank of $N_- \cap N_+$ is equal to $l$, generated by the $l$ different cycles $\eta_i^-$ for $i=0..l-1$. As the topology of $Z_\pm$ is the same as that of the building blocks discussed above, we have $h^{2,1}(Z_\pm)=|K(Z_\pm)|=4$, so that we recover \eqref{eq:bettig2l}.

Before we discuss the mirror maps in the context, let us make two observations. The first observation is that we can use the formula \eqref{eq:bettinumbers} to also learn about torsion in cohomology in this setting. First note that $N_- + N_{+}$ contains $8-l$ pairs of cycles $2 \eta_{1klm}$ and  $2 \eta_{\bar{1}klm}$ as linear combinations $\eta_i^+ + \eta_i^-$. Besides the forms $\eta_i^\pm$ written above, $N_- + N_{+}$ also contains some of the half integral combinations \eqref{eq:halfintcombosKummer}. However, no integral linear combinations of these can ever be used to find recover of the $\eta_{jklm}$. Hence all $8-l$ pairs of forms $\eta_{1klm}$ and $\eta_{\bar{1}klm}$ which are resolved on one side and deformed on the other cannot be contained in $N_- + N_{+}$. As $N_- + N_{+}$ contains $2\eta_{1klm}$ and $2\eta_{\bar{1}klm}$, as well as $\eta^\pm_{jklm}$, we find that
\begin{equation}
\text{tor}\,\, H^3(J_l) = \text{tor}\,\, \Gamma^{3,19}/(N_+ + N_-) = \mathbb{Z}_2^{8-l} \, . 
\end{equation}
Note that $\eta_{1klm} \sim \eta_{\bar{1}klm}$ in the quotient. The number of independent torsion cycles precisely equals the number of discrete torsion phases which are used in the CFT description of \cite{Gaberdiel:2004vx}. 
The appearance of torsion in cohomology in relation to discrete torsion phases is discussed for Calabi-Yau manifolds in the context of mirror symmetry in \cite{Aspinwall:1995rb,1998math......9072G}. For Calabi-Yau manifolds, the central observation are non-trivial torsional cycles in $\text{tor} H_2(X,\mathbb{Z})\cong \text{tor} H^3(X,\mathbb{Z})$ for a resolution $X$ of an orbifold theory including discrete torsion in the CFT \cite{Aspinwall:1995rb}. The discrete torsion in the CFT can be thought of as non-trivial periods of the B-field over cycles in $\text{tor} H_2(X,\mathbb{Z})$. As the relation $\text{tor} H_2(X,\mathbb{Z})\cong \text{tor} H^3(X,\mathbb{Z})$ does not depend on the dimension of $X$, discrete torsion phases are found in $H^3(X,\mathbb{Z})$ for $G_2$ manifolds as well. 

The second observation is that both building blocks admit elliptic fibration with a holomorphic section after resolving or deforming the nodes. To see this, let us explicitly write down the hyper K\"ahler structure of $S_-$ after we have deformed $l$ pairs of nodes and resolved the remaining $8-l$ pairs. In the same notation as above, we then have
\begin{equation}\label{eq:HKforS_-}
\begin{aligned}
\omega_- & = c_{31} dx_{31} + c_{75}  dx_{75} + \sum_{i=l+1}^8 c_i \eta_i^+ \\
R_- & = -c_{15} dx_{15} + c_{37}  dx_{37} \\
I_- & = c_{35} dx_{35} +  c_{17}  dx_{17}  + \sum_{i=1}^{l} c_i \eta_i^+ 
\end{aligned}
\end{equation}
where the $c_{ij}$ and $c_i$ are real coefficients, which should be chosen such that these forms are appropriately normalized. Note that the $8-l$ cycles $ \eta_i^+$ for $i=l+1..8$ are even under $\alpha$ and the $l$ cycles $\eta_i^+$ for $i=1..l$ are odd under $\alpha$, as required for the fibration of $S_-$ over $z_3^-$ to be a Calabi-Yau threefold. From this presentation, it follows that the $2\cdot(8-l)$ nodes corresponding to the $\eta_i^+$ for $i=l+1..8$ are resolved and the remaining nodes are deformed. 

We may now pick the elliptic curve $E$ and its dual $E_0$, which is related to the section $\sigma = E_0 - E$ as
\begin{equation}
\begin{aligned}
 E & =  dx_{57} \\
 E_0& = dx_{57} + \tfrac12\left(dx_{13} + \eta_{1357} \pm  \eta_{\bar{1}357} + \eta_{1\bar{3}57} \pm  \eta_{\bar{1}\bar{3}57}\right)
\end{aligned}
\end{equation}
which are both integral cycles on a smoothed Kummer surface. We have to choose the two signs in $E_0$ such that $E_0$ is even under $\alpha$ in order for $E_0$ to become a cycle on the building block $Z_-$ as well. With this choice, $E_0 \cdot \Omega_- = 0$ follows, as the expansion of $I_-$ only contains two-forms which are odd under $\alpha$ and hence perpendicular to forms even under $\alpha$. The Picard lattice of any fibre $S_-$ hence contains a lattice $U$ so that $S_-$ is elliptically fibered. The same argument can be applied to $S_+$ and the analogue of the expansion above can be found by simply applying the matching condition to \eqref{eq:HKforS_-}.

\subsection{TCS Mirror Maps}

Knowing the topology of the resolved building block, and the fact that for this example $Z_-$ and $Z_+$ are isomorphic, we can make an easy argument which verifies that the actions in \eqref{eq:GKmaps} indeed can be realized using our construction. The building blocks $Z_\pm$ satisfy 
\begin{equation}
\begin{aligned}
|N_\pm|=|N_\pm^\vee|=10. \\
h^{2,1}(Z_\pm) = |K_\pm| \, .
\end{aligned}
\end{equation}
For orthogonal embeddings, such as the one relevant here, the lattices $N$ and $\tilde{T}$ (recall that $\tilde{T}$ is defined by $T = U \oplus \tilde{T}$) satisfy
\begin{equation}
|N_-| + |N_+| - |N_- \cap N_+| + |\tilde{T}_-| + |\tilde{T}_+| - |\tilde{T}_- \cap \tilde{T}_+|  - |N_- \cap \tilde{T}_+|  - |\tilde{T}_- \cap N_+| = 20.
\end{equation}
Now, from $|K_\pm|=4$ and $8+\ell = b_2 = |K_-| + |K_+| + |N_- \cap N_+|$, it follows that
\begin{equation}
|N_- \cap N_+| = \ell
\end{equation}
for the different resolutions. Then
\begin{equation}
20 =  \ell +   |\tilde{T}_- \cap \tilde{T}_+ | +|N_- \cap \tilde{T}_+|  + |\tilde{T}_- \cap N_+|.
\end{equation}
On the other hand, 
\begin{equation}
b_3 = 1 + 22 - (|N_-| + |N_+| - |N_- \cap N_+|) + |N_- \cap \tilde{T}_+| + |\tilde{T}_- \cap N_+| + 24
\end{equation}
implies
\begin{equation}\label{eq:sym_sum}
|N_- \cap \tilde{T}_+| + |\tilde{T}_- \cap N_+ | = 20-2\ell \, , 
\end{equation}
so that
\begin{equation}
|\tilde{T}_- \cap \tilde{T}_+| = \ell
\end{equation}
follows. Swapping $Z_\pm$ with $Z^\vee_\pm$ in our construction produces the map ${\mathcal I}_4^+$, and indeed the roles of $|N_- \cap N_+|$ and $|\tilde{T}_- \cap \tilde{T}_+|$ are swapped, which is hence not expected to lead to any change in Betti numbers for these cases. 

Going back to the explicit identification of holomorphic coordinates on $X_\pm$, \eqref{eq:holcoordsZ_-} and \eqref{eq:holcoordsZ_+}, our prescription of performing four T-dualities on the SYZ fibres of $X_\pm$ together with the extra $S^{\, 1}$ corresponds to T-dualities along the directions $(2,3,4,5)$, which by \eqref{tab:orbif_caliber} gives the mirror map ${\mathcal I}_4^+$. 

We now turn to the mirror maps related with associative fibrations. As argued in the last section, the $K3$ fibres of both building blocks $Z_\pm$ are elliptically fibered, so that there are two such maps. Let us first exploit the elliptic fibration in $Z_+$. The mirror map is then given by swapping $Z_-$ with $(Z_-)^\vee$ while leaving $Z_+$ unchanged. This has the effect of swapping $N_-$ with $\widetilde{T}_-$, while switching $N_+$ with $\widetilde{N}_+$ and $\widetilde{T}_+$ with $T_+$, so that
\begin{equation}
b_2(J^\wedge) = |K_-| + |K_+| + |\widetilde{T}_- \cap \widetilde{N}_+| = 8 + 8 - \ell = 16 - \ell
\end{equation}
where we have used the symmetry of Equation \eqref{eq:sym_sum} to determine $|\widetilde{T}_- \cap N_+| = 10 - \ell$ and that $\widetilde{T}_- \cap \widetilde{N}_+$ differs from the former by a unimodular lattice of rank 2. Moreover,
\be
\begin{aligned}
b_3(J^\wedge) &= 1 + 22 - (|\widetilde{T}_-| + |\widetilde{N}_+| - |\widetilde{T}_- \cap \widetilde{N}_+|) + |\widetilde{T}_- \cap T_+| + |N_- \cap N_+| + 24\\
&=39+\ell
\end{aligned}
\ee
which indeed is the result expected by applying the map ${\mathcal I}_3^-$. The same results are found by working out the rank of $N_+\cap N_-$ after the mirror map. As the mirror map on $Z_-$ changes which nodes are deformed and which ones are blown up, we now have
\begin{equation}\label{eq:ex2n0n14_2}
\begin{aligned}
N_-^\vee \supset \langle \eta_i^- \rangle_{i = l+1..8} +  \langle \eta_i^+ \rangle_{i = 1.. l} \, ,\\
N_{+}\supset\langle \eta_i^- \rangle_{i = l+1..8} +  \langle \eta_i^- \rangle_{i= 1 .. l} \, .
\end{aligned}
\end{equation}
so that $N_+\cap N_-^\vee$ is now generated by $8-l$ lattice vectors instead of $l$. By \eqref{eq:holcoordsZ_-} and \eqref{eq:holcoordsZ_+}, this mirror map corresponds to performing three T-dualities along the directions $1,4,5$, which indeed corresponds to one of the ${\mathcal I}_3^-$ maps in Table \ref{tab:orbif_caliber}.

Finally, we can also exploit the elliptic fibration in $Z_-$. consider swapping only $Z_+$ with $(Z_+)^\vee$ while leaving $Z_-$ unchanged, which should capture the mirror map ${\mathcal I}_3^-$  associated with T-dualities along the directions $1,2,3$. As we have assumed above that all sixteen nodes on $Z_+$ are deformed, we now have
\begin{equation}\label{eq:ex2n0n14_3}
\begin{aligned}
N_- \supset \langle \eta_i^+ \rangle_{i = l+1..8} +  \langle \eta_i^- \rangle_{i = 1.. l} \, ,\\
N_{+}^\vee\supset\langle \eta_i^+ \rangle_{i = l+1..8} +  \langle \eta_i^+ \rangle_{i= 1 .. l} \, .
\end{aligned}
\end{equation}
Again, the rank of the intersection of $N_+ \cap N_-$ changes from $l$ to $8-l$ as expected for ${\mathcal I}_3^-$.

\section{Examples from dual pairs of tops}\label{sec:EXAMPLS}

A wide variety of generalized building blocks which can be used in the construction of TCS $G_2$ manifolds is obtained from pairs of four-dimensional projecting tops $\Diamond, \Diamond^\circ$ \cite{Braun:2016igl}. For such building blocks there is an analogue of Batyrev mirror symmetry, namely the dual pairs of building blocks are obtained swapping the roles of the two projecting tops in the construction
\begin{equation}
Z = (\Diamond, \Diamond^\circ) \leftrightarrow Z^\vee = (\Diamond^\circ, \Diamond)
\end{equation}
This fact can be exploited to generate a wide variety of concrete explicit examples for both types of mirror maps we discussed above.

Here we also notice that the examples we construct can involve also singular fibrations. Also in this case the spectra of the corresponding KK reduced supergravities always match, which gives a zeroth order consistency check for our duality. We are going to discuss such examples in more detail below.

\subsection{Different $G_2$ manifolds from a single matching}\label{sect:diffg2samematching}

For a fixed building block constructed from a dual pair of projecting tops, we may find `nearby' building blocks for which the K3 fibre is from the same algebraic family, and has the same lattice $N$, by considering different tops with the same $\Delta_F^\circ = \Diamond^\circ \cap F$\footnote{In general, it is not enough to have the same $\Delta_F^\circ$ for two tops to give rise to the same lattice $N$, as there may be monodromy acting on the algebraic cycles of the fibre depending on the choice of $\Diamond^\circ$.}. A particularly easy way to find such tops is to start from a given top $\Diamond^\circ_+$ with few lattice points above the plane $F$ and add further vertices $\{v\}$ to it which do not lie in the plane $F$. As the new top is supposed to be projecting, this means such new vertices must lie above $\Delta_F$, i.e. the choices are restricted to be integral points of $\Delta^\circ_{F,+}$, displaced by some positive integer $n$ in the fourth direction. If the new polyhedron constructed in this way is a projecting top (i.e. half a reflexive polytope), a new example $\Diamond^\circ_{+,\{v\}}$ with the same $N_+$, but different $K_+$ and $h^{2,1}_+$ is found. Any matching between the building block $X_+ = X_{(\Diamond,\Diamond^\circ_{+})}$ and any other building block $X_-$ is now also a matching for the building block $X_{\{v\}}$ associated with $\Diamond^\circ_{+,\{v\}}$ and $X_-$. Hence we can use this method to easily find a large set of examples with different values of $b_2$ and $b_3$. We will present examples of this method below. 

The building blocks $X_{+,\{v\}}$ are related to $X_{+}$ by singular transitions in which three-cycles are collapsed by a deformation of complex structure, followed by a crepant resolution of singularities. Hence $K(X_{+,\{v\}})$ is generally larger than $K(X_{+})$, so that employing $X_{+,\{v\}}$ leads to a set of $G_2$ manifolds with larger values of $b_2$ as compared to $X_{+}$ (recall that the matching, and hence $N_+ \cap N_-$, is fixed). From the point of view of the K3 fibration, increasing the number of lattice points above $F$ will give us a reducible fibre with more reducible components.

\subsection{Smooth $G_2$ manifolds with smooth $\Tf$ mirrors}

One can easily construct examples for which $N_+ \cap N_- =0$, so that it in particular has no roots, by employing building blocks for which the asymptotic K3 fibres are a mirror pair. Under our mirror construction, this property will be preserved for such examples, so that $N_+^\vee \cap N_-^\vee =0$ holds as well. 

\subsubsection{Building blocks fibered by a quartic K3 or its mirror}

As the simplest algebraic realization of a $K3$ is given by a quartic hypersurface in $\P^3$, the simplest building block can be found as a hypersurface in $\P^3 \times \P^1$ of bidegree $(4,1)$. In the language of tops, this means we consider a pair of dual tops with vertices
\begin{equation}\label{eq:topsquarticfibregeneric}
\Diamond^\circ_+ = \left(\begin{array}{rrrrr}
-1 & 0 & 0 & 0 & 1 \\
-1 & 0 & 0 & 1 & 0 \\
-1 & 0 & 1 & 0 & 0 \\
0 & 1 & 0 & 0 & 0
\end{array}\right) \, , \hspace{1cm} \Diamond_+ =\left(\begin{array}{rrrrrrrr}
-1 & -1 & 3 & 3 & -1 & -1 & -1 & -1 \\
-1 & -1 & -1 & -1 & 3 & 3 & -1 & -1 \\
-1 & -1 & -1 & -1 & -1 & -1 & 3 & 3 \\
-1 & 0 & 0 & -1 & 0 & -1 & 0 & -1
\end{array}\right) 
\end{equation}
The ambient toric variety of any building block has a fan with the extra ray $\nu_0 = (0,0,0,-1)$ besides lattice points on $\Diamond^\circ$. Applying \eqref{HypSurf} then reproduces a hypersurface of bidegree $(4,1)$ in $\P^3 \times \P^1$. Using \eqref{eq:bbtopology}, the Hodge numbers of $X_+ = X_{(\Diamond_+,\Diamond^\circ_+)}$ are found to be
\begin{equation}
 h^{1,1}(X_+) = 2 \, ,\hspace{1cm} h^{2,1}(X_+) = 33 \, ,  \hspace{1cm} |N(X_+)| = 1
\end{equation}
which can easily be verified using the standard index and vanishing theorems. This building block can also be found by degenerating a $K3$ fibered Calabi-Yau threefold, given by a hypersurface of bidegree $(4,2)$ in $\P^3 \times \P^1$.  

The lattice $N(X_+)$ is simply $(4)$ in this case (generated by the hyperplane class of $\P^3$) and the lattice $T$ is 
\begin{equation}
T_+ = (-4) \oplus U^{\oplus 2} \oplus (-E_8)^{\oplus 2} \, .
\end{equation}
It follows that $K(X_+)=0$, which corresponds to the $K3$ fibration having no reducible fibres and hence no localized divisors.

Computing the Hodge numbers for $X_+^\vee$ results in
\begin{equation}
 h^{1,1}(X_-^\vee) = 53 \hspace{1cm} h^{2,1}(X_-^\vee) = 0 \, ,  \hspace{1cm} |N(X_-^\vee)| = 19\, .
\end{equation}

Let us make a choice for the building block $X_-$ for which the fibre is the mirror of the quartic K3 surface. A particularly simple choice of such a top is given by 
\begin{equation}\label{eq:mirrorgenquartic}
\Diamond^\circ_- = \left(\begin{array}{rrrrr}
-1 & -1 & -1 & 0 & 3 \\
-1 & -1 & 3 & 0 & -1 \\
-1 & 3 & -1 & 0 & -1 \\
0 & 0 & 0 & 1 & 0
\end{array}\right) \, , \hspace{1cm} \Diamond_- = \left(\begin{array}{rrrrrrrr}
-1 & -1 & 1 & 1 & 0 & 0 & 0 & 0 \\
-1 & -1 & 0 & 0 & 0 & 0 & 1 & 1 \\
-1 & -1 & 0 & 0 & 1 & 1 & 0 & 0 \\
-1 & 0 & 0 & -1 & -1 & 0 & -1 & 0
\end{array}\right) 
\end{equation}
The hodge numbers of $X_-$ are
\begin{equation}
h^{1,1}(X_-) = 32 \, ,\hspace{.5cm} h^{2,1}(X_-) = 3 \, , \hspace{1cm} |N(X_-)| = 19.
\end{equation}
and the $N$ is the orthogonal complement of $(4)$ in $\Gamma^{2,18}$. Hence we have $|K_-| = 12$.

Computing the Hodge numbers for $X_-^\vee$ results in
\begin{equation}
 h^{1,1}(X_-^\vee) = 5 \hspace{1cm} h^{2,1}(X_-^\vee) = 12 \, ,  \hspace{1cm} |N(X_-^\vee)| = 1\, .
\end{equation}

We may perpendicularly glue $S_+$ and $S_-$ to a $G_2$ manifold with Betti Numbers
\begin{equation}
\begin{aligned}
 b_2(J) & = 12\,, && b_3(J) &= 107
\end{aligned}
\end{equation}
For the mirror, $|K|$ and $h^{2,1}$ are exchanged, so that
\begin{equation}
\begin{aligned}
 b_2(J^\vee) & = 36\,, && b_3(J^\vee) &= 83 \, .
\end{aligned}
\end{equation}

We may now employ the method explained in Section \ref{sect:diffg2samematching} above to find more examples which employ the same gluing as in the last section above. Let us focus on the building block associated with a generic fibration of a quartic K3 surface (a hypersurface in $\P^3 \times \P^1$), which can be constructed from the dual pair of tops in \eqref{eq:topsquarticfibregeneric}. By adding further vertices $\{v\}$, we can find new building blocks with the same $N = (4)$ which then can be matched with the building block $X_-$ associated with the pair of tops in \eqref{eq:mirrorgenquartic}. 

By making a simple scan through possible sets $\{v\}$, it turns out that all inequivalent cases can already be found by adding a single new vertex $v$ to $\Diamond^\circ$. We find the following inequivalent examples of smooth $G_2$ manifolds along with their smooth mirrors:
\begin{equation}
\begin{array}{ccccccc}
v & |K_+| & h^{2,1}_+ & b_2(J) & b_3(J) & b_2(J^\vee) & b_3(J^\vee)  \\
\hline
(1, 0, 0, 0, 1) & 1 & 22 & 13 & 86 & 25 & 74 \\ 
(1, 0, 0, 0, 2) & 2 & 15 & 14 & 73 & 18 & 69 \\ 
(1, 0, 0, 0, 3) & 3 & 12 & 15 & 68 & 15 & 68  
\end{array}
\end{equation}

\subsubsection{Building blocks fibered by an elliptic $K3$ surface}\label{sect:smoothJwithellK3}

We now consider examples of building blocks for which the fibre is an elliptic $K3$ surface. Let us consider elliptic K3 surfaces with a single reducible fibre of type $II^*$, so that $N = U \oplus (-E_8)$. We may perpendicularly glue any pair of building blocks with such K3 fibres to find a $G_2$ manifold $J$ with a smooth mirror $J^\vee$. 

Let us consider the building block as $Z$ constructed from the dual pair of tops
\begin{equation}\label{eq:ue8topsimple}
\Diamond^\circ = \left(\begin{array}{rrrrr}
-1 & 0 & 2 & 2 & 2 \\
0 & -1 & 3 & 3 & 3 \\
0 & 0 & -1 & 6 & 6 \\
0 & 0 & 0 & 0 & 1
\end{array}\right) \, ,\hspace{1cm} 
\Diamond = \left(\begin{array}{rrrrrr}
-2 & 1 & 1 & 1 & 1 & 1 \\
1 & 1 & 1 & 1 & 1 & -1 \\
0 & 6 & 6 & -1 & -1 & 0 \\
0 & 0 & -6 & 0 & -6 & 0
\end{array}\right)
\end{equation}
The Hodge numbers of $Z$ and $Z^\vee$ are
\begin{equation}
\begin{aligned}
h^{1,1}(Z) &= 11 \hspace{1cm} h^{2,1}(Z) &=& 240 \hspace{1cm} |N(Z)| &=10 \\
h^{1,1}(Z^\vee) &= 251 \hspace{1cm} h^{2,1}(Z^\vee) &=& 0  \hspace{1cm} |N(Z^\vee)| &=10 \\
\end{aligned}
\end{equation}
so that $|K(Z)| = 0 $ and $|K(Z^\vee)| = 240$.

As before, we may find many new building blocks with the same lattice $N = U \oplus E_8$ by using the technique described in Section \ref{sect:diffg2samematching}. Any two such building blocks $Z_a$ and $Z_b$ can be made into a $G_2$ manifold by gluing such that $N_a \cap N_b = 0$. The mirror building blocks again have $N = U \oplus E_8$, and the mirror gluing is such that $N^\vee_a \cap N^\vee_b = 0$ again. For any pair of such building blocks we find the Betti numbers
\begin{equation}
\begin{aligned}
 b_2(J) &= K_a + K_b \\
 b_3(J) &= 23 + (K_a+K_b) + 2(h^{2,1}_a + h^{2,1}_b) \, .
\end{aligned}
\end{equation}

Below we list a few building blocks with $N = U \oplus E_8$. For the sake of simplicity we focus on tops $\Diamond^\circ_{v}$ which contain a single extra vertex in comparison to the top $\Diamond^\circ$ of \eqref{eq:ue8topsimple}.
\begin{equation}
 \begin{array}{c|c|c}
v & |K| = (h^{2,1})^{\vee} & h^{2,1} = |K^\vee| \\
\hline
(2, 3, -1, 1) & 7 & 37 \\ 
(2, 3, 0, 1) & 6 & 66 \\ 
(2, 3, 1, 1) & 5 & 95 \\ 
(2, 3, 2, 1) & 4 & 124 \\ 
(2, 3, 3, 1) & 3 & 153 \\ 
(2, 3, 4, 1) & 2 & 182 \\
(2, 3, 5, 1) & 1 & 211 \\ 
(2, 3, 6, 2) & 6 & 126 \\ 
(2, 3, 6, 3) & 12 & 84 \\ 
(2, 3, 6, 6) & 36 & 36 \\ 
(2, 3, 6, 7) & 36 & 36 \\ 
(2, 3, 6, 14) & 84 & 12 \\ 
(2, 3, 6, 21) & 126 & 6 \\ 
(2, 3, 6, 42) & 240 & 0 
 \end{array}
\end{equation}
As indicated above, the mirror building blocks are such that $|K|$ and $h^{2,1}$ are swapped. We may glue any of the building blocks above (or its mirror) with any other building block (or its mirror) to find a $G_2$ manifold $J$.

\subsection{Smooth $G_2$ manifolds with singular $\Tf$ mirrors}

As discussed in Section \ref{sect:sing_mirror}, the matching condition forces every K3 fibre of $J$ to have the associated ADE singularities, whenever the lattice $N_+\cap N_-$ contains roots. This implies that $J$ has an $S^{\, 3}$ worth of ADE singularities. As the $\Tf$ mirror map swaps the lattices
\begin{equation}
N_+\cap N_- \,\, \leftrightarrow \,\, \tilde{T}^\vee_+ \cap \tilde{T}^\vee_- 
\end{equation}
a smooth $G_2$ manifold may have a singular mirror under $\Tf$ if $\tilde{T}^\vee_+ \cap \tilde{T}^\vee_- $ contains roots. 

\subsubsection{Quartic K3 surfaces}

An example of this type already appeared in \cite{Braun:2017ryx}, so we can be brief here. Using the building block $X$ associated with the pair of dual tops introduce in \eqref{eq:topsquarticfibregeneric} twice, we find a $G_2$ manifold with Betti numbers 
\begin{equation}
 b_2(J) = 0 \,,\hspace{.5cm} b_3(J) = 155 \,,
\end{equation}
and 
\begin{equation}
\begin{aligned}
N_+ \cap N_- &= 0 \\
T^\vee_+ \cap T^\vee_- &= (-4)^{\oplus 2} \oplus (-E_8)^{\oplus 2} \, .
\end{aligned}
\end{equation}

The mirror building blocks obtained by swapping the roles of $\Diamond$ and $\Diamond^\circ$ in \eqref{eq:topsquarticfibregeneric} have
\begin{equation}
  h^{1,1}(X^\vee) = 53 \hspace{1cm} h^{2,1}(X^\vee) = 0 \hspace{1cm} |K(Z^\vee)| = 33 \, ,
\end{equation}
and 
\begin{equation}
N(X^\vee) = T^\vee(X) = (-4) \oplus U \oplus (-E_8)^{\oplus 2}
\end{equation}
and the mirror $J^\vee$ of $J$ under $\Tf$ is found by gluing two such building blocks with 
\begin{equation}
N(X_+) \cap N(X_-) = (-4)^{\oplus 2} \oplus (-E_8)^{\oplus 2} \, ,
\end{equation}
so that every K3 fibre has two $E_8$ singularities. 

A naive application of \eqref{eq:bettinumbers} results in 
\begin{equation}
\begin{aligned}
b_2 = 18 + 33 + 33 &= 84 \\
b_3 = 1 + 2 + 1 + 1 + 33 + 33 &=  71 \, .
\end{aligned}
\end{equation}

\subsubsection{Elliptic K3 surfaces}\label{sect:ellbb}

A large class of building blocks with interesting mirrors can be found using elliptic K3 surfaces as the fibre. Although the mirrors will sometimes be smooth, as in the example discussed above in Section \ref{sect:smoothJwithellK3}, they will give us many interesting examples with geometric singularities.

Before construction the building blocks, let us discuss possible choices of elliptic K3 surfaces for the fibre and possible hyper K\"ahler rotations as well as their mirrors. A family of elliptic K3 surfaces with a section can be characterized in terms of a decomposition of its Picard lattice,
\begin{equation}
Pic = W \oplus U  \, ,
\end{equation}
where $W$ is the frame lattice containing information about reducible fibres and extra sections. For frame lattices $W$ sitting inside one of the two even unimodular sixteen-dimensional lattices, $\Gamma^{16}_1 = E_8 \oplus E_8$ and $\Gamma^{16}_2 = \widetilde{D}_{16}$, \footnote{This lattice can be constructed as the root lattice $D_{16}$ with an additional `glue vector' $v$ for which $2v\in D_{16}$ \cite{conway1998sphere}.} we may assume that
\begin{equation}
T = \left( W^\perp \right) \oplus U^{\oplus 2} \, ,
\end{equation}
where $W^\perp$ is the orthogonal complement of $W$ in one of the $\Gamma^{16}_i$. It is well-known how to construct such elliptic K3 surfaces using reflexive polytopes \cite{Candelas:1996su,Perevalov:1997vw}. 

We can now employ the elliptic K3 surfaces described above to engineer interesting building blocks. In particular, we may consider building blocks $X$ for which $N = Pic = U \oplus W$. This means that there is no monodromy acting on the components of reducible elliptic fibres, i.e. we are in the `split' case of \cite{Bershadsky:1996nh}. It is a simple task to generalize this to more complicated cases. 

Below are some examples of elliptic building blocks over $\P^1\times \P^1$ for which $W \subset E_8 \times E_8$ and $U \oplus W = N$. 
\begin{equation}
 \begin{array}{c|c|c|c}
W & W^\perp & |K| & h^{2,1}  \\
\hline
E_8\oplus E_8 & 0 & 12 & 20 \\
0 & E_8 \oplus E_8 & 0 & 112 \\
A_2 & E_8 \oplus E_6 & 0 & 90 \\
A_4 & E_8 \oplus A_4 & 0 & 76 \\
E_7 & E_8 \oplus A_1 & 1 & 70
 \end{array}
\end{equation}
It is easy to extend this list by further examples. Furthermore, there are many associated building blocks which can be obtained by the method explained in Section \ref{sect:diffg2samematching}.

Let us consider the case $W = A_4$. The tops $\Diamond^\circ$ and $\Diamond$ used to construct the corresponding building block have vertices
\begin{equation}\label{eq:simplesu5top}
\begin{aligned}
\Diamond^\circ_{A4} & = \left(\begin{array}{rrrrrrrr}
-1 & 0 & 0 & 0 & 1 & 2 & 2 & 2 \\
0 & -1 & 0 & 1 & 1 & 3 & 3 & 3 \\
0 & 0 & 1 & 1 & 1 & 1 & -1 & 0 \\
0 & 0 & 0 & 0 & 0 & 0 & 0 & 1
\end{array}\right)\\
\Diamond_{A4} &= 
\left(\begin{array}{rrrrrrrrrrrr}
-2 & -1 & 1 & -1 & 1 & 1 & 1 & 1 & 1
& 1 & 0 & 0 \\
1 & 1 & 1 & 1 & 1 & 1 & 1 & 0 & 0 &
-1 & 0 & 0 \\
0 & -1 & 6 & -1 & 6 & -1 & -1 & -1 & -1
& 0 & -1 & -1 \\
0 & -2 & 0 & 0 & -6 & 0 & -6 & 0 & -3
& 0 & 0 & -1
\end{array}\right) 
\end{aligned}
\end{equation}
in this case. We can now find many more examples with the same lattice $N$ by adding further vertices to $\Diamond^\circ$ such that the new polytope is also a projecting top. The following cases exist for a single new vertex $v$:
\begin{equation}
 \begin{array}{c|c|c}
  v & |K| & h^{2,1} \\
  \hline
(2, 3, -1, 1) & 1 & 47 \\
(1, 2, 0, 1) & 1 & 57 \\
(2, 3, 1, 1) & 1 & 72 \\
(0, 1, 0, 1) & 2 & 38 \\
(-1, 0, 0, 1) & 3 & 19 \\
 \end{array}
\end{equation}
and for two new vertices $v_1,v_2$:
\begin{equation}
 \begin{array}{c|c|c}
  v_1.v_2 & |K| & h^{2,1} \\ 
  \hline
  (-1, 0, 0, 1) , (2, 3, -1, 1) & 4 & 14 \\ 
(0, -1, 0, 1) , (-1, 0, 0, 1) & 7 & 11 \\ 
(0, 1, 0, 1) , (2, 3, -1, 1) & 3 & 29 \\ 
(0, 1, 0, 1) , (-1, 0, 0, 1) & 3 & 19 \\ 
(0, 1, 0, 1) , (0, -1, 0, 1) & 5 & 13 \\ 
(1, 1, 0, 1) , (0, 1, 0, 1) & 3 & 35 \\ 
(1, 1, 0, 1) , (2, 3, -1, 2) & 4 & 23 \\ 
(1, 2, 0, 1) , (2, 3, -1, 1) & 2 & 40 \\ 
(1, 2, 0, 1) , (-1, 0, 0, 1) & 3 & 19 \\ 
(1, 2, 0, 1) , (0, 1, 0, 1) & 2 & 38 \\ 
(1, 2, 0, 1) , (1, 1, 0, 1) & 2 & 45 \\ 
(2, 3, 0, 1) , (2, 3, -1, 1) & 1 & 47 \\ 
(2, 3, 0, 1) , (-1, 0, 0, 1) & 3 & 19 \\ 
(2, 3, 0, 1) , (0, 1, 0, 1) & 2 & 38 \\ 
(2, 3, 0, 1) , (1, 2, 0, 1) & 1 & 57 \\ 
(2, 3, 1, 1) , (2, 3, -1, 1) & 2 & 43 \\ 
(2, 3, 1, 1) , (1, 2, 0, 1) & 2 & 55 \\ 
(2, 3, 1, 1) , (2, 3, 0, 1) & 1 & 72 \\ 
(2, 3, 1, 1) , (0, 1, 1, 1) & 3 & 54 \\ 
(2, 3, 1, 1) , (1, 2, 1, 1) & 2 & 65 \\ 
(2, 3, 1, 2) , (1, 1, 1, 1) & 4 & 48 
 \end{array}
\end{equation}
A thorough classification of possibilities will require to use the classification of semi stable degenerations of K3 fibres \cite{Kulikov,Persson_Pinkham} together with a generalization to fibre components of multiplicity $>1$, see also \cite{1307.6514,Braun:2016sks}.

For any pair of the examples of building blocks tabulated above, we may find hyper K\"ahler rotations for the elliptic K3 fibres such that
\begin{equation}
\begin{aligned}
N_+ \cap N_- & = W_+ \cap W_- \\
T_+ \cap N_- &= U \oplus  W_+^\perp \cap W_- \\ 
N_+ \cap T_- &= U \oplus W_+ \cap W_-^\perp \\  
T_+ \cap T_- &= U \oplus W_+^\perp \cap W_-^\perp \, .
\end{aligned}
\end{equation}
If $W_+ \cap W_- $ (or $W_+^\perp \cap W_-^\perp$ contains vectors of length $-2$, $J$ (or $J^\vee$) has the corresponding ADE singularities. 

If the $W_\pm$ are furthermore orthogonally embedded in $\Gamma^{16}_i$, this gives us an orthogonal gluing. In these cases one can easily compute
\begin{equation}
\begin{aligned}
b_2(J) &= |K_+| + |K_-| + W_+ \cap W_- \\
b_3(J) &= 23 + |K_+| + |K_-| + 2\left(h^{2,1}(X_+) + h^{2,1}(X_-)\right) -  W_+ \cap W_- 
\end{aligned}
\end{equation}
as well as 
\begin{equation}
\begin{aligned}
b_2(J^{\vee}) &= h^{2,1}(X_+)+ h^{2,1}(X_-) + W_+^\perp \cap W_-^\perp \\
b_3(J^{\vee}) &= 23 + h^{2,1}(X_+) + h^{2,1}(X_-) + 2\left|K_+| + |K_-|\right) -  W_+^\perp \cap W_-^\perp
\end{aligned}
\end{equation}
Further examples of building blocks which can be used here are given in Sections \ref{sect:smoothJwithellK3} (for $W=-E_8$ ) and \ref{sect:singi3mirrors} (for $W = 0$).

\subsection{Smooth $G_2$ manifolds with smooth $\Tt$ mirrors}\label{sect:singi3mirrors}

In order to find smooth $G_2$ manifolds with smooth $\Tt$ mirrors, we need to find pairs of building blocks $Z_+$, $Z_-$ for which at least one of the two ($Z_+$ say) is elliptic and a matching for which both $N_+\cap N_-$ and $N_+\cap \widetilde{T}_-$ contain no roots. A simple class of examples is given by using building blocks with $N = U$. Starting from the dual pair of tops
\begin{equation}
\Diamond^\circ \left(\begin{array}{rrrrr}
-1 & 0 & 2 & 2 & 2 \\
0 & -1 & 3 & 3 & 3 \\
0 & 0 & -1 & 0 & 1 \\
0 & 0 & 0 & 1 & 0
\end{array}\right)\hspace{1cm}
\Diamond = \left(\begin{array}{rrrrrr}
-2 & 1 & 1 & 1 & 1 & 1 \\
1 & 1 & 1 & 1 & 1 & -1 \\
0 & 6 & 6 & -6 & -6 & 0 \\
0 & 0 & -6 & 0 & -6 & 0
\end{array}\right)
\end{equation}
which gives a building block $Z$ with $N(Z) = U$, $K=0$ and $h^{2,1} = 112$. Adding a single new vertex $v$ to $\Diamond^\circ$ results in building blocks with the properties
\begin{equation}
 \begin{array}{c|c|c}
 v & |K| & h^{2,1} \\
 \hline
(2, 3, -1, 1) & 1 & 83 \\ 
(2, 3, -1, 2) & 2 & 54 \\ 
(2, 3, -1, 3) & 5 & 33 \\ 
(2, 3, -1, 4) & 8 & 24 \\ 
(2, 3, -1, 6) & 12 & 16 \\ 
(2, 3, -1, 12) & 20 & 12 \\ 
(-1, 0, 0, 1) & 3 & 25 \\ 
(0, 1, 0, 1) & 2 & 54 \\ 
(1, 2, 0, 1) & 1 & 83 \\ 
(2, 3, 0, 2) & 2 & 66 \\ 
(2, 3, 0, 3) & 4 & 44 \\ 
(2, 3, 0, 6) & 20 & 12 \\ 
(2, 3, 1, 1) & 1 & 83 \\ 
(2, 3, 1, 2) & 2 & 54 \\ 
(2, 3, 1, 3) & 5 & 33 \\ 
(2, 3, 1, 4) & 8 & 24 \\
(2, 3, 1, 6) & 12 & 16 \\ 
(2, 3, 1, 12) & 20 & 12  
 \end{array}
\end{equation}
Any such pair may be glued such that $N_a \cap N_b = 0$ and we find a smooth $G_2$ manifold. As both building blocks are elliptic, there are two mirror maps of the type $\Tt$. For any such building block $\widetilde{T}(Z)= U \oplus (-E_8)^{\oplus 2}$ and $\widetilde{T}_a \cap N_b = N_a \cap \widetilde{T}_b = 0$, so that we find two (in general different) smooth $\Tt$ mirrors. The Betti numbers of these along with their two mirrors ${\Tt}_{,a}$ and ${\Tt}_{,b}$  are
\begin{equation}
\begin{aligned}
 b_2(J) &= K_a + K_b \\
 b_3(J) &= 23 + (K_a+K_b) + 2(h^{2,1}_a + h^{2,1}_b) \\
 & \\
 b_2(J^{\wedge_a}) &= h^{2,1}_a + K_b \\
 b_3(J^{\wedge_a}) &= 23 + (2K_a+ K_b) + (h^{2,1}_a + 2h^{2,1}_b) \\
 & \\
 b_2(J^{\wedge_b}) &= K_a + h^{2,1}_b \\
 b_3(J^{\wedge_b}) &= 23 + (K_a+2K_b) + (2h^{2,1}_a + h^{2,1}_b) 
 \end{aligned}
\end{equation}

Note that applying both ${\Tt}_{,a}$ and ${\Tt}_{,b}$ means we swap both $N_a \leftrightarrow \widetilde{T}_a$ and $N_b \leftrightarrow \widetilde{T}_b$ as done for the mirror map $\Tf$. This leads to $N_a \cap N_b = (-E_8)^{\oplus 2}$ so that we end up with a singular TCS $G_2$ manifold. 

\subsection{Smooth $G_2$ manifolds with singular $\Tt$ mirrors}

Finally, let us present a set of examples with potentially singular $\Tt$ mirrors. For this, we can use any of the sets of building blocks introduced in  Sections \ref{sect:smoothJwithellK3}, \ref{sect:ellbb} and \ref{sect:singi3mirrors}. Depending on frame lattices $W$ and the matching, various patterns can be realized. Again, whenever $W_a \cap W_b$ contains roots, the corresponding $G_2$ manifold has the associated ADE singularities. Recall that this does not imply a non-Abelian gauge if we make a generic choice of the $B$-field inside $N_a \cap N_b$. 

The Betti numbers, along with the two potential mirrors are given by 
\begin{equation}
\begin{aligned}
 b_2(J) &= K_a + K_b + |W_a \cap W_b| \\
 b_3(J) &= 23 + (K_a+K_b) + 2(h^{2,1}_a + h^{2,1}_b) - |W_a \cap W_b| \\
 & \\
 b_2(J^{\wedge_a}) &= h^{2,1}_a + K_b + |W_a^\perp \cap W_b|  \\
 b_3(J^{\wedge_a}) &= 23 + (2K_a+ K_b) + (h^{2,1}_a + 2h^{2,1}_b) - |W_a^\perp \cap W_b|  \\
 & \\
 b_2(J^{\wedge_b}) &= K_a + h^{2,1}_b + |W_a \cap W_b^\perp| \\
 b_3(J^{\wedge_b}) &= 23 + (K_a+2K_b) + (2h^{2,1}_a + h^{2,1}_b) - |W_a \cap W_b^\perp|  
 \end{aligned}
\end{equation}
One may also perform both maps ${\Tt}_{,a}$ and ${\Tt}_{,b}$, which gives a manifold with the same Betti numbers as the image under $\Tf$:
\begin{equation}
 \begin{aligned}
 b_2(J^{\vee}) &= K_a + h^{2,1}_b + |W_a^\perp \cap W_b^\perp| \\
 b_3(J^{\vee}) &= 23 + (K_a+2K_b) + (2h^{2,1}_a + h^{2,1}_b) - |W_a^\perp \cap W_b^\perp|  
 \end{aligned}
\end{equation}

\section{Acknowledgements}

We thank Bobby Acharya, Marc-Antoine Fiset, Sergei Gukov, Jim Halverson, Zohar Komargodski, Magdalena Larfors, Yang Li, David Morrison, Sakura Sch\"afer-Nameki, Nathan Seiberg, and Samson Shatashvili for helpful discussions. This work was in part performed at the Aspen Center for Physics, which is supported by National Science Foundation grant PHY-1607611. This work was partially supported by Simons Foundation grant \# 488629. 
APB thanks the SCGP of Stony Brook University for hospitality while part of this work was done. The work of APB was supported by the SCFT grant ST/L000474/1 and the ERC grant 682608 (HIGGSBNDL).

\appendix

\section{K3 Mirror Symmetry and SYZ}\label{sect:k3mirrorsyz}

In this Appendix, we review the results of \cite{1998math......9072G} concerning the relation of the mirror map for K3 surfaces with T-duality along a calibrated fibration, comparing them with the more traditional approach of \cite{Aspinwall:1994rg,2002math.....10219H}. We then give the $K3$ mirror maps associated with two T-dualities along $T^2$ fibres calibrated by the different forms of the hyper-K\"ahler structure of the $K3$ surface. 

The charge lattice of IIA superstrings on K3 coincides with $H^\bullet(S,\mathbb{Z})$ which is an even unimodular lattice of signature $(4,20)$, $\Gamma_{4,20}$. A  point in the moduli space of the corresponding 2d $\cn=(4,4)$ $c=6$ SCFT is given by fixing a four-plane $\hat{\Sigma}$ in $\Gamma_{4,20}\otimes \R$. Any such point in moduli space can have many different interpretations in terms of geometry and a $B$-field. A choice of a geometric realization amounts to choosing an hyperbolic $(1,1)$ sublattice $U_v$ of $\Gamma_{4,20}$, and different choices of such sublattices give different geometric interpretations. Denoting the generators of $U_v$ by $v_0$ and $v$ such that
\begin{equation}
v^2 = 0 = v_0^2 \qquad\qquad v \cdot v_0 = 1.
\end{equation}
we may think of $v_0$ and $v$ as corresponding to a generator of $H^0(S,\mathbb{Z})$ and $H^4(S,\mathbb{Z})$, respectively. This splits $\Gamma_{4,20} = U_v \oplus \Gamma_{3,19}$. A choice of K\"ahler form $\omega$, complex structure $\Omega = R + i I$ and B-field $B$ can then be specified in terms of the fourplane $\hat{\Sigma}$ as follows:
\begin{equation}
\begin{aligned}
\hat{\omega} &= \omega - (\omega \cdot B) v  \\
\hat{\Omega} &= \Omega - (\Omega \cdot B) v \\
\hat{B} &= B + v_0 +\tfrac12(\omega^2-B^2) v
\end{aligned}
\end{equation}
Of course, choosing a geometric interpretation does not entail choosing a complex structure, i.e. the identification we have written above is invariant under hyper-K\"ahler rotations acting on $R,I,\omega$, which fits with the fact that a Ricci-flat metric on a $K3$ surface is determined by a positive-norm three-plane $\Sigma$ in $\Gamma^{3,19} \otimes \R = H^2(K3,\mathbb{Z})\otimes \R$. 

Assuming that the K3 surface $S$ furthermore has a sLag $T^2$ fibration calibrated by $R$ entails that there is a 2-cycle $E \in \Gamma_{3,19} \simeq H^2(S,\mathbb{Z})$ such that i.e. $I\cdot E=0$ and $\omega \cdot E = 0$ but $R\cdot E \neq 0$. We can choose an $E_0 \in \Gamma_{3,19}$ such that
\begin{equation}
E^2 = 0 = E_0^2 \qquad\qquad E\cdot E_0 = 1,
\end{equation}
which entails that $\Gamma_{4,20} = U_v \oplus U_E \oplus \Gamma_{2,18}$. We can hence expand the generators of $\hat{\Sigma}$ with respect to $U_v$ and $U_E$ as:
\begin{equation}\label{eq:decompsigmahat}
\begin{aligned}
\hat{\omega} & = \omega_2 -(R_2 \cdot \omega_2 ) E - (B_2 \cdot \omega_2) v \\
\hat{R} & = R_2 + E_0 + \tfrac12(I_2^2 - R_2^2) E - (B_2 \cdot R_2 + \alpha) v\\
\hat{I} & = I_2  -(I_2 \cdot R_2 ) E - (B_2 \cdot I_2) v \\
\hat{B} & = B_2 + v_0 +\tfrac12(\omega_2^2-B_2^2) v + \alpha E
\end{aligned}
\end{equation}
where the subscript $_2$ indicates the projection to the sublattice $\Gamma^{2,18}$ and $\omega_2 \cdot R_2 = 0$, as well as $\omega_2^2 = R_2^2$. The various coefficients above are fixed by demanding that the generators of $\hat{\Sigma}$ and $\Sigma$ are orthogonal and normalized such that
\begin{equation}
\hat{R}^2 = \hat{I}^2 = \hat{B}^2 =  \hat{\omega}^2 =  R^2 = I^2 = \omega^2  = B^2\, .
\end{equation}

Following the arguments of SYZ, the mirror map arises from performing two T-dualities along the fibres $E$ of a sLag fibration calibrated by $Re(\Omega)$ and acts as \cite{1998math......9072G}:
\begin{equation}\label{eq:grossmirror}
\begin{aligned}
 \omega^\vee & = I + I \cdot (B-E_0)\, E \\
  R^\vee & = - B + E_0 + \left(\tfrac12(\omega_2^2-B_2^2) + (B\cdot E_0)\right)\, E \\
 I^\vee & = -\omega - \omega\cdot (B-E_0) \, E \\
 B^\vee & =  R - E_0  + R\cdot\left(B - E_0 \right)\,E
\end{aligned}
\end{equation}
One can think of $\sigma = E_0 - E$ as a section of the sLag fibration with fibre $E$ and we have assumed that $B \in E^\perp$. 

As mirror symmetry leaves the SCFT invariant, there is an alternative description of the mirror map as the automorphism of $\Gamma^{4,20}$ which swaps $U_E \leftrightarrow U_v$ and identifies 
\begin{equation}
\begin{aligned}
\hat{\omega}^\vee &= \hat{I}  \hspace{1cm}& \hat{I}^\vee &=  \hat{\omega} \\
\hat{R}^\vee &= \hat{B} \hspace{1cm}& \hat{B}^\vee &=  \hat{R} \, .
\end{aligned}
\end{equation}
This implies the new geometrical identification \cite{2002math.....10219H}
\begin{equation}\label{eq:mirror_map_master}
\begin{aligned}
\omega^\vee & = I_2 - (B_2 \cdot I_2) E & = &\,\,I - I\cdot(B + E_0)\, E \\
R^\vee &= B_2 + E_0 +\tfrac12(\omega_2^2-B_2^2) E & = &\,\, B + E_0 +\left(\tfrac12(\omega^2-B^2) - (B\cdot E_0) \right) E\\
I^\vee & = \omega_2  - (B_2 \cdot \omega_2) E& = &\,\, \omega - \omega\cdot(B + E_0)\, E \\
B^\vee & =   R_2 - (B_2 \cdot R_2 + \alpha) E &=& \,\,R - E_0  - R\cdot\left(B + E_0 \right)E
\end{aligned}
\end{equation}
Note that this yields the map \eqref{eq:grossmirror} upon replacing $B \rightarrow -B$ and $\omega \rightarrow -\omega$, which can again be realized by the automorphism
\begin{equation}
\left(\begin{array}{cc}
       v \\
       v_0
      \end{array}\right) \mapsto - \left(\begin{array}{cc}
       v \\
       v_0
      \end{array}\right) \hspace{1cm}
      \hat{B} \mapsto - \hat{B}\hspace{1cm}
      \hat{\omega}\mapsto - \hat{\omega} \,.
\end{equation}
In the following, we will adopt the map resulting from this automorphism, \eqref{eq:grossmirror}, as the mirror map.

In the simple situation where $\omega = \omega_2$, $I = I_2$ and $R \in U_E$, $B \in U_v$, i.e. $B=0$ both before and after the mirror map, the above identification becomes
\begin{equation}\label{eq:mirrorashk}
\begin{aligned}
\omega^\vee &= I &\hspace{1cm}&
R^\vee &= R &\hspace{1cm}&
I^\vee &= -\omega 
\end{aligned}
\end{equation}
which is just a hyper-K\"ahler rotation. Note that the sign change in $\omega$ is crucial here. From the point of view of the SYZ picture of mirror symmetry, this sign change is related to keeping the orientation of the fibres of the SYZ fibration intact \cite{1998math......9072G}.

In our applications to $G_2$ manifolds, the $B$-field in the asymptotic K3 fibres $S_{0\pm}$ must become a globally well-defined two-form on $J$. This means that is must be confined to the intersection $N_+\cap N_-$ in the absence of torsion, i.e. $B$ is in particular contained in the Picard lattice of both $S_{0+}$ and $S_{0-}$. Furthermore, in order to have a well-defined mirror map on $J$, this must hold both before and after an application of the mirror map, which gives us several constraints. Note that in particular the matching of $S_\pm$ forces that 
\begin{equation}
\omega_\pm \cdot B_\pm =  R_\mp \cdot B_\mp = 0 \, . 
\end{equation}

Let us first discuss the case in which $E$ is calibrated bt $R$ considered above. In this case $B \cdot \Omega = B^\vee \cdot \Omega^\vee = 0$ gives  
$B_2 \cdot \Omega_2 =  \omega_2 \cdot \Omega_2 = 0$, as well as $R_2 \cdot I_2 = \omega_2 \cdot B_2 = 0$ and the mirror map becomes
\begin{equation}
\begin{aligned}\label{eq:RmirrorBorth}
\omega^\vee & = I_2 &=& I \\
R^\vee & = -B_2 + E_0 + \tfrac12 (\omega_2^2 - B_2^2) E &=& -B_2 + R-R_2 \\
I^\vee & = -\omega_2 &=& -\omega \\
B^\vee & &=& R_2 
\end{aligned}
\end{equation}

Let us now consider the case where $E$ is calibrated by $\pm I$. We can find the mirror map by replacing 
\begin{equation}
\left( \begin{array}{c}
        R \\
        I
       \end{array}\right) \rightarrow 
 \left( \begin{array}{c}
        \pm I \\
        \mp R
       \end{array}\right) \, .
\end{equation}
in \eqref{eq:decompsigmahat} and \eqref{eq:mirror_map_master}. The constraint $B \cdot \Omega = B^\vee \cdot \Omega^\vee = 0$ now gives, as for the case where $E$ is calibrated by $R$, $B_2 \cdot \Omega_2 =  \omega_2 \cdot \Omega_2 = 0$, as well as $R_2 \cdot I_2 = \omega_2 \cdot B_2 = 0$ and the mirror map becomes
\begin{equation}\label{eq:ImirrorBorth}
\begin{aligned}
 \omega^\vee & = \mp R_2  &=& \mp R \\
 R^\vee & = \pm \omega_2 &=& \pm \omega \\
 I^\vee & = \mp B_2 + E_0 + \tfrac12(\omega_2^2 - B_2^2) E &=& \mp B + I - I_2 \\
 B^\vee & = \pm I_2 &=& \pm I_2
\end{aligned}
\end{equation}

Finally, let us consider the case where $E$ is calibrated by $\omega$. We can find this from \eqref{eq:decompsigmahat} and \eqref{eq:mirror_map_master} by replacing 
\begin{equation}
\left( \begin{array}{c}
        R \\
        \omega
       \end{array}\right) \rightarrow 
 \left( \begin{array}{c}
        \omega \\
        -R
       \end{array}\right) \, .
\end{equation}
Implementing $B \cdot \Omega = B^\vee \cdot \Omega^\vee = 0$ gives only $B_2 \cdot \Omega_2 =  \omega_2 \cdot \Omega_2 = 0$. However, in case we want to match this with another building block for which $B_2 \cdot \Omega_2 = 0$, we also need $B_2 \cdot \omega_2 = 0$ and $I_2 \cdot R_2 = 0$. Using this  
we find the mirror map
\begin{equation}
\begin{aligned}\label{eq:omegamirrorBorth}
\omega^\vee & = -B_2 + E_0 + \tfrac12 (R_2^2 - B_2^2) E & = & -B_2 + \omega - \omega_2  \\
I^\vee & = R_2 &=&  R\\
R^\vee & = -I_2 &=& -I\\
B^\vee & = \omega_2 &=& \omega_2  
\end{aligned}
\end{equation}
in this case.

\bibliography{G2MIR.bib}

\providecommand{\href}[2]{#2}\begingroup\raggedright\begin{thebibliography}{10}

\bibitem{Acharya:2004qe}
B.~S. Acharya and S.~Gukov, ``{M theory and singularities of exceptional
  holonomy manifolds},''
  \href{http://dx.doi.org/10.1016/j.physrep.2003.10.017}{{\em Phys. Rept.}
  {\bfseries 392} (2004) 121--189},
\href{http://arxiv.org/abs/hep-th/0409191}{{\ttfamily arXiv:hep-th/0409191
  [hep-th]}}.
%%CITATION = HEP-TH/0409191;%%.

\bibitem{Shatashvili:1994zw}
S.~L. Shatashvili and C.~Vafa, ``{Superstrings and manifold of exceptional
  holonomy},'' \href{http://dx.doi.org/10.1007/BF01671569}{{\em Selecta Math.}
  {\bfseries 1} (1995) 347},
\href{http://arxiv.org/abs/hep-th/9407025}{{\ttfamily arXiv:hep-th/9407025
  [hep-th]}}.
%%CITATION = HEP-TH/9407025;%%.

\bibitem{Acharya:1996fx}
B.~S. Acharya, ``{Dirichlet Joyce manifolds, discrete torsion and duality},''
  \href{http://dx.doi.org/10.1016/S0550-3213(97)00163-6}{{\em Nucl. Phys.}
  {\bfseries B492} (1997) 591--606},
\href{http://arxiv.org/abs/hep-th/9611036}{{\ttfamily arXiv:hep-th/9611036
  [hep-th]}}.
%%CITATION = HEP-TH/9611036;%%.

\bibitem{Figueroa-OFarrill:1996tnk}
J.~M. Figueroa-O'Farrill, ``{A Note on the extended superconformal algebras
  associated with manifolds of exceptional holonomy},''
  \href{http://dx.doi.org/10.1016/S0370-2693(96)01506-7}{{\em Phys. Lett.}
  {\bfseries B392} (1997) 77--84},
\href{http://arxiv.org/abs/hep-th/9609113}{{\ttfamily arXiv:hep-th/9609113
  [hep-th]}}.
%%CITATION = HEP-TH/9609113;%%.

\bibitem{Acharya:1997rh}
B.~S. Acharya, ``{On mirror symmetry for manifolds of exceptional holonomy},''
  \href{http://dx.doi.org/10.1016/S0550-3213(98)00140-0}{{\em Nucl. Phys.}
  {\bfseries B524} (1998) 269--282},
\href{http://arxiv.org/abs/hep-th/9707186}{{\ttfamily arXiv:hep-th/9707186
  [hep-th]}}.
%%CITATION = HEP-TH/9707186;%%.

\bibitem{Acharya:2000gb}
B.~S. Acharya, ``{On Realizing N=1 superYang-Mills in M theory},''
\href{http://arxiv.org/abs/hep-th/0011089}{{\ttfamily arXiv:hep-th/0011089
  [hep-th]}}.
%%CITATION = HEP-TH/0011089;%%.

\bibitem{Gukov:2002jv}
S.~Gukov, S.-T. Yau, and E.~Zaslow, ``{Duality and fibrations on $G_2$
  manifolds},''
\href{http://arxiv.org/abs/hep-th/0203217}{{\ttfamily arXiv:hep-th/0203217
  [hep-th]}}.
%%CITATION = HEP-TH/0203217;%%.

\bibitem{Roiban:2002iv}
R.~Roiban, C.~Romelsberger, and J.~Walcher, ``{Discrete torsion in singular
  G(2) manifolds and real LG},'' {\em Adv. Theor. Math. Phys.} {\bfseries 6}
  (2003) 207--278,
\href{http://arxiv.org/abs/hep-th/0203272}{{\ttfamily arXiv:hep-th/0203272
  [hep-th]}}.
%%CITATION = HEP-TH/0203272;%%.

\bibitem{Gaberdiel:2004vx}
M.~R. Gaberdiel and P.~Kaste, ``{Generalized discrete torsion and mirror
  symmetry for g(2) manifolds},''
  \href{http://dx.doi.org/10.1088/1126-6708/2004/08/001}{{\em JHEP} {\bfseries
  08} (2004) 001},
\href{http://arxiv.org/abs/hep-th/0401125}{{\ttfamily arXiv:hep-th/0401125
  [hep-th]}}.
%%CITATION = HEP-TH/0401125;%%.

\bibitem{deBoer:2005pt}
J.~de~Boer, A.~Naqvi, and A.~Shomer, ``{The Topological G(2) string},''
  \href{http://dx.doi.org/10.4310/ATMP.2008.v12.n2.a2}{{\em Adv. Theor. Math.
  Phys.} {\bfseries 12} no.~2, (2008) 243--318},
\href{http://arxiv.org/abs/hep-th/0506211}{{\ttfamily arXiv:hep-th/0506211
  [hep-th]}}.
%%CITATION = HEP-TH/0506211;%%.

\bibitem{Becker:2014rea}
K.~Becker, D.~Robbins, and E.~Witten, ``{The $\alpha'$ Expansion On A Compact
  Manifold Of Exceptional Holonomy},''
  \href{http://dx.doi.org/10.1007/JHEP06(2014)051}{{\em JHEP} {\bfseries 06}
  (2014) 051},
\href{http://arxiv.org/abs/1404.2460}{{\ttfamily arXiv:1404.2460 [hep-th]}}.
%%CITATION = ARXIV:1404.2460;%%.

\bibitem{MR2024648}
A.~Kovalev, ``Twisted connected sums and special {R}iemannian holonomy,''
  \href{http://dx.doi.org/10.1515/crll.2003.097}{{\em J. Reine Angew. Math.}
  {\bfseries 565} (2003) 125--160}.
  \url{http://dx.doi.org/10.1515/crll.2003.097}.

\bibitem{MR3109862}
A.~Corti, M.~Haskins, J.~Nordstr{\"o}m, and T.~Pacini, ``Asymptotically
  cylindrical {C}alabi-{Y}au 3-folds from weak {F}ano 3-folds,''
  \href{http://dx.doi.org/10.2140/gt.2013.17.1955}{{\em Geom. Topol.}
  {\bfseries 17} no.~4, (2013) 1955--2059}.
  \url{http://dx.doi.org/10.2140/gt.2013.17.1955}.

\bibitem{Corti:2012kd}
A.~Corti, M.~Haskins, J.~Nordström, and T.~Pacini,
  ``{$\mathrm{G}_{2}$-manifolds and associative submanifolds via semi-Fano
  $3$-folds},'' \href{http://dx.doi.org/10.1215/00127094-3120743}{{\em Duke
  Math. J.} {\bfseries 164} no.~10, (2015) 1971--2092},
\href{http://arxiv.org/abs/1207.4470}{{\ttfamily arXiv:1207.4470 [math.DG]}}.
%%CITATION = ARXIV:1207.4470;%%.

\bibitem{Halverson:2014tya}
J.~Halverson and D.~R. Morrison, ``{The landscape of M-theory compactifications
  on seven-manifolds with G$_{2}$ holonomy},''
  \href{http://dx.doi.org/10.1007/JHEP04(2015)047}{{\em JHEP} {\bfseries 04}
  (2015) 047},
\href{http://arxiv.org/abs/1412.4123}{{\ttfamily arXiv:1412.4123 [hep-th]}}.
%%CITATION = ARXIV:1412.4123;%%.

\bibitem{Halverson:2015vta}
J.~Halverson and D.~R. Morrison, ``{On gauge enhancement and singular limits in
  G$_{2}$ compactifications of M-theory},''
  \href{http://dx.doi.org/10.1007/JHEP04(2016)100}{{\em JHEP} {\bfseries 04}
  (2016) 100},
\href{http://arxiv.org/abs/1507.05965}{{\ttfamily arXiv:1507.05965 [hep-th]}}.
%%CITATION = ARXIV:1507.05965;%%.

\bibitem{Braun:2017ryx}
A.~P. Braun and M.~Del~Zotto, ``{Mirror Symmetry for $G_2$-Manifolds: Twisted
  Connected Sums and Dual Tops},''
  \href{http://dx.doi.org/10.1007/JHEP05(2017)080}{{\em JHEP} {\bfseries 05}
  (2017) 080},
\href{http://arxiv.org/abs/1701.05202}{{\ttfamily arXiv:1701.05202 [hep-th]}}.
%%CITATION = ARXIV:1701.05202;%%.

\bibitem{Guio:2017zfn}
T.~C. d.~C. Guio, H.~Jockers, A.~Klemm, and H.-Y. Yeh, ``{Effective action from
  M-theory on twisted connected sum $G_2$-manifolds},''
\href{http://arxiv.org/abs/1702.05435}{{\ttfamily arXiv:1702.05435 [hep-th]}}.
%%CITATION = ARXIV:1702.05435;%%.

\bibitem{Braun:2017uku}
A.~P. Braun and S.~Schafer-Nameki, ``{Compact, Singular G2-Holonomy Manifolds
  and M/Heterotic/F-Theory Duality},''
\href{http://arxiv.org/abs/1708.07215}{{\ttfamily arXiv:1708.07215 [hep-th]}}.
%%CITATION = ARXIV:1708.07215;%%.

\bibitem{joyce1996I}
D.~D. Joyce, ``Compact riemannian 7-manifolds with holonomy $g_2$. i,'' {\em J.
  Differential Geom.} {\bfseries 43} no.~2, (1996) 291--328.

\bibitem{joyce1996II}
D.~D. Joyce, ``Compact riemannian 7-manifolds with holonomy $g\sb 2$. ii,''
  {\em J. Differential Geom.} {\bfseries 43} no.~2, (1996) 329--375.

\bibitem{joyce2000compact}
D.~Joyce, {\em Compact Manifolds with Special Holonomy}.
\newblock Oxford mathematical monographs. Oxford University Press, 2000.

\bibitem{Braun:2016igl}
A.~P. Braun, ``{Tops as building blocks for G$_{2}$ manifolds},''
  \href{http://dx.doi.org/10.1007/JHEP10(2017)083}{{\em JHEP} {\bfseries 10}
  (2017) 083},
\href{http://arxiv.org/abs/1602.03521}{{\ttfamily arXiv:1602.03521 [hep-th]}}.
%%CITATION = ARXIV:1602.03521;%%.

\bibitem{Affleck:1982as}
I.~Affleck, J.~A. Harvey, and E.~Witten, ``{Instantons and (Super)Symmetry
  Breaking in (2+1)-Dimensions},''
\href{http://dx.doi.org/10.1016/0550-3213(82)90277-2}{{\em Nucl. Phys.}
  {\bfseries B206} (1982) 413--439}.
%%CITATION = NUPHA,B206,413;%%.

\bibitem{Katz:1996th}
S.~H. Katz and C.~Vafa, ``{Geometric engineering of N=1 quantum field
  theories},'' \href{http://dx.doi.org/10.1016/S0550-3213(97)00283-6}{{\em
  Nucl. Phys.} {\bfseries B497} (1997) 196--204},
\href{http://arxiv.org/abs/hep-th/9611090}{{\ttfamily arXiv:hep-th/9611090
  [hep-th]}}.
%%CITATION = HEP-TH/9611090;%%.

\bibitem{deBoer:1997ka}
J.~de~Boer, K.~Hori, Y.~Oz, and Z.~Yin, ``{Branes and mirror symmetry in N=2
  supersymmetric gauge theories in three-dimensions},''
  \href{http://dx.doi.org/10.1016/S0550-3213(97)00444-6}{{\em Nucl. Phys.}
  {\bfseries B502} (1997) 107--124},
\href{http://arxiv.org/abs/hep-th/9702154}{{\ttfamily arXiv:hep-th/9702154
  [hep-th]}}.
%%CITATION = HEP-TH/9702154;%%.

\bibitem{Nicolai:2003bp}
H.~Nicolai and H.~Samtleben, ``{Chern-Simons versus Yang-Mills gaugings in
  three-dimensions},''
  \href{http://dx.doi.org/10.1016/S0550-3213(03)00569-8}{{\em Nucl. Phys.}
  {\bfseries B668} (2003) 167--178},
\href{http://arxiv.org/abs/hep-th/0303213}{{\ttfamily arXiv:hep-th/0303213
  [hep-th]}}.
%%CITATION = HEP-TH/0303213;%%.

\bibitem{Gukov:2002er}
S.~Gukov and D.~Tong, ``{D-brane probes of G(2) holonomy manifolds},''
  \href{http://dx.doi.org/10.1103/PhysRevD.66.087901}{{\em Phys. Rev.}
  {\bfseries D66} (2002) 087901},
\href{http://arxiv.org/abs/hep-th/0202125}{{\ttfamily arXiv:hep-th/0202125
  [hep-th]}}.
%%CITATION = HEP-TH/0202125;%%.

\bibitem{Gukov:2002es}
S.~Gukov and D.~Tong, ``{D-brane probes of special holonomy manifolds, and
  dynamics of N = 1 three-dimensional gauge theories},''
  \href{http://dx.doi.org/10.1088/1126-6708/2002/04/050}{{\em JHEP} {\bfseries
  04} (2002) 050},
\href{http://arxiv.org/abs/hep-th/0202126}{{\ttfamily arXiv:hep-th/0202126
  [hep-th]}}.
%%CITATION = HEP-TH/0202126;%%.

\bibitem{delaOssa:2016ivz}
X.~de~la Ossa, M.~Larfors, and E.~E. Svanes, ``{Infinitesimal moduli of G2
  holonomy manifolds with instanton bundles},''
  \href{http://dx.doi.org/10.1007/JHEP11(2016)016}{{\em JHEP} {\bfseries 11}
  (2016) 016},
\href{http://arxiv.org/abs/1607.03473}{{\ttfamily arXiv:1607.03473 [hep-th]}}.
%%CITATION = ARXIV:1607.03473;%%.

\bibitem{delaOssa:2017pqy}
X.~de~la Ossa, M.~Larfors, and E.~E. Svanes, ``{The infinitesimal moduli space
  of heterotic $G_2$ systems},''
  \href{http://dx.doi.org/10.1007/s00220-017-3013-8}{{\em Commun. Math. Phys.}
  (2017) },
\href{http://arxiv.org/abs/1704.08717}{{\ttfamily arXiv:1704.08717 [hep-th]}}.
%%CITATION = ARXIV:1704.08717;%%.

\bibitem{delaOssa:2017gjq}
X.~de~la Ossa, M.~Larfors, and E.~E. Svanes, ``{Restrictions of Heterotic $G_2$
  Structures and Instanton Connections},''
\newblock 2017.
\newblock \href{http://arxiv.org/abs/1709.06974}{{\ttfamily arXiv:1709.06974
  [math.DG]}}.
\newblock
\url{https://inspirehep.net/record/1624570/files/arXiv:1709.06974.pdf}.
\newblock
%%CITATION = ARXIV:1709.06974;%%.

\bibitem{Fiset:2017auc}
M.-A. Fiset, C.~Quigley, and E.~E. Svanes, ``{Marginal deformations of
  heterotic $G_2$ sigma models},''
\href{http://arxiv.org/abs/1710.06865}{{\ttfamily arXiv:1710.06865 [hep-th]}}.
%%CITATION = ARXIV:1710.06865;%%.

\bibitem{Bergshoeff:2001pv}
E.~Bergshoeff, R.~Kallosh, T.~Ortin, D.~Roest, and A.~Van~Proeyen, ``{New
  formulations of D = 10 supersymmetry and D8 - O8 domain walls},''
  \href{http://dx.doi.org/10.1088/0264-9381/18/17/303}{{\em Class. Quant.
  Grav.} {\bfseries 18} (2001) 3359--3382},
\href{http://arxiv.org/abs/hep-th/0103233}{{\ttfamily arXiv:hep-th/0103233
  [hep-th]}}.
%%CITATION = HEP-TH/0103233;%%.

\bibitem{Papadopoulos:1995da}
G.~Papadopoulos and P.~K. Townsend, ``{Compactification of D = 11 supergravity
  on spaces of exceptional holonomy},''
  \href{http://dx.doi.org/10.1016/0370-2693(95)00929-F}{{\em Phys. Lett.}
  {\bfseries B357} (1995) 300--306},
\href{http://arxiv.org/abs/hep-th/9506150}{{\ttfamily arXiv:hep-th/9506150
  [hep-th]}}.
%%CITATION = HEP-TH/9506150;%%.

\bibitem{Bershadsky:1996nh}
M.~Bershadsky, K.~A. Intriligator, S.~Kachru, D.~R. Morrison, V.~Sadov, and
  C.~Vafa, ``{Geometric singularities and enhanced gauge symmetries},''
  \href{http://dx.doi.org/10.1016/S0550-3213(96)90131-5}{{\em Nucl. Phys.}
  {\bfseries B481} (1996) 215--252},
\href{http://arxiv.org/abs/hep-th/9605200}{{\ttfamily arXiv:hep-th/9605200
  [hep-th]}}.
%%CITATION = HEP-TH/9605200;%%.

\bibitem{Acharya:2001gy}
B.~S. Acharya and E.~Witten, ``{Chiral fermions from manifolds of G(2)
  holonomy},''
\href{http://arxiv.org/abs/hep-th/0109152}{{\ttfamily arXiv:hep-th/0109152
  [hep-th]}}.
%%CITATION = HEP-TH/0109152;%%.

\bibitem{Witten:2001uq}
E.~Witten, ``{Anomaly cancellation on G(2) manifolds},''
\href{http://arxiv.org/abs/hep-th/0108165}{{\ttfamily arXiv:hep-th/0108165
  [hep-th]}}.
%%CITATION = HEP-TH/0108165;%%.

\bibitem{Katz:1997eq}
S.~Katz, P.~Mayr, and C.~Vafa, ``{Mirror symmetry and exact solution of 4-D N=2
  gauge theories: 1.},'' {\em Adv. Theor. Math. Phys.} {\bfseries 1} (1998)
  53--114,
\href{http://arxiv.org/abs/hep-th/9706110}{{\ttfamily arXiv:hep-th/9706110
  [hep-th]}}.
%%CITATION = HEP-TH/9706110;%%.

\bibitem{Cecotti:1988qn}
S.~Cecotti, S.~Ferrara, and L.~Girardello, ``{Geometry of Type II Superstrings
  and the Moduli of Superconformal Field Theories},''
\href{http://dx.doi.org/10.1142/S0217751X89000972}{{\em Int. J. Mod. Phys.}
  {\bfseries A4} (1989) 2475}.
%%CITATION = IMPAE,A4,2475;%%.

\bibitem{Hitchin:1986ea}
N.~J. Hitchin, A.~Karlhede, U.~Lindstrom, and M.~Rocek, ``{Hyperkahler Metrics
  and Supersymmetry},''
\href{http://dx.doi.org/10.1007/BF01214418}{{\em Commun. Math. Phys.}
  {\bfseries 108} (1987) 535}.
%%CITATION = CMPHA,108,535;%%.

\bibitem{Batyrev:1994hm}
V.~V. Batyrev, ``{Dual polyhedra and mirror symmetry for Calabi-Yau
  hypersurfaces in toric varieties},'' {\em J. Alg. Geom.} {\bfseries 3} (1994)
  493--545,
\href{http://arxiv.org/abs/alg-geom/9310003}{{\ttfamily arXiv:alg-geom/9310003
  [alg-geom]}}.
%%CITATION = ALG-GEOM/9310003;%%.

\bibitem{Vafa:1986wx}
C.~Vafa, ``{Modular Invariance and Discrete Torsion on Orbifolds},''
\href{http://dx.doi.org/10.1016/0550-3213(86)90379-2}{{\em Nucl. Phys.}
  {\bfseries B273} (1986) 592--606}.
%%CITATION = NUPHA,B273,592;%%.

\bibitem{Harvey:1982xk}
R.~Harvey and H.~B. Lawson, Jr., ``{Calibrated geometries},''
\href{http://dx.doi.org/10.1007/BF02392726}{{\em Acta Math.} {\bfseries 148}
  (1982) 47}.
%%CITATION = ACMTA,148,47;%%.

\bibitem{Mclean96}
R.~C. Mclean, ``Deformations of calibrated submanifolds,'' {\em Commun. Analy.
  Geom} {\bfseries 6} (1996) 705--747.

\bibitem{Morrison:2010vf}
D.~R. Morrison, ``{On the structure of supersymmetric T**3 fibrations},''
\href{http://arxiv.org/abs/1002.4921}{{\ttfamily arXiv:1002.4921 [math.AG]}}.
%%CITATION = ARXIV:1002.4921;%%.

\bibitem{Gross:2012rw}
M.~Gross, ``{Mirror symmetry and the Strominger-Yau-Zaslow conjecture},''
  \href{http://dx.doi.org/10.4310/CDM.2012.v2012.n1.a3}{{\em Current
  Developments in Mathematics} {\bfseries 1} (2012) 133--191},
  \href{http://arxiv.org/abs/1212.4220}{{\ttfamily arXiv:1212.4220 [math.AG]}}.

\bibitem{Aspinwall:1995rb}
P.~S. Aspinwall, D.~R. Morrison, and M.~Gross, ``{Stable singularities in
  string theory},'' \href{http://dx.doi.org/10.1007/BF02104911}{{\em Commun.
  Math. Phys.} {\bfseries 178} (1996) 115--134},
\href{http://arxiv.org/abs/hep-th/9503208}{{\ttfamily arXiv:hep-th/9503208
  [hep-th]}}.
%%CITATION = HEP-TH/9503208;%%.

\bibitem{1998math......9072G}
M.~{Gross}, ``{Special Lagrangian Fibrations II: Geometry},'' {\em ArXiv
  Mathematics e-prints} (1998) ,
  \href{http://arxiv.org/abs/math/9809072}{{\ttfamily math/9809072}}.

\bibitem{Aspinwall:1994rg}
P.~S. Aspinwall and D.~R. Morrison, ``{String theory on K3 surfaces},''
\href{http://arxiv.org/abs/hep-th/9404151}{{\ttfamily arXiv:hep-th/9404151
  [hep-th]}}.
%%CITATION = HEP-TH/9404151;%%.

\bibitem{MR525944}
V.~V. Nikulin, ``Integer symmetric bilinear forms and some of their geometric
  applications,'' {\em Izv. Akad. Nauk SSSR Ser. Mat.} {\bfseries 43} no.~1,
  (1979) 111--177, 238.

\bibitem{Becker:1996ay}
K.~Becker, M.~Becker, D.~R. Morrison, H.~Ooguri, Y.~Oz, and Z.~Yin,
  ``{Supersymmetric cycles in exceptional holonomy manifolds and Calabi-Yau 4
  folds},'' \href{http://dx.doi.org/10.1016/S0550-3213(96)00491-9}{{\em Nucl.
  Phys.} {\bfseries B480} (1996) 225--238},
\href{http://arxiv.org/abs/hep-th/9608116}{{\ttfamily arXiv:hep-th/9608116
  [hep-th]}}.
%%CITATION = HEP-TH/9608116;%%.

\bibitem{Blumenhagen:2001jb}
R.~Blumenhagen and V.~Braun, ``{Superconformal field theories for compact G(2)
  manifolds},'' \href{http://dx.doi.org/10.1088/1126-6708/2001/12/006}{{\em
  JHEP} {\bfseries 12} (2001) 006},
\href{http://arxiv.org/abs/hep-th/0110232}{{\ttfamily arXiv:hep-th/0110232
  [hep-th]}}.
%%CITATION = HEP-TH/0110232;%%.

\bibitem{Roiban:2001cp}
R.~Roiban and J.~Walcher, ``{Rational conformal field theories with G(2)
  holonomy},'' \href{http://dx.doi.org/10.1088/1126-6708/2001/12/008}{{\em
  JHEP} {\bfseries 12} (2001) 008},
\href{http://arxiv.org/abs/hep-th/0110302}{{\ttfamily arXiv:hep-th/0110302
  [hep-th]}}.
%%CITATION = HEP-TH/0110302;%%.

\bibitem{Gutowski:2001fm}
J.~Gutowski and G.~Papadopoulos, ``{Moduli spaces and brane solitons for M
  theory compactifications on holonomy G(2) manifolds},''
  \href{http://dx.doi.org/10.1016/S0550-3213(01)00419-9}{{\em Nucl. Phys.}
  {\bfseries B615} (2001) 237--265},
\href{http://arxiv.org/abs/hep-th/0104105}{{\ttfamily arXiv:hep-th/0104105
  [hep-th]}}.
%%CITATION = HEP-TH/0104105;%%.

\bibitem{Atiyah:2001qf}
M.~Atiyah and E.~Witten, ``{M theory dynamics on a manifold of G(2)
  holonomy},'' {\em Adv. Theor. Math. Phys.} {\bfseries 6} (2003) 1--106,
\href{http://arxiv.org/abs/hep-th/0107177}{{\ttfamily arXiv:hep-th/0107177
  [hep-th]}}.
%%CITATION = HEP-TH/0107177;%%.

\bibitem{Cecotti:2010fi}
S.~Cecotti, A.~Neitzke, and C.~Vafa, ``{R-Twisting and 4d/2d
  Correspondences},''
\href{http://arxiv.org/abs/1006.3435}{{\ttfamily arXiv:1006.3435 [hep-th]}}.
%%CITATION = ARXIV:1006.3435;%%.

\bibitem{Cecotti:2012gh}
S.~Cecotti and M.~Del~Zotto, ``{4d N=2 Gauge Theories and Quivers: the
  Non-Simply Laced Case},''
  \href{http://dx.doi.org/10.1007/JHEP10(2012)190}{{\em JHEP} {\bfseries 10}
  (2012) 190},
\href{http://arxiv.org/abs/1207.7205}{{\ttfamily arXiv:1207.7205 [hep-th]}}.
%%CITATION = ARXIV:1207.7205;%%.

\bibitem{Fiol:2000pd}
B.~Fiol, ``{The BPS spectrum of N=2 SU(N) SYM and parton branes},''
  \href{http://dx.doi.org/10.1088/1126-6708/2006/02/065}{{\em JHEP} {\bfseries
  02} (2006) 065},
\href{http://arxiv.org/abs/hep-th/0012079}{{\ttfamily arXiv:hep-th/0012079
  [hep-th]}}.
%%CITATION = HEP-TH/0012079;%%.

\bibitem{Cecotti:2012va}
S.~Cecotti, ``{Categorical Tinkertoys for N=2 Gauge Theories},''
  \href{http://dx.doi.org/10.1142/S0217751X13300068}{{\em Int. J. Mod. Phys.}
  {\bfseries A28} (2013) 1330006},
\href{http://arxiv.org/abs/1203.6734}{{\ttfamily arXiv:1203.6734 [hep-th]}}.
%%CITATION = ARXIV:1203.6734;%%.

\bibitem{Katz:1996xe}
S.~H. Katz and C.~Vafa, ``{Matter from geometry},''
  \href{http://dx.doi.org/10.1016/S0550-3213(97)00280-0}{{\em Nucl. Phys.}
  {\bfseries B497} (1997) 146--154},
\href{http://arxiv.org/abs/hep-th/9606086}{{\ttfamily arXiv:hep-th/9606086
  [hep-th]}}.
%%CITATION = HEP-TH/9606086;%%.

\bibitem{Cecotti:2013lda}
S.~Cecotti, M.~Del~Zotto, and S.~Giacomelli, ``{More on the N=2 superconformal
  systems of type $D_p(G)$},''
  \href{http://dx.doi.org/10.1007/JHEP04(2013)153}{{\em JHEP} {\bfseries 04}
  (2013) 153},
\href{http://arxiv.org/abs/1303.3149}{{\ttfamily arXiv:1303.3149 [hep-th]}}.
%%CITATION = ARXIV:1303.3149;%%.

\bibitem{Kovalev2010}
A.~Kovalev and J.~Nordstr{\"o}m, ``Asymptotically cylindrical 7-manifolds of
  holonomy g2 with applications to compact irreducible g2-manifolds,''
  \href{http://dx.doi.org/10.1007/s10455-010-9210-8}{{\em Annals of Global
  Analysis and Geometry} {\bfseries 38} no.~3, (2010) 221--257},
  \href{http://arxiv.org/abs/0907.0497}{{\ttfamily arXiv:0907.0497 [math.DG]}}.
  \url{http://dx.doi.org/10.1007/s10455-010-9210-8}.

\bibitem{nordstroem_thesis}
J.~Nordstr{\"o}m, {\em "Deformations and gluing of asymptotically cylindrical
  manifolds with exceptional holonomy"}.
\newblock Ph.D. Thesis. Cambridge University, 2008.

\bibitem{Curio:1997rn}
G.~Curio and D.~Lust, ``{A Class of N=1 dual string pairs and its modular
  superpotential},'' \href{http://dx.doi.org/10.1142/S0217751X97003066}{{\em
  Int. J. Mod. Phys.} {\bfseries A12} (1997) 5847--5866},
\href{http://arxiv.org/abs/hep-th/9703007}{{\ttfamily arXiv:hep-th/9703007
  [hep-th]}}.
%%CITATION = HEP-TH/9703007;%%.

\bibitem{Nikulin86discretereflection}
V.~V. Nikulin, ``Discrete reflection groups in lobachevsky spaces and algebraic
  surfaces,'' in {\em IN: PROCEDINGS OF THE INTERNATIONAL CONGRESS OF
  MATHEMATICIANS}, pp.~654--671.
\newblock 1986.

\bibitem{Nahm:1999ps}
W.~Nahm and K.~Wendland, ``{A Hiker's guide to K3: Aspects of N=(4,4)
  superconformal field theory with central charge c = 6},''
  \href{http://dx.doi.org/10.1007/PL00005548}{{\em Commun. Math. Phys.}
  {\bfseries 216} (2001) 85--138},
\href{http://arxiv.org/abs/hep-th/9912067}{{\ttfamily arXiv:hep-th/9912067
  [hep-th]}}.
%%CITATION = HEP-TH/9912067;%%.

\bibitem{Braun:2009wh}
A.~P. Braun, R.~Ebert, A.~Hebecker, and R.~Valandro, ``{Weierstrass meets
  Enriques},'' \href{http://dx.doi.org/10.1007/JHEP02(2010)077}{{\em JHEP}
  {\bfseries 02} (2010) 077},
\href{http://arxiv.org/abs/0907.2691}{{\ttfamily arXiv:0907.2691 [hep-th]}}.
%%CITATION = ARXIV:0907.2691;%%.

\bibitem{conway1998sphere}
J.~Conway and N.~Sloane, {\em Sphere Packings, Lattices and Groups}.
\newblock Grundlehren der mathematischen Wissenschaften. Springer New York,
  1998.

\bibitem{Candelas:1996su}
P.~Candelas and A.~Font, ``{Duality between the webs of heterotic and type II
  vacua},'' \href{http://dx.doi.org/10.1016/S0550-3213(96)00410-5}{{\em
  Nucl.Phys.} {\bfseries B511} (1998) 295--325},
\href{http://arxiv.org/abs/hep-th/9603170}{{\ttfamily arXiv:hep-th/9603170
  [hep-th]}}.
%%CITATION = HEP-TH/9603170;%%.

\bibitem{Perevalov:1997vw}
E.~Perevalov and H.~Skarke, ``{Enhanced gauged symmetry in type II and F theory
  compactifications: Dynkin diagrams from polyhedra},''
  \href{http://dx.doi.org/10.1016/S0550-3213(97)00477-X}{{\em Nucl. Phys.}
  {\bfseries B505} (1997) 679--700},
\href{http://arxiv.org/abs/hep-th/9704129}{{\ttfamily arXiv:hep-th/9704129
  [hep-th]}}.
%%CITATION = HEP-TH/9704129;%%.

\bibitem{Kulikov}
V.~S. Kulikov, ``Degenerations of k 3 surfaces and enriques surfaces,'' {\em
  Mathematics of the USSR-Izvestiya} {\bfseries 11} no.~5, (1977) 957.

\bibitem{Persson_Pinkham}
U.~Persson and H.~Pinkham, ``Degeneration of surfaces with trivial canonical
  bundle,'' {\em Annals of Mathematics} {\bfseries 113} no.~1, (1981) 45--66.

\bibitem{1307.6514}
R.~Davis, C.~Doran, A.~Gewiss, A.~Novoseltsev, D.~Skjorshammer, A.~Syryczuk,
  and U.~Whitcher, ``Short tops and semistable degenerations,''
  \href{http://arxiv.org/abs/arXiv:1307.6514}{{\ttfamily arXiv:1307.6514}}.

\bibitem{Braun:2016sks}
A.~P. Braun and T.~Watari, ``{Heterotic-Type IIA Duality and Degenerations of
  K3 Surfaces},'' \href{http://dx.doi.org/10.1007/JHEP08(2016)034}{{\em JHEP}
  {\bfseries 08} (2016) 034},
\href{http://arxiv.org/abs/1604.06437}{{\ttfamily arXiv:1604.06437 [hep-th]}}.
%%CITATION = ARXIV:1604.06437;%%.

\bibitem{2002math.....10219H}
D.~{Huybrechts}, ``{Moduli spaces of hyperkaehler manifolds and mirror
  symmetry},'' {\em ArXiv Mathematics e-prints} (Oct., 2002) ,
  \href{http://arxiv.org/abs/math/0210219}{{\ttfamily math/0210219}}.

\end{thebibliography}\endgroup

\end{document}